\documentclass[a4paper,11pt]{article}
\usepackage[utf8]{inputenc}
\usepackage{jheppub}

\usepackage{amsmath,amssymb,bm,mathtools,mathrsfs}
\usepackage{booktabs}
\usepackage{adjustbox}
\usepackage{needspace}
\usepackage{placeins}
\usepackage{float}
\usepackage{xcolor}
\usepackage{tikz}
\usepackage{physics}
\usepackage{cancel}
\usetikzlibrary{positioning}
\hypersetup{colorlinks=true,linkcolor=blue,citecolor=blue,urlcolor=blue}

\newcommand{\ii}{\mathrm{i}}
\newcommand{\ee}{\mathrm{e}}
\newcommand{\eps}{\epsilon}
\newcommand{\one}{\mathbf{1}}
\allowdisplaybreaks[3]
\renewcommand{\afterTocRuleSpace}{\bigskip\bigskip\clearpage}

\title{Canonical Differential Equations of Two-site Bubble Cosmological Correlator Family}

\author[a,b]{Jiaqi Chen}
\affiliation[a]{Beijing Key Laboratory of Optical Detection Technology for Oil and Gas, China University of Petroleum-Beijing, Beijing 102249, China}
\affiliation[b]{Basic Research Center for Energy Interdisciplinary, College of Science, China University of Petroleum-Beijing, Beijing 102249, China}
\emailAdd{jiaqichen@cup.edu.cn}
\author[c,d,e]{Bo Feng}
\affiliation[c]{State Key Laboratory of Nuclear Physics and Technology, Institute of Quantum Matter, South China Normal University, Guangzhou 510006, China}
\affiliation[d]{Guangdong Basic Research Center of Excellence for Structure and Fundamental Interactions of Matter, Guangdong Provincial Key Laboratory of Nuclear Science, Guangzhou 510006, China}
\affiliation[e]{Peng Huanwu Center for Fundamental Theory, Hefei, Anhui, 230026, China}
\emailAdd{fengbo@scnu.edu.cn}
\author[f]{Yingze Su}
\affiliation[f]{Nagoya University, Department of Physics, Furo-cho, Chikusa-ku, Nagoya 464-8602, Japan}
\emailAdd{su.yingze.v2@s.mail.nagoya-u.ac.jp}

\abstract{
Using the $\mathrm{d}\log$ integrand construction in de Sitter (dS) space, we obtain the canonical
differential equations of the massless bubble correlator family in dimensional regularization,
together with the analytic boundary vector. These determine the expansion of every master integral
in the family in terms of multiple polylogarithms to arbitrary order in $\epsilon$.
}


\begin{document}
\maketitle
\newpage
\section{Introduction}

In cosmological collider physics, new massive particles present during inflation can leave characteristic signatures in the correlation functions of primordial perturbations, providing a possible probe of new physics at the inflationary scale.
These squeezed-limit signatures were established in quasi-single-field inflation models \cite{Chen:2009we,Chen:2009zp,Baumann:2011nk,Noumi:2012vr} and in the framework of cosmological collider physics \cite{Arkani-Hamed:2015bza}, with their phenomenology developed in \cite{Lee:2016vti}.
Correlation functions of primordial fluctuations connect inflationary physics to observations of the cosmic microwave background and large-scale structure, and the corresponding bispectrum and trispectrum shapes are subject to observational constraints \cite{Planck:2019kim}.
Loop diagrams carry further information: they can expose the Standard Model spectrum during inflation, from the electroweak and Higgs thresholds \cite{Chen:2016nrs,Chen:2016uwp,Chen:2016hrz} to the origin of neutrino masses \cite{Chen:2018xck}, and loops of Standard Model fermions can generate oscillatory signals in the inflaton bispectrum \cite{Hook:2019vcn}.
This motivates analytic control over the momentum dependence of cosmological  correlators.

Considerable analytic progress has been made on these correlators and on the wavefunction coefficients of the universe, through the cosmological bootstrap \cite{Arkani-Hamed:2018kmz,Baumann:2019oyu,Baumann:2020dch,Pimentel:2022fsc,Baumann:2022jpr,ChenTao:2023weightshifting}, the cosmological polytope \cite{Arkani-Hamed:2017fdk}, kinematic flow \cite{Arkani-Hamed:2023kig}, boundary differential equations for massive exchanges \cite{Arkani-Hamed:2015bza,Qin:2022fbv,Qin:2023ejc,Jazayeri:2022kjy,Aoki:2024uyi,Liu:2024str,Xianyu:2025lbk,Gasparotto:2024bku,Baumann:2026atn}, Mellin--Barnes representations \cite{Sleight:2019mgd,Sleight:2019hfp,Sleight:2021plv,Qin:2022lva,Xianyu:2023ytd}, dispersive methods \cite{Liu:2024xyi,Liu:2026jzn}, and spectral methods \cite{Xianyu:2022jwk,Marolf:2010zp,Loparco:2023rug,Werth:2024mjg,Belrhali:2026ktb,Grafe:2026qsm,Belrhali:2026rkn,Grafe:2026avi}. Further approaches include the de Sitter S-matrix \cite{Melville:2023kgd,Melville:2024ove}, the Grassmannian kinematic space \cite{Arundine:2026fbr,Arundine:2026myr}, and unitarity and locality constraints \cite{Goodhew:2020hob,Melville:2021lst,Cespedes:2020xqq,AguiSalcedo:2023nds,Das:2025qsh}. Numerical methods include direct Schwinger--Keldysh integration \cite{Wang:2021qez} and the cosmological flow \cite{Pinol:2023oux,Werth:2023pfl,Werth:2024aui}.

Recent results include flat-space benchmarks and de Sitter (dS) loop integrals, which involve both bulk-time and loop-momentum integrations.
Benchmark bubble and triangle integrals were computed for equal-time correlators in flat space \cite{Pimentel:2026oneloop}, and a compact formula was derived for the one-loop triangle correlator of conformally coupled scalars \cite{Henn:2026triangle}, exploiting the reduction of conformally coupled loop integrands to flat-space integrands \cite{Chowdhury:2025simplicity,Heckelbacher:2022hbq}.
For flat-space wavefunction coefficients, elliptic contributions cancel in the triangle but survive in the box \cite{McCune:2026elliptic}.
For massive internal lines, spectral decomposition reconstructs one-loop correlators from tree-level input \cite{Xianyu:2022jwk}, while dispersive methods reconstruct them from their nonlocal signals up to local counterterms \cite{Liu:2024xyi}.
Spectral finite-difference methods have also been developed for dS/AdS loop integrals \cite{Bassett:2026spectral} and applied to the two-loop effective potential in dS \cite{Bassett:2026potential}.
The partial Mellin--Barnes representation has been applied to massive bubble correlators \cite{Qin:2024gtr}, and split representations resum the massive bubble \cite{Grafe:2026qsm}. Exact fermionic bubble signals follow from spectral and Mellin--Barnes methods \cite{Aoki:2026vbc}, and the leading analytical signals of massive triangle loops have been extracted \cite{Qin:2026yuo}. More generally, an on-shell bootstrap with spectral dispersion covers loop diagrams of various topologies \cite{Liu:2026jzn}.
For one-loop correlators with massive exchange, Ref.~\cite{Qin:2023bjk} derives a cutting rule for nonlocal signals and relates their factorization to the operator-product expansion.
Related cutting rules have been derived from bulk evolution \cite{Tong:2021wai} and in the in-in formalism \cite{Ema:2024hkj}. The on-shell factorization of nonlocal signals extends to arbitrary loop order \cite{Qin:2023nhv}.
For massless fields, the infrared structure of loop corrections has been analyzed with the method of regions \cite{Beneke:2023wmt}, and $\lambda\phi^4$ has been explored as an effective theory in de Sitter \cite{Cespedes:2025ple}.
Here we study the full regulated bubble time-integral family and obtain its master integrals in terms of multiple polylogarithms (MPLs).

For flat-space Feynman integrals, a systematic computational framework is available.
Integration-by-parts (IBP) relations \cite{Chetyrkin:1981qh,Laporta:2000dsw} reduce every integral of a family to a finite set of master integrals, and differential equations \cite{Kotikov:1990kg,Kotikov:1991pm,Gehrmann:1999as} compute the masters from boundary data.
For polylogarithmic integrals, $\dd\log$ representations and leading-singularity analysis guide the choice of a basis with canonical $\eps$-factorized differential equations \cite{Henn:2013pwa}, allowing a systematic solution in terms of MPLs.
Together with efficient reduction and evaluation frameworks such as the auxiliary mass flow method \cite{Liu:2017jxz,Liu:2022mfb,Liu:2022chg}, these techniques provide systematic high-precision numerical methods for the Feynman integrals entering scattering amplitudes.

This program has been extended to cosmological correlators.
IBP reduction and differential equations have been applied to conformally coupled scalar integrals in power-law FRW backgrounds, including the dS limit \cite{De:2023xue,Arkani-Hamed:2023kig}. For integrands involving Hankel functions, these methods apply to tree diagrams with massive internal lines \cite{ChenFeng:2024dsibp,Chen:2024glu} and to massive loop integrals \cite{Chen:2026dqp}.
IBP reductions and differential equations for conformally coupled in-in correlators have also been derived in dressed flat-space representations \cite{Sachs:2026dressed}.
Universal iterative reduction and $\dd\log$-form differential-equation formulas for the vertex time integrals were developed in \cite{ChenFeng:2024dsibp,Chen:2024glu}. The MadStree package implements this approach for tree-level correlators and for the time integrations of loop graphs at fixed loop momenta \cite{Chen:2026iwf}.

In this paper, we construct a canonical differential equation (CDE) for the massless bubble time-integral family, including the spatial loop integration and generic vertex powers.
The construction follows flat-space methods. Master integrands with constant leading singularities can be selected a priori \cite{Arkani-Hamed:2010pyv,Arkani-Hamed:2012zlh,Henn:2020lye,DlapaLiZhang:2021UT}, with the Baikov representation \cite{Baikov:1996rk,Baikov:1996iu} providing a convenient construction of $\dd\log$ integrands \cite{Chen:2020uyk,Chen:2022lzr}. Intersection theory \cite{Mastrolia:2018uzb,Frellesvig:2019uqt,Frellesvig:2020qot,Mizera:2019vvs,Weinzierl:2020xyy,Caron-Huot:2021xqj} explains why such integrands lead to canonical differential equations and gives the selection rules of the CDE matrix \cite{Chen:2023kgw,Chen:2024ovh}.
Motivated by fibration intersection theory, Ref.~\cite{Chen:2026dqp} proposed extending this Baikov construction to dS integrands involving Hankel functions and verified it for the one-loop bubble family.
For the massless two-site bubble, we determine both the canonical system and its analytic boundary data. Together they fix the master integrals order by order in $\eps$.
For a single time ordering, mapping the topsector to an auxiliary momentum-integral family simplifies the $\dd\log$ construction.
The resulting single-$\theta$ basis gives the dual-$\theta$ canonical basis and differential equation, presented in Section~\ref{sec:dual-theta}.

We study the two-site one-loop bubble directly in its time-integral representation, following \cite{ChenFeng:2024dsibp} in introducing analytic regulators for the time powers so that the IBP relations can be derived with controlled endpoint contributions.
We use $d=3-2\eps$ for the spatial loop integration and parametrize the total time-power shifts by $\rho_1=\rho_2=\delta\eps$, with $\delta$ independent of $\eps$.
We determine the boundary vector from asymptotic regions of the original integrals and obtain the $\eps$ expansion of every master integral in the family to arbitrary order in terms of MPLs through iterated integration \cite{Chen:1977ena}.

The paper is organized as follows.
Section~\ref{sec:background} reviews the scalar modes and propagators
used in the calculation, and Section~\ref{sec:bubble-setup} defines
the two-site bubble and its kinematics.
Section~\ref{sec:pm-companion} studies the mixed-sign component as
a preliminary example of the logarithmic-basis construction.
In Section~\ref{sec:pp-sector}, we derive the IBP relations for the
ordered component and construct its canonical differential equation.
Section~\ref{sec:pp-solution} determines the boundary vector,
gives the MPL solution and presents the numerical checks.
Section~\ref{sec:dual-theta} converts these results to the dual-$\theta$
family and its parity-blocked canonical system.
Section~\ref{sec:physical-integrals} applies the integral-family solution
to the conformally coupled $\phi^4$ and $\phi^3$ bubbles and to the
massless minimally coupled cubic bubble. The sum over Schwinger--Keldysh
components cancels the time-regulator poles in the conformal $\phi^3$
example; these poles remain in the massless example.
We conclude in Section~\ref{sec:conclusions}.
The appendices collect the IBP identities, dimension-recurrence derivations,
connection matrices and higher-weight MPL coefficients.

\section{Review of cosmological correlators}
\label{sec:background}

We consider a self-interacting massless scalar field with curvature coupling
$\xi$ in a spatially flat FLRW background. The action is
\begin{equation}
 S[\phi]=-
 \int\dd^{d}x \,\dd \tau\sqrt{-g}\left[
 \frac12(\partial_\mu\phi)^2
 +\frac12\xi R\phi^2
 +\sum_{n\geq3}\frac{\lambda_n}{n!}\phi^n
 \right],
 \label{frw-scalar-action}
\end{equation}
where the $(d+1)$-dimensional metric is
\begin{equation}
 \dd s^2=a^2(\tau)\left(-\dd\tau^2+\dd\bm x^2\right),
 \label{frw-conformal-metric}
\end{equation}
with scale factor $a(\tau)$.

Let $f_k(\tau)$ denote the mode function of $\phi$.  The quadratic part of Eq.~\eqref{frw-scalar-action} implies that the rescaled mode $u_k=a^{(d-1)/2}f_k$ obeys
\begin{equation}
 u_k''+\left[k^2+(\xi-\xi_c)a^2R\right]u_k=0,
 \qquad
 \xi_c=\frac{d-1}{4d},
 \label{eq:rescaled-mode}
\end{equation}
where $R$ is the Ricci scalar and primes denote conformal-time derivatives.

We now specialize Eq.~\eqref{frw-conformal-metric} to de Sitter space,
\begin{equation}
 a(\tau)=-\frac{1}{H\tau},
 \qquad -\infty<\tau<0,
\end{equation}
where $H$ is the Hubble parameter and $R=d(d+1)H^2$. For the original scalar field, write
\begin{equation}
 \phi(\tau,\bm x)
 =\int\frac{\dd^d\bm k}{(2\pi)^d}
 \left[
 f_k(\tau)b_{\bm k}\ee^{\ii\bm k\cdot\bm x}
 +f_k^*(\tau)b_{\bm k}^{\dagger}\ee^{-\ii\bm k\cdot\bm x}
 \right].
\end{equation}
The massless mode equation following from Eq.~\eqref{frw-scalar-action} is
\begin{equation}
 f_k''(\tau)-\frac{d-1}{\tau}f_k'(\tau)
 +\left[k^2+\frac{\xi d(d+1)}{\tau^2}\right]f_k(\tau)=0.
\end{equation}
The positive-frequency condition at early times selects the Bunch--Davies mode, which can be written, up to a constant phase, as \cite{ChenWangXianyu:2017sk}
\begin{equation}
\begin{aligned}
 f_{k,\nu}(\tau)
 &=-\ii\frac{\sqrt{\pi}}{2}
 \ee^{\ii\pi(\nu/2+1/4)}
 H^{(d-1)/2}(-\tau)^{d/2}H_{\nu}^{(1)}(-k\tau),
 \\
 \nu^2&=\frac{d^2}{4}-\xi d(d+1).
\end{aligned}
 \label{ds-general-mode}
\end{equation}
For any free mode, the Wightman functions are
\begin{align}
 G_{>}(k;\tau_1,\tau_2)
 &=f_k(\tau_1)f_k^*(\tau_2),
 \nonumber\\
 G_{<}(k;\tau_1,\tau_2)
 &=f_k^*(\tau_1)f_k(\tau_2).
 \label{mode-wightman}
\end{align}
We consider the minimally coupled and conformally coupled scalars.

\paragraph{Massless minimally coupled scalar}
For $\xi=0$, one has $\nu=d/2$. In four spacetime dimensions $\nu=3/2$, and the mode and Wightman function reduce to
\begin{align}
 f_k(\tau)
 &=\frac{H}{\sqrt{2k^3}}(1+\ii k\tau)\ee^{-\ii k\tau},
 \\
 G_{>}(k;\tau_1,\tau_2)
 &=\frac{H^2}{2k^3}
 (1+\ii k\tau_1)(1-\ii k\tau_2)
 \ee^{-\ii k(\tau_1-\tau_2)}.
 \label{massless-prop}
\end{align}

\paragraph{Conformally coupled scalar}
For $\xi=\xi_c$, Eq.~\eqref{ds-general-mode} has $\nu=1/2$ in any dimension. It is convenient to work with the rescaled field $\varphi=a^{(d-1)/2}\phi$, whose mode is $u_k=a^{(d-1)/2}f_k$. Equation~\eqref{eq:rescaled-mode} then becomes
\begin{equation}
 u_k''+k^2u_k=0,
 \qquad
 u_k(\tau)=\frac{\ee^{-\ii k\tau}}{\sqrt{2k}}.
 \label{frw-conformal-mode}
\end{equation}
The conformal vacuum selects this positive-frequency solution for a
spatially flat FRW scale factor.

Substituting $\phi=a^{-(d-1)/2}\varphi$ with $\xi=\xi_c$ into Eq.~\eqref{frw-scalar-action} gives, up to a total derivative, the free action of a massless scalar in flat spacetime and transfers the background dependence to the vertices. For a general FLRW scale factor, a $\lambda_n\phi^n$ vertex acquires the factor $a^{d+1-n(d-1)/2}(\tau)$. In de Sitter space the resulting bulk action is
\begin{equation}
 S[\varphi]=-
 \int\dd\tau\,\dd^d\bm x\left[
 \frac12(\partial_\mu\varphi)^2
 +\sum_{n\geq3}\frac{\lambda_n}{n!}(-\tau)^{\frac{d-1}{2}n-(d+1)}\varphi^n
 \right],
 \label{eq:conformal-action}
\end{equation}
where powers of $H$ have been absorbed into the couplings.  In particular, the $\varphi^4$ vertex is time independent for $d=3$, as expected from classical conformal invariance.
Similarly, a power-law FLRW background \cite{FanXianyu:2024powerlaw} gives monomial vertex weights, so it can be treated by the same integral family.
The concept of integral family will be further clarified in Section~\ref{sec:bubble-setup}.

\paragraph{Schwinger--Keldysh Formalism}
The following introduction of Schwinger--Keldysh formalism is based on Ref. \cite{ChenWangXianyu:2017sk}.
For a final time $\tau_*$, the equal-time correlator is computed by the path integral
\begin{equation}
\begin{aligned}
 \bigl\langle\phi(\tau_*,\bm k_1)\cdots\phi(\tau_*,\bm k_n)\bigr\rangle
 &={}\int\!\mathcal D\phi_+\mathcal D\phi_-\,
 \phi(\tau_*,\bm k_1)\cdots\phi(\tau_*,\bm k_n)
 \\
 &\times\ee^{\ii S[\phi_+]-\ii S[\phi_-]}
 \prod_{\bm k}\delta\!\left[\phi_+(\tau_*,\bm k)-\phi_-(\tau_*,\bm k)\right],
\end{aligned}
\end{equation}
where $\pm$ label the two contour branches and the overall
momentum-conserving delta function is suppressed. For either choice
of Wightman functions in Eq.~\eqref{mode-wightman}, the four
bulk-to-bulk propagators follow from the same contour ordering:
\begin{align}
 G_{++}(k;\tau_1,\tau_2)
 &=G_{>}(k;\tau_1,\tau_2)\theta(\tau_1-\tau_2)
  +G_{<}(k;\tau_1,\tau_2)\theta(\tau_2-\tau_1),
 \nonumber\\
 G_{+-}(k;\tau_1,\tau_2)
 &=G_{<}(k;\tau_1,\tau_2),
 \nonumber\\
 G_{-+}(k;\tau_1,\tau_2)
 &=G_{>}(k;\tau_1,\tau_2),
 \nonumber\\
 G_{--}(k;\tau_1,\tau_2)
 &=G_{<}(k;\tau_1,\tau_2)\theta(\tau_1-\tau_2)
  +G_{>}(k;\tau_1,\tau_2)\theta(\tau_2-\tau_1).
 \label{sk-wightman}
\end{align}

The bulk-to-boundary propagators are
$K_+(k;\tau,\tau_*)=f_k(\tau_*)f_k^*(\tau)$ and
$K_-(k;\tau,\tau_*)=K_+^*(k;\tau,\tau_*)$.

\paragraph{Massless minimally coupled scalar}
In four dimensions, Eqs.~\eqref{massless-prop} and~\eqref{sk-wightman} give
\begin{align*}
 G^\chi_{++}(k;\tau_1,\tau_2) &=\frac{H^2}{2k^3}\Big[ (1+\ii k\tau_1)(1-\ii k\tau_2)\ee^{-\ii k(\tau_1-\tau_2)}\theta(\tau_1-\tau_2)\\
 &\qquad + (1-\ii k\tau_1)(1+\ii k\tau_2)\ee^{+\ii k(\tau_1-\tau_2)}\theta(\tau_2-\tau_1)\Big],\\
 G^\chi_{+-}(k;\tau_1,\tau_2) &=\frac{H^2}{2k^3}(1-\ii k\tau_1)(1+\ii k\tau_2)\ee^{+\ii k(\tau_1-\tau_2)},\\
 G^\chi_{-+}(k;\tau_1,\tau_2) &=\frac{H^2}{2k^3}(1+\ii k\tau_1)(1-\ii k\tau_2)\ee^{-\ii k(\tau_1-\tau_2)},\\
 G^\chi_{--}(k;\tau_1,\tau_2) &=\frac{H^2}{2k^3}\Big[ (1-\ii k\tau_1)(1+\ii k\tau_2)\ee^{+\ii k(\tau_1-\tau_2)}\theta(\tau_1-\tau_2)\\
 &\qquad + (1+\ii k\tau_1)(1-\ii k\tau_2)\ee^{-\ii k(\tau_1-\tau_2)}\theta(\tau_2-\tau_1)\Big].
\end{align*}
At $\tau_*=0$, the bulk-to-boundary propagators are
\begin{equation}
 K_\pm^\chi(k;\tau)
 =\frac{H^2}{2k^3}(1\mp\ii k\tau)\ee^{\pm\ii k\tau}.
 \label{massless-bulk-boundary}
\end{equation}

\paragraph{Conformally coupled scalar}
For the rescaled conformal field $\varphi$, replacing $f_k$ by $u_k$
in Eq.~\eqref{mode-wightman} gives
\begin{align}
 G_{>}(k;\tau_1,\tau_2)
 &=\frac{1}{2k}\ee^{-\ii k(\tau_1-\tau_2)},
 &
 G_{<}(k;\tau_1,\tau_2)
 &=\frac{1}{2k}\ee^{\ii k(\tau_1-\tau_2)}.
 \label{conf-wightman}
\end{align}
At $\tau_*=0$, the corresponding bulk-to-boundary propagators are
\begin{equation}
    K_\pm(k;\tau)=\frac{1}{2k}\ee^{\pm\ii k\tau},
    \qquad k=|\bm k|.
    \label{bulk-boundary-propagators}
\end{equation}
These are obtained by setting $\tau_2=0$ in Eq.~\eqref{conf-wightman}.

Substituting these Wightman functions into \eqref{sk-wightman} gives the four conformal-scalar propagators used below:
\begin{subequations}
\label{conf-sk}
\begin{align}
 G_{++}(k;\tau_1,\tau_2)
 &={}\frac{1}{2k}\left[
 \ee^{-\ii k(\tau_1-\tau_2)}\theta(\tau_1-\tau_2)
 +\ee^{\ii k(\tau_1-\tau_2)}\theta(\tau_2-\tau_1)
 \right],
 \\
 G_{+-}(k;\tau_1,\tau_2)
 &={}\frac{1}{2k}\ee^{\ii k(\tau_1-\tau_2)},
 \\
 G_{-+}(k;\tau_1,\tau_2)
 &={}\frac{1}{2k}\ee^{-\ii k(\tau_1-\tau_2)},
 \\
 G_{--}(k;\tau_1,\tau_2)
 &={}\frac{1}{2k}\left[
 \ee^{\ii k(\tau_1-\tau_2)}\theta(\tau_1-\tau_2)
 +\ee^{-\ii k(\tau_1-\tau_2)}\theta(\tau_2-\tau_1)
 \right].
\end{align}
\end{subequations}
The early-time $\ii0$ prescription is implicit in the exponentials and selects the damped vacuum contour.  It is distinct from the analytic regulators introduced below.

\paragraph{Massive scalar}
Although massive fields are not used in this paper, we also provide the
corresponding massive propagator for reference.
For a scalar of mass $m$ and $\xi=0$, the index is
$\nu^2=d^2/4-m^2/H^2$, and the Wightman functions are
\begin{align}
 G_{>}(k;\tau_1,\tau_2)
 &={}\frac{\pi}{4}\ee^{-\pi\operatorname{Im}\nu}H^{d-1}
 \bigl[(-\tau_1)(-\tau_2)\bigr]^{d/2}
 H_{\nu}^{(1)}(-k\tau_1)H_{\nu^*}^{(2)}(-k\tau_2),
 \\
 G_{<}(k;\tau_1,\tau_2)
 &={}G_{>}(k;\tau_2,\tau_1).
\end{align}
The corresponding SK components follow from Eq.~\eqref{sk-wightman}.

\section{General two-site one-loop bubble}
\label{sec:bubble-setup}

\begin{figure}[htbp]
 \centering
 \begin{tikzpicture}[
   x=1cm,
   y=1cm,
   every node/.style={font=\small},
   external/.style={draw=black!72, line width=0.65pt},
   internal/.style={draw=blue!52!black, line width=1.05pt},
   vertex/.style={circle, draw=blue!52!black, fill=white, line width=0.9pt,
                  inner sep=0pt, minimum size=5.5pt},
   boundary/.style={rectangle, draw=black!85, fill=white, line width=0.8pt,
                    inner sep=0pt, minimum size=6.5pt}
 ]
   \coordinate (v1) at (-1.55,0);
   \coordinate (v2) at ( 1.55,0);
   \coordinate (l1) at (-4.75, 1.05);
   \coordinate (l2) at (-4.75,-1.05);
   \coordinate (r1) at ( 4.75, 1.05);
   \coordinate (r2) at ( 4.75,-1.05);

   \draw[external] (l1) -- (v1);
   \draw[external] (l2) -- (v1);
   \draw[external] (r1) -- (v2);
   \draw[external] (r2) -- (v2);

   \draw[internal] (v1)
     arc[start angle=180,end angle=0,radius=1.55cm];
   \draw[internal] (v1)
     arc[start angle=180,end angle=360,radius=1.55cm];

   \node[boundary] at (l1) {};
   \node[boundary] at (l2) {};
   \node[boundary] at (r1) {};
   \node[boundary] at (r2) {};
   \node[vertex] at (v1) {};
   \node[vertex] at (v2) {};
   \node[inner sep=1pt] at (-1.12,0) {$\tau_1$};
   \node[inner sep=1pt] at ( 1.12,0) {$\tau_2$};

   \node at (-4.05,0) {$E_1$};
   \node at ( 4.05,0) {$E_2$};
   \node at (-3.32,0) {$\vdots$};
   \node at ( 3.32,0) {$\vdots$};

   \node[text=blue!52!black] at (0, 1.80) {$z_1$};
   \node[text=blue!52!black] at (0,-1.80) {$z_2$};
 \end{tikzpicture}
 \caption{The two-site one-loop bubble, with external energy sums $E_1,E_2$ and internal momentum magnitudes $z_1,z_2$.}
 \label{bubble-fig}
\end{figure}
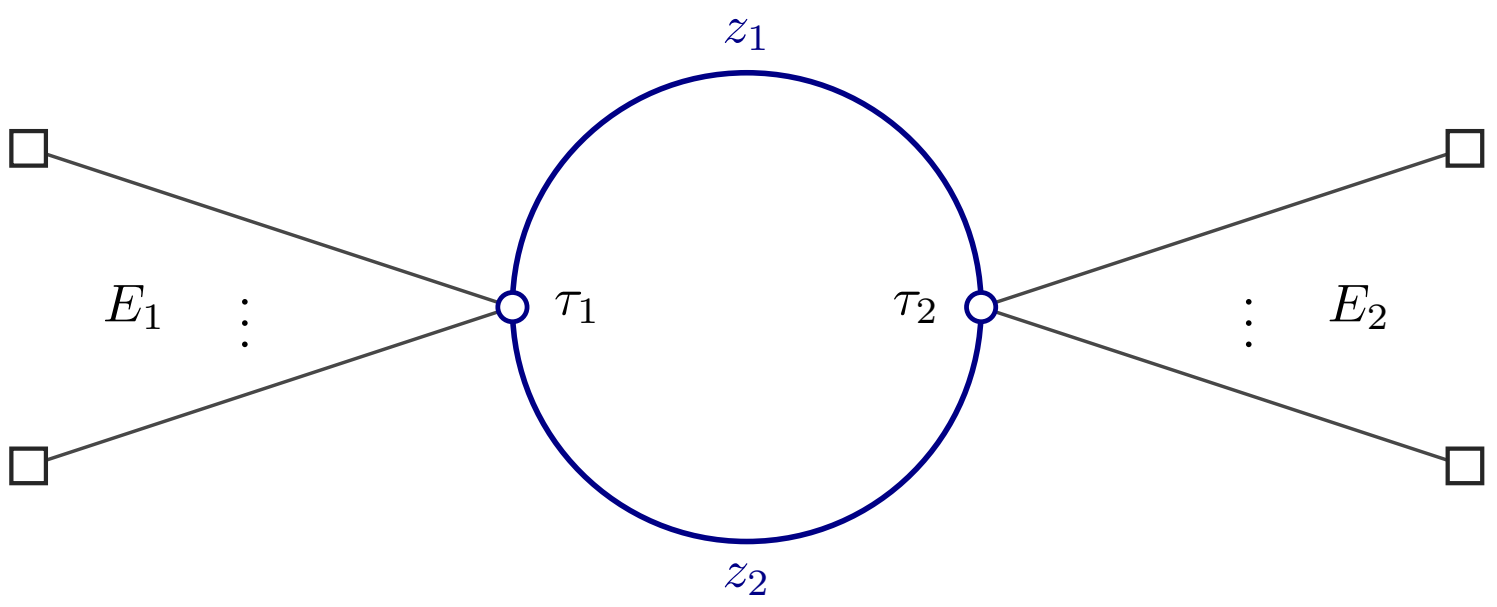

We consider two vertices $v_1,v_2$ joined by two massless internal lines.
Each vertex carries a monomial time weight fixed by the interaction
and the background. Under the all-incoming convention, the energy sums
and the spatial momentum transferred between the vertices are
\begin{equation}
    E_1=\sum_{r\in v_1}|\bm k_r|,
    \qquad
    E_2=\sum_{r\in v_2}|\bm k_r|,
    \qquad
    \bm p=\sum_{r\in v_1}\bm k_r=-\sum_{r\in v_2}\bm k_r.
\end{equation}
For loop momentum $\bm\ell$ and external momentum $\bm p$, we write
\begin{equation}
    z_1=\ell=|\bm\ell|,
    \qquad
    z_2=q=|\bm\ell+\bm p|,
    \qquad
    p=|\bm p|,
\end{equation}
The notation $z_i$ anticipates the Baikov representation introduced below.
The triangle inequality implies
\begin{equation}
    \qquad E_1,E_2\geq p.
\end{equation}
Equality at either vertex requires its nonzero external momenta to be parallel and point in the same direction. 
The energy sums $E_1,E_2$ are not identified with the magnitude $p$. 
The kinematic configuration is illustrated in Figure~\ref{bubble-fig}.

\Needspace{11\baselineskip}
Let $\sigma_1,\sigma_2\in\{+,-\}$ label the two contour (SK) branches,
with the signs read as $\pm1$ in the exponentials. For massless minimally
coupled fields, the bubble contribution has the form
\begin{align}
 \mathcal A_{\sigma_1\sigma_2}
 &={}N_{\sigma_1\sigma_2}
 \int_{-\infty}^{0}\dd\tau_1\dd\tau_2\,
 (-\tau_1)^{a_1}(-\tau_2)^{a_2}
 \nonumber\\
 &\quad\times\prod_{r\in v_1}K_{\sigma_1}^\chi(k_r;\tau_1)
 \prod_{r\in v_2}K_{\sigma_2}^\chi(k_r;\tau_2)
 \nonumber\\
 &\quad\times\int\dd^d\bm\ell\,
 G_{\sigma_1\sigma_2}^\chi(z_1;\tau_1,\tau_2)
 G_{\sigma_1\sigma_2}^\chi(z_2;\tau_1,\tau_2).
 \label{general-bubble}
\end{align}
Here $K^\chi$ and $G^\chi$ are the massless propagators specified in
Section~\ref{sec:background}, and $N_{\sigma_1\sigma_2}$ contains
the couplings, symmetry factors and branch signs. The time integrals
are understood with the analytic regulators introduced below;
the future-boundary expression need not have a finite late-time limit.

Expanding the polynomial factors in these propagators leaves
exponential kernels with shifted time and momentum powers. We define
the corresponding integral families by
\begin{align}
 \tilde J^{(+-)}[a_1,a_2;b_1,b_2;D]
 &={}\int\dd^D\bm\ell
 \int_{-\infty}^{0}\dd\tau_1\dd\tau_2\,
 \frac{(-\tau_1)^{a_1}(-\tau_2)^{a_2}}{z_1^{b_1}z_2^{b_2}}
 \nonumber\\
 &\quad\times
 \ee^{\ii(E_1+z_1+z_2)\tau_1-\ii(E_2+z_1+z_2)\tau_2},
 \label{bubble-mixed-component}\\[1ex]
 \tilde J^{(++)}[a_1,a_2;b_1,b_2;D]
 &={}\int\dd^D\bm\ell
 \int_{-\infty}^{0}\dd\tau_1\dd\tau_2\,
 \frac{(-\tau_1)^{a_1}(-\tau_2)^{a_2}}{z_1^{b_1}z_2^{b_2}}
 \ee^{\ii(E_1\tau_1+E_2\tau_2)}
 \nonumber\\
 &\quad\times\bigl[
 \ee^{-\ii(z_1+z_2)(\tau_1-\tau_2)}
 \theta(\tau_1-\tau_2)
 +\ee^{\ii(z_1+z_2)(\tau_1-\tau_2)}
 \theta(\tau_2-\tau_1)\bigr].
 \label{bubble-ordered-component}
\end{align}
At $b_1=b_2=1$, these have the same kernels as the rescaled conformal
bubble. The indices $b_i$ count the full denominator powers.
The regulator and normalization conventions are specified in the next
two sections, where $d=3-2\eps$ and dimension-shifted integrals with
$D=d-2$ will also be used.

The massless bubble is then a finite linear combination of integrals in these families,
\begin{equation}
 \mathcal A_{\sigma_1\sigma_2}
 =\sum_\alpha C_{\sigma_1\sigma_2,\alpha}\,
 \tilde J^{(\sigma_1\sigma_2)}
 [a_{1,\alpha},a_{2,\alpha};b_{1,\alpha},b_{2,\alpha};d].
 \label{bubble-family-assembly}
\end{equation}
The coefficients depend on external momenta and couplings, but not on
the integration variables. For example, the constant polynomial term
from two massless internal lines gives $b_1=b_2=3$; the other terms
shift the time and momentum powers. Massless external legs and local
derivative interactions give further index shifts and numerator
insertions.

The mixed family contains no Heaviside function, so its two time
integrations factorize at fixed loop momentum. The two terms in the
ordered family correspond to the complementary regions $\tau_1>\tau_2$
and $\tau_2>\tau_1$.

The full bubble is obtained by summing the SK components,
\begin{equation}
    \mathcal{A}_{\mathrm{bub}}
    =\mathcal{A}_{++}+\mathcal{A}_{+-}+\mathcal{A}_{-+}+\mathcal{A}_{--}.
\end{equation}
For real couplings and vertex powers, the remaining SK components follow
by complex conjugation,
$\mathcal A_{-+}=\mathcal A_{+-}^{*}$ and
$\mathcal A_{--}=\mathcal A_{++}^{*}$, with the corresponding conjugate
vacuum contours. These relations apply term by term to
Eq.~\eqref{bubble-family-assembly}. The second term of
\eqref{bubble-ordered-component} is obtained from the first by
exchanging $(E_1,a_1,\tau_1)\leftrightarrow(E_2,a_2,\tau_2)$, leaving
$p$ unchanged.  We therefore study only the $(+-)$ component and the
$\tau_1>\tau_2$ part of the $(++)$ component.  The former provides the
preliminary example in Section~\ref{sec:pm-companion}; the latter,
together with its contact terms, is treated in
Section~\ref{sec:pp-sector}.

\section{The \texorpdfstring{$(+-)$}{(+-)} component: a preliminary example}
\label{sec:pm-companion}

The mixed component contains no Heaviside function, so its two time
integrations factorize at fixed loop momentum.  We use this simpler
component to introduce the IBP reduction and CDE construction.

We regulate both the loop-momentum and time integrals before deriving
the IBP relations.  For the spatial loop integration, we use dimensional
regularization with 
\begin{equation}
      d=3-2\eps
\end{equation}
This leaves $\nu=1/2$ for a
conformally coupled scalar, while the massless minimally coupled index
becomes $\nu=d/2=3/2-\eps$.  To retain the simpler form 
of the Hankel function, we keep $\nu=3/2$, which amounts to add a d-dependent
mass term, and the mode function will be modified with $a(\tau)^\eps$.\footnote{For details of the dimensional regularization in dS, see Section~2 of Ref.~\cite{Beneke:2023wmt}.}

We here equivalently introduce another independent analytic regulator for the time integrations.  The
total time-power shifts $\rho_i$ include both the dimensional and
analytic contributions, and we choose
\begin{equation}
      \rho\equiv\rho_1=\rho_2=\delta\eps,
 \label{pm-regulators}
\end{equation}
The parameter $\delta$ is independent of $\eps$.
The vertex weights are shifted according to
\begin{equation}
 (-\tau_i)^{a_i}\longrightarrow(-\tau_i)^{a_i+\rho_i}.
\end{equation}
For these families, we define the normalized loop measure by
\begin{equation}
 \int[\dd^D\bm\ell]F(\bm\ell)
 =\ee^{(1+2\delta)\gamma_E\eps}
 \int\frac{\dd^D\bm\ell}{\pi^{(D-1)/2}}F(\bm\ell),
 \qquad D\in\{d,d-2\}.
 \label{eq:loop-measure}
\end{equation}
Here $\gamma_E$ is the Euler constant; the exponential removes the
$\gamma_E$ terms common to the $\eps$ expansions, as in the standard
normalization of dimensionally regulated Feynman integrals. And the factor
$\pi^{(D-1)/2}$ is included to cancel the $\pi$ factors in the angular integration.

\subsection{Momentum family and \texorpdfstring{$\dd\log$}{dlog} forms}

The elementary time integration is
\begin{equation}
 \int_{-\infty}^0\dd\tau\,(-\tau)^{n-1+\delta\eps}
 \ee^{\pm\ii E\tau}
 =\frac{\ee^{\mp\ii\pi(n+\delta\eps)/2}\Gamma[n+\delta\eps]}
 {(E\mp\ii0)^{n+\delta\eps}}.
 \label{pm-time-integrals}
\end{equation}
With early-time damping, this integral converges for
\begin{equation}
 \Re(n+\delta\eps)>0.
\end{equation}
From the $(+-)$ family \eqref{bubble-mixed-component}, we define the momentum family after the time integration as
\begin{equation}
 J^{(+-)}[n_1,n_2;b_1,b_2; D]
 =\int[\dd^D\bm\ell]\,
 \frac{1}{z_3^{n_1+\delta\eps}z_4^{n_2+\delta\eps}z_1^{b_1}z_2^{b_2}}.
 \label{pm-regulated-J-family}
\end{equation}
where the two new energy denominators are defined as
\begin{equation}
      z_3=E_1+z_1+z_2,\qquad z_4=E_2+z_1+z_2.
      \label{baikov+-}
\end{equation}
Here and below, the spatial dimension is written as a semicolon-separated
argument in the brackets; an omitted dimension argument denotes the default
$d=3-2\eps$. The momentum family omits the Gamma functions and phases
of the time transforms in Eq.~\eqref{pm-time-integrals}. Its contribution
to a correlator therefore includes
\begin{equation}
 \Gamma[a_1+1+\delta\eps]\Gamma[a_2+1+\delta\eps]\,
 J^{(+-)}[a_1+1,a_2+1;1,1].
\end{equation}
The remaining time-transform phases follow from Eq.~\eqref{pm-time-integrals},
and the overall factors for the conformal examples are specified in
Eq.~\eqref{phys-SK-sum}.
The Gamma functions and phases of the time transforms are not part of
the momentum IBP coefficients. 

We'd like to first determine the number of master integrals of this integral family. We claim that there exist 7 master integrals for the $(+-)$ component, which means
we need to construct seven independent canonical basis integrals.
The detail of reduction program will be explained in next section. 
And we will shortly give the explanation in terms of twisted cohomology.

\paragraph{Master integrals from twisted cohomology} In the variables
\begin{equation}
 s=z_1+z_2,\qquad r=z_1-z_2,
\end{equation}
the Baikov polynomial factorizes into linear factors, and the energy
denominators are linear in the integration variables:
\begin{equation}
 \mathcal P=(s-p)(s+p)(p-r)(p+r),\qquad z_3=E_1+s,\quad z_4=E_2+s .
\end{equation}
For $D=d=3-2\eps$ the integrand of Eq.~\eqref{pm-baikov-family} is a rational
function times the multivalued factor
\begin{equation}
 u=\mathcal P^{-\eps}z_3^{-\delta\eps}z_4^{-\delta\eps}.
 \label{pm-twist}
\end{equation}
For generic regulator values and kinematics, we count the master integrals
sector by sector using the critical points of the corresponding twist.

In the sector without $z_1$ and $z_2$ cut, $u$ factorizes into a
function of $s$ and a function of $r$, and the critical-point equations
give
\begin{equation}
\begin{aligned}
 \partial_s\log u&=-\eps\Bigl(\frac1{s-p}+\frac1{s+p}\Bigr)
 -\delta\eps\Bigl(\frac1{s+E_1}+\frac1{s+E_2}\Bigr)=0,\\
 \partial_r\log u&=\frac{2\eps r}{p^2-r^2}=0 .
\end{aligned}
\end{equation}
The first equation is cubic in $s$ and the second has the single root
$r=0$, so this sector has 3 master integrals. 

On the cut $z_2=0$
one has $s=r=z_1$, the two Baikov factors coincide,
$\mathcal P|_{z_2=0}=-(z_1^2-p^2)^2$, and
\begin{equation}
 \partial_{z_1}\log u\big|_{z_2=0}=-\frac{4\eps z_1}{z_1^2-p^2}
 -\delta\eps\Bigl(\frac1{z_1+E_1}+\frac1{z_1+E_2}\Bigr)=0
\end{equation}
is again cubic, giving 3 master integrals.  The cut $z_1=0$ gives
three by the same argument.  The double cut $z_1=z_2=0$ is a single
point, where $\mathcal P=-p^4\neq0$, and gives one master integral.  In
total there are $3+3+3+1=10$ independent integrals.

But it should be noticed that there exist the symmetry between exchange $z_1 \leftrightarrow z_2$,
so the $z_1$ cut and $z_2$ cut actually give us the same master, so the total number should be $3+3+1=7$,
which is exactly the same as we obtained from computer reduction.

\begin{figure}[t]
 \centering
 \includegraphics[width=0.42\textwidth]{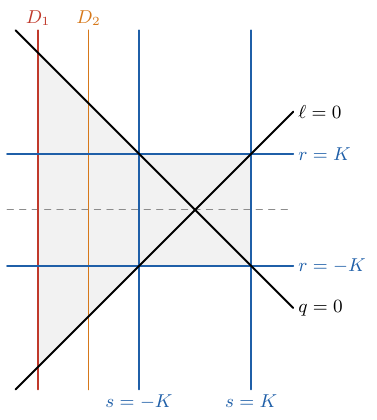}
 \caption{Line arrangement of the $(+-)$ family in the variables
 $s=z_1+z_2$, $r=z_1-z_2$.  Blue: the Baikov lines $s=\pm p$, $r=\pm p$;
 red and orange: $z_3=0$ and $z_4=0$; black: $z_1=0$ and $z_2=0$.  The ten
 bounded chambers (shaded) reduce to seven under the exchange
 $z_1\leftrightarrow z_2$, the reflection in the dashed line $r=0$.}
 \label{fig:pm-arrangement}
\end{figure}

The same number follows from the Euler characteristic.  Regulating the
propagators by $z_1^{\lambda_1}z_2^{\lambda_2}$ makes all eight lines
twisted.  The number of critical points then
equals to $|\chi(\mathbb C^2\setminus\mathcal A)|$, which also equals to the number of bounded chambers,
shown in Fig.~\ref{fig:pm-arrangement}.

The number of bounded regions is obviously 10, but again because of the symmetry $z_1\leftrightarrow z_2$, i.e.\ $r\to-r$.
The region should be counted after involving the reflection by the dashed line. 
Again we counted seven bounded regions, which is the number of master integrals of the $(+-)$
component.

\paragraph{The dlog criterion}
In the polylogarithmic class, integrals with a dlog representation at
$\eps=0$ are candidates for a uniform transcendental (UT) basis.
Schematically, this representation is
\begin{equation}
 \int\sum_i c_i\bigwedge_j\dd\log g_{i,j},
 \qquad c_i\in\mathbb C.
 \label{eq:dlog-form}
\end{equation}
The functions $g_{i,j}$ may depend on the integration variables and the
external kinematics, whereas the coefficients $c_i$ are constants,
called leading singularities.

To identify candidate canonical integrals, we search for dlog forms in
the Baikov representation
\cite{Baikov:1996rk,Mastrolia:2018uzb,Weinzierl:2022feynman,Chen:2022lzr,DlapaLiZhang:2021UT}, which replaces loop-momentum
components by independent scalar products.  For this family, expressing
those scalar products in terms of $z_1,z_2$ makes the energy denominators
linear and exposes the factorization of the Gram polynomial.  This
provides the variables used in the logarithmic construction below.
Appendix~\ref{app:dimension-recurrence} gives the Baikov measure
and its normalization.
The resulting Baikov representation is
\begin{equation}
 J^{(+-)}[n_1,n_2;b_1,b_2; D]=\mathcal N_D p^{2-D}
 \int_{\mathcal D_p}\dd z_1\,\dd z_2\,
 \frac{z_1 z_2\,\mathcal P^{(D-3)/2}}
 {z_3^{n_1+\delta\eps}z_4^{n_2+\delta\eps}z_1^{b_1}z_2^{b_2}},
 \label{pm-baikov-family}
\end{equation}
The normalization factor is
\begin{equation}
 \mathcal N_D=
 \frac{2^{4-D}\ee^{(1+2\delta)\gamma_E\eps}}{\Gamma[(D-1)/2]}.
\end{equation}
The Baikov polynomial and integration domain are, respectively,
\begin{equation}
 \mathcal P=[(z_1+z_2)^2-p^2][p^2-(z_1-z_2)^2]
\end{equation}
and
\begin{equation}
 \mathcal D_p=\bigl\{(z_1,z_2):z_1,z_2\geq0,\quad
 |z_1-z_2|\leq p\leq z_1+z_2\bigr\}.
\end{equation}

We first construct $\dd \log$ forms in the physical spatial dimension $d=3-2\eps$, where there's 
no Baikov Polynomial in the denominator. What we can directly construct are the following three
\begin{subequations}
\label{eq:mixed-forms}
\begin{equation}
\begin{aligned}
 \Phi_4&=2\dd\log z_3\wedge\dd\log z_2,\\
 \Phi_5&=2\dd\log z_4\wedge\dd\log z_2,\\
 \Phi_7&=2\dd\log z_1\wedge\dd\log z_2.
\end{aligned}
\end{equation}
We can obtain further $\dd \log$ forms by lowering the spatial dimension
to $d-2$. Then there will be an extra inverse Baikov polynomial.
We introduce variables that factorize the Baikov polynomial:
\begin{equation}
 s=z_1+z_2,\qquad r=z_1-z_2,\qquad
 \mathcal P=(s^2-p^2)(p^2-r^2).
\end{equation}
The roots of the
two factors suggest the logarithmic one-forms
\begin{equation}
 \omega_+=\dd\log\frac{s-p}{s+p},\qquad
 \omega_-=\dd\log\frac{p-r}{p+r}.
 \label{eq:root-forms}
\end{equation}
which allows us to construct four additional $\dd \log$ forms  
\begin{equation}
\begin{aligned}
 \Phi_1&=\omega_+\wedge\omega_-,
 &\Phi_2&=2\omega_-\wedge\dd\log z_3,\\
 \Phi_3&=-2\omega_-\wedge\dd\log z_4,
 &\Phi_6&=-4\omega_-\wedge\dd\log z_1.
\end{aligned}
\end{equation}
\end{subequations}
All seven forms have constant leading singularities and we need to transform them back to the indices family \eqref{pm-regulated-J-family}.
The normalization of the associated integrals is
given below.
The dimension recurrence is derived in
Appendix~\ref{app:dimension-recurrence}.

\subsection{Canonical basis and connection}

After momentum IBP reduction the seven basis integrals have the short forms
\begin{equation}
\begin{aligned}
 f^{(+-)}_1&=-\frac p\eps J^{(+-)}[0,0;1,3]
       -\delta pJ^{(+-)}[1,0;1,2]-\delta pJ^{(+-)}[0,1;1,2],\\
 f^{(+-)}_2&=-\frac{2(2\eps-1)}\eps J^{(+-)}[1,0;1,1],\\
 f^{(+-)}_3&=\frac{2(2\eps-1)}\eps J^{(+-)}[0,1;1,1],\\[0.3em]
 f^{(+-)}_4&=pJ^{(+-)}[1,0;1,2],\\
 f^{(+-)}_5&=pJ^{(+-)}[0,1;1,2],\\
 f^{(+-)}_6&=-\frac1\eps J^{(+-)}[0,0;0,3]
       +\frac{4\eps-1}\eps J^{(+-)}[0,0;1,2],\\[0.3em]
 f^{(+-)}_7&=pJ^{(+-)}[0,0;2,2].
\end{aligned}
\label{eq:mixed-basis}
\end{equation}
To match the UT weights between different basis, the four integrals defined in the lowered dimension include
the additional relative factor
\begin{equation}
 -\frac1\eps.
\end{equation}
With this choice, all seven integrals share the representation
\begin{equation}
\begin{pmatrix}f^{(+-)}_1\\f^{(+-)}_2\\f^{(+-)}_3\\f^{(+-)}_4\\f^{(+-)}_5\\f^{(+-)}_6\\f^{(+-)}_7\end{pmatrix}
=\frac{\ee^{(1+2\delta)\gamma_E\eps}}{\Gamma[1-\eps]}
\int_{\mathcal D_p}\left(\frac{4p^2}{\mathcal P}\right)^\eps
z_3^{-\delta\eps}z_4^{-\delta\eps}
\begin{pmatrix}\Phi_1\\\Phi_2\\\Phi_3\\\Phi_4\\\Phi_5\\\Phi_6\\\Phi_7\end{pmatrix}.
\label{eq:mixed-periods}
\end{equation}
which is in the form we required as \eqref{eq:dlog-form}.
Collecting them into the vector
\begin{equation}
 \bm f^{(+-)}=(f^{(+-)}_1,\ldots,f^{(+-)}_7)^T,
\end{equation}
we obtain the seven-dimensional canonical differential equation
\begin{equation}
 \dd\bm f^{(+-)}=\eps\sum_{W\in\mathbb L_{+-}}R^{(+-)}_W(\delta)\dd\log W\,\bm f^{(+-)}.
 \label{eq:mixed-cde}
\end{equation}
Its alphabet is
\begin{equation}
 \mathbb L_{+-}=\{p,E_1,E_2,p-E_1,p+E_1,p-E_2,p+E_2,E_1-E_2\}.
 \label{eq:mixed-letters}
\end{equation}
The connection matrices are given in Appendix~\ref{app:mixed-residues}.
Their sum satisfies
\begin{equation}
 \sum_{W\in\mathbb L_{+-}}R^{(+-)}_W(\delta)=-2(1+\delta)\one_7,
\end{equation}
as required by the common scaling degree.

\section{The \texorpdfstring{$(++)$}{(++)} component}
\label{sec:pp-sector}

\subsection{Regulated families and IBP identities}

For the
time-ordered component, the total energy and the ordered energy denominators
(the sign-flipped $z_3$) are
\begin{equation}
 k_t=E_1+E_2,\qquad
 z_3=E_1-z_1-z_2,\qquad z_4=E_2+z_1+z_2
 \label{pp-energy-denominators}
\end{equation}
and it should not be confused with Eq.~ \eqref{baikov+-}.

With the regulator choice of Eq.~\eqref{pm-regulators}, we denote the
normalized single-ordering topsector family by $J^{\mathrm{top}}$.
It retains both the time integrations and the loop-momentum integral,
and is distinct from the auxiliary momentum family $J$ introduced below:
\begin{equation}
 J^{\mathrm{top}}_{a_1,a_2;b_1,b_2}=\ii^{a_1+a_2}\int_{-\infty}^0\dd\tau_1\dd\tau_2
 (-\tau_1)^{a_1+\delta\eps}(-\tau_2)^{a_2+\delta\eps}
 \theta(\tau_1-\tau_2)\int[\dd^d\bm\ell]
 \frac{\ee^{\ii z_3\tau_1+\ii z_4\tau_2}}{z_1^{b_1}z_2^{b_2}},
 \label{pp-ordered-integral}
 \end{equation}
When a time IBP differentiates the ordering function, it produces a
delta function supported at coincident times.  These equal-time terms
define the contact subsector family $J^{\mathrm{sub}}$
 \begin{equation}
 J^{\mathrm{sub}}[n;b_1,b_2;D]
 =\ii^{n-1}\int_{-\infty}^0\dd\tau\,(-\tau)^{n+2\delta\eps}
 \ee^{\ii k_t\tau}\int[\dd^D\bm\ell]\frac1{z_1^{b_1}z_2^{b_2}}.
 \label{pp-contact-family}
\end{equation}
Here the phases in front of the integrals make the IBP coefficients real rational functions of the
parameters. The analytic time powers regulate the late-time endpoint, while the
vacuum $\ii0$ prescription fixes the early-time contour.  The IBP identities
are derived in a convergence domain and continued meromorphically to
the chosen regulator values.

Time and loop-momentum derivatives give two sets of IBP identities.
The time derivatives act on the Heaviside function and generate
equal-time contact integrals; the momentum derivatives leave the Heaviside
function invariant and produce no subsector integrals.  The contact family
closes under its own identities, so it reduces independently, whereas
ordered integrals can reduce onto both top and contact masters.
The explicit IBP relations and symmetry relation are collected in
Appendices~\ref{app:pp-time-ibps} and~\ref{app:pp-spatial-ibps}.

In practice, solving the large system of IBP identities and reducing the
integrals to a set of master integrals is usually carried out with dedicated
computer programs. Several public implementations of the Laporta algorithm~\cite{Laporta:2000dsw}
are available, including FIRE~\cite{Smirnov:2008iw,SmirnovZeng:2025fire7},
Reduze~\cite{vonManteuffel:2012np}, and
Kira~\cite{Maierhofer:2017gsa,Klappert:2020nbg,LangeUsovitschWu:2025kira3}.
In the present calculation, we use Kira to perform the reduction.
For the regulated family, the resulting master vector is
\begin{equation}
\boxed{\begin{aligned}
 \bm f^{(++)}_{\mathrm{raw}}=\bigl(&
 J^{\mathrm{sub}}[0;1,1],J^{\mathrm{sub}}[0;1,2],J^{\mathrm{sub}}[0;2,2],
 J^{\mathrm{top}}_{0,0;0,0},J^{\mathrm{top}}_{0,0;0,1},
 J^{\mathrm{top}}_{0,-1;0,0},\\
 &J^{\mathrm{top}}_{0,0;0,2},J^{\mathrm{top}}_{0,0;0,3},
 J^{\mathrm{top}}_{0,0;1,2},J^{\mathrm{top}}_{0,0;2,2}
 \bigr)^T .
\end{aligned}}
\label{pp-raw-master-vector}
\end{equation}
The first three entries are contact masters and the remaining seven are
topsector masters. 

External differentiation is performed as follows:
\begin{equation}
 \partial_{E_1}J^{\mathrm{top}}_{a_1,a_2;b_1,b_2}
 =-J^{\mathrm{top}}_{a_1+1,a_2;b_1,b_2},
 \qquad
 \partial_{E_2}J^{\mathrm{top}}_{a_1,a_2;b_1,b_2}
 =-J^{\mathrm{top}}_{a_1,a_2+1;b_1,b_2}.
 \label{pp-energy-derivatives}
\end{equation}
With the phase convention above, neither the basis nor these differentiations contain an explicit $\ii$.
The $p$ derivative is also taken directly, at fixed direction of
$\bm p$.  Its explicit form is given in
Appendix~\ref{app:derivatives}.

\subsection{\texorpdfstring{$\dd \log$}{dlog} form selection}
\label{sec:pp-dlog-forms}

We seek a basis whose differential equation has both an overall $\eps$
and a logarithmic connection.  We use the constant-leading-singularity
criterion introduced in Eq.~\eqref{eq:dlog-form}.

\subsubsection{Contact-sector logarithmic forms}
\label{sec:pp-contact-dlogs}

Before constructing the contact dlog forms, we perform the single time
integration in \eqref{pp-contact-family} which is  similar to \eqref{pm-time-integrals}
\begin{equation}
 J^{\mathrm{sub}}[n;b_1,b_2; D]
 =-\ee^{-\ii\pi\delta\eps}
 \frac{\Gamma[1+n+2\delta\eps]}
 {(k_t-\ii0)^{1+n+2\delta\eps}}\,
 J_C[b_1,b_2; D],
 \label{eq:contact-factor}
\end{equation}
where the remaining momentum family is the massless bubble
\begin{equation}
 J_C[b_1,b_2; D]
 =\int[\dd^D\bm\ell]\,\frac1{z_1^{b_1}z_2^{b_2}}
 =p^{D-b_1-b_2}\mathcal B_{b_1b_2}(D).
 \label{eq:contact-bubble}
\end{equation}
Factoring out the power of the external momentum defines the
dimensionless bubble coefficient.  In the loop-measure convention of
Eq.~\eqref{eq:loop-measure}, it is
\begin{equation}
 \mathcal B_{b_1b_2}(D)
 =\sqrt\pi\,\ee^{(1+2\delta)\gamma_E\eps}
 \frac{\Gamma[(b_1+b_2-D)/2]\Gamma[(D-b_1)/2]\Gamma[(D-b_2)/2]}
 {\Gamma[b_1/2]\Gamma[b_2/2]\Gamma[D-(b_1+b_2)/2]}.
 \label{eq:bubble-coeff}
\end{equation}
The prefactor in \eqref{eq:contact-factor} is independent of the loop
variables.  We therefore apply the $\dd \log$ criterion to the spatial kernel
of $J_C$, which contains no energy denominators.
We again use the Baikov representation
\begin{equation}
 J_C[b_1,b_2; D]
 =\mathcal N_D p^{2-D}\int_{\mathcal D_p}\dd z_1\,\dd z_2\,
 z_1^{1-b_1}z_2^{1-b_2}\mathcal P^{(D-3)/2}.
\end{equation}

A constant-leading-singularity choice is
\begin{align}
 \Phi_1&=-\omega_+\wedge\omega_-
       =-\frac{8p^2\dd z_1\wedge\dd z_2}{\mathcal P},\notag\\
 \Phi_2&=-4\omega_-\wedge\dd\log z_2
       =\frac{8p(s^2-p^2)\dd z_1\wedge\dd z_2}{z_2\mathcal P},\notag\\
 \Phi_3&=2\dd\log z_1\wedge\dd\log z_2
       =\frac{2\dd z_1\wedge\dd z_2}{z_1 z_2}.
 \label{eq:contact-forms}
\end{align}
The first two forms are represented in $d-2$ dimensions and the third
in $d$ dimensions.  We choose $n=0$ for all three contact integrals and
multiply by $k_t$ to remove the integer power.  Matching the complete Baikov kernels to
Eq.~\eqref{eq:contact-forms}, with the relative normalization of
Eq.~\eqref{eq:mixed-periods} for the dimension-shifted forms,
defines the contact basis:
\begin{align}
 f^{(++)}_1
 &=\frac{p k_t}{\eps}J^{\mathrm{sub}}[0;1,1; d-2]
 =\frac{(1-2\eps)^2k_t}{\eps^2p}J^{\mathrm{sub}}[0;1,1],\notag\\[0.3em]
 f^{(++)}_2
 &=-\frac{k_t}{\eps}\left[
 J^{\mathrm{sub}}[0;-1,2; d-2]-p^2J^{\mathrm{sub}}[0;1,2; d-2]\right]\notag\\
 &=\frac{(4\eps-1)k_t}{\eps}J^{\mathrm{sub}}[0;1,2],\notag\\[0.3em]
 f^{(++)}_3&=p k_tJ^{\mathrm{sub}}[0;2,2].
 \label{eq:contact-basis}
\end{align}
Here the last argument of $J^{\mathrm{sub}}$ specifies its spatial dimension,
and the second equalities for $f^{(++)}_1$ and $f^{(++)}_2$ follow
from dimension recurrence. In the second line, the terms proportional to $J_C[0,1;d-2]$ and
$J_C[1,0;d-2]$ vanish as scaleless integrals. The dimension shifts above can also be checked directly by inserting
the bubble coefficient in Eq.~\eqref{eq:bubble-coeff}
into the factorized expression~\eqref{eq:contact-factor}.

\subsubsection{From the auxiliary momentum family to the topsector}

If we first want to study the topsector block alone, we can first remove the Heaviside function
from the integral family in \eqref{pp-ordered-integral}.  In the
original time IBPs, the contact subsector is generated only when a time
derivative acts on this function:
\begin{equation}
 \partial_{\tau_1}\theta(\tau_1-\tau_2)
 =\delta(\tau_1-\tau_2),\qquad
 \partial_{\tau_2}\theta(\tau_1-\tau_2)
 =-\delta(\tau_1-\tau_2).
\end{equation}
Omitting the Heaviside function removes these equal-time terms, which just generate the
contact integrals belonging to subsector.  The
remaining time-IBP coefficients and the momentum-IBP relations are
unchanged.  Thus, at generic regulator values, the family without time
ordering reproduces the reduction of the ordered topsector modulo
contact integrals, see \eqref{pp-time-ibp-1}--\eqref{pp-time-ibp-2}.

We therefore construct the topsector basis in this auxiliary family
first and then map it back to the original family.  Without the
Heaviside function, the two time integrations factorize and give the
momentum family
\begin{equation}
 J[n_1,n_2;b_1,b_2; D]
 =\int[\dd^D\bm\ell]\,
 \frac1{(z_3)^{n_1+\delta\eps}(z_4)^{n_2+\delta\eps}
 z_1^{b_1}z_2^{b_2}}.
 \label{pp-J-family}
\end{equation}
The normalization relating the two sets of indices follows directly from
the time integrations.  With branches fixed by the vacuum contour, each
vertex gives
\begin{equation}
 \ii^{a_i}\int_{-\infty}^0\dd\tau_i\,
 (-\tau_i)^{a_i+\delta\eps}\ee^{\ii z_{i+2}\tau_i}
 =\ee^{-\ii\pi(1+\delta\eps)/2}
 \frac{\Gamma[1+a_i+\delta\eps]}
 {(z_{i+2})^{1+a_i+\delta\eps}}.
 \label{eq:time-transform}
\end{equation}
Using the Pochhammer symbol, we factor the index dependence as
\begin{equation}
 \Gamma[1+a_i+\delta\eps]
 =\Gamma[1+\delta\eps](1+\delta\eps)_{a_i},\qquad
 (z)_a=\frac{\Gamma[z+a]}{\Gamma[z]}.
\end{equation}
The two time integrations supply a common normalization independent of
the indices and external scales.  It can be omitted from the homogeneous
IBP and differential equations.  But the Pochhammer factors must be retained,
because their ratios reproduce the coefficients of time-index shifts.
The map between the original topsector and the auxiliary family is therefore
\begin{equation}
 J^{\mathrm{top}}_{a_1,a_2;b_1,b_2}
 \longleftrightarrow (1+\delta\eps)_{a_1}(1+\delta\eps)_{a_2}
 J[a_1+1,a_2+1;b_1,b_2].
 \label{pp-top-pairing}
\end{equation}
With the contact terms omitted, the time IBPs of the original family
are equivalent to the partial-fraction relations of the momentum family
$J$, with the index normalization fixed by Eq.~\eqref{pp-top-pairing}.
The time-index shifts become algebraic relations among the energy
denominators and momentum magnitudes.  Thus the time IBPs in
Appendix~\ref{app:pp-time-ibps} are implemented by partial-fraction
reduction in the auxiliary family.  Explicitly, these relations are
\begin{equation}
\begin{aligned}
 J[n_1-1,n_2;b_1,b_2]
 &={}E_1J[n_1,n_2;b_1,b_2]\\
 &\quad-J[n_1,n_2;b_1-1,b_2]-J[n_1,n_2;b_1,b_2-1],\\[0.4em]
 J[n_1,n_2-1;b_1,b_2]
 &={}E_2J[n_1,n_2;b_1,b_2]\\
 &\quad+J[n_1,n_2;b_1-1,b_2]+J[n_1,n_2;b_1,b_2-1].
\end{aligned}
 \label{eq:partial-fractions}
\end{equation}
They follow directly from the linear energy denominators in
Eq.~\eqref{pp-energy-denominators}.

The momentum IBPs give the same relations in the two families.  They do
not differentiate the Heaviside function and therefore generate no
contact subsector.  Performing the time integrals converts the
loop-momentum derivatives of the time exponentials into derivatives of
the energy denominators of $J$, while the spatial propagator factors
and IBP vectors are unchanged.  The explicit momentum identities are
given in Appendix~\ref{app:pp-spatial-ibps}.

Differential equations are obtained by differentiating a basis with
respect to external kinematics and then reducing the resulting
integrals by IBP.  The index map also preserves these external
derivatives.  Since the two families have equivalent reduction
relations in the topsector, the differentiated basis elements reduce
with the same coefficients.  We can therefore construct the
differential equation in the momentum family first and use
Eq.~\eqref{pp-top-pairing} to obtain the equation for the original
time-integral topsector.  Denoting the auxiliary canonical basis by
$\bm f_{\rm aux}$ and its lift by $\bm f^{(++)}_T$, we obtain
\begin{equation}
\begin{aligned}
 \dd\bm f_{\rm aux}&=\eps\sum_W R_W\,\dd\log W\,\bm f_{\rm aux},\\
 \dd\bm f^{(++)}_T&=\eps\sum_W R_W\,\dd\log W\,\bm f^{(++)}_T
 \qquad\bmod\ J^{\mathrm{sub}}.
\end{aligned}
\label{pp-top-cde-lift}
\end{equation}
The contact couplings of the full system are determined separately by
reduction in the original ordered family.

\subsubsection{Topsector logarithmic forms}
\label{sec:pp-top-dlogs}

The mixed-sector construction of Section~\ref{sec:pm-companion} now
guides the ordered topsector.  At the level of the auxiliary momentum
family, reversing the first external energy introduces an index-dependent
sign:
\begin{equation}
 E_1\longmapsto-E_1,\qquad (-1)^{n_1}.
\end{equation}
The common non-integer phase is independent of the kinematics and drops
out of the connection.  This correspondence relates the momentum IBP
systems; the ordered integrals still require the time pairing and the
contact terms described above.

For the ordered family, the three contact forms already occupy the first
three indices.  We therefore label the seven topsector forms from four
to ten.  Starting from the mixed forms in
Eq.~\eqref{eq:mixed-forms}, we replace their energy denominators
by the ordered ones of Eq.~\eqref{pp-energy-denominators} and increase
each form's label by three.  The replacement also acts on their
differentials.  The resulting ordered forms are
\begin{equation}
\begin{aligned}
 \Phi_4&=\omega_+\wedge\omega_-,&
 \Phi_5&=2\omega_-\wedge\dd\log z_3,\\
 \Phi_6&=-2\omega_-\wedge\dd\log z_4,&
 \Phi_7&=2\dd\log z_3\wedge\dd\log z_2,\\
 \Phi_8&=2\dd\log z_4\wedge\dd\log z_2,&
 \Phi_9&=-4\omega_-\wedge\dd\log z_1,\\
 \Phi_{10}&=2\dd\log z_1\wedge\dd\log z_2.&&
\end{aligned}
\label{pp-seven-dlogs}
\end{equation}

Expanding the wedge products at fixed external kinematics gives
\begin{equation}
\begin{aligned}
 \Phi_4&=\frac{8p^2\dd z_1\wedge\dd z_2}{\mathcal P},&
 \Phi_5&=\frac{8p(s^2-p^2)\dd z_1\wedge\dd z_2}{\mathcal P z_3},\\
 \Phi_6&=\frac{8p(s^2-p^2)\dd z_1\wedge\dd z_2}{\mathcal P z_4},&
 \Phi_7&=-\frac{2\dd z_1\wedge\dd z_2}{z_3z_2},\\
 \Phi_8&=\frac{2\dd z_1\wedge\dd z_2}{z_4z_2},&
 \Phi_9&=\frac{8p(s^2-p^2)\dd z_1\wedge\dd z_2}{z_1\mathcal P},\\
 \Phi_{10}&=\frac{2\dd z_1\wedge\dd z_2}{z_1 z_2}.&&
\end{aligned}
 \label{eq:dlog-kernels}
\end{equation}

\paragraph{From $\dd \log$ forms to integrals}
Matching the rational kernels in Eq.~\eqref{eq:dlog-kernels}
to the Baikov representation gives the three physical-dimensional
representatives, with regulator-dependent normalization treated
separately:
\begin{equation}
\begin{aligned}
 \Phi_7&\longleftrightarrow-pJ[1,0;1,2],\\
 \Phi_8&\longleftrightarrow pJ[0,1;1,2],\\
 \Phi_{10}&\longleftrightarrow pJ[0,0;2,2].
\end{aligned}
 \label{pp-direct-periods}
\end{equation}
The other four forms have the following bare combinations in the lowered
spatial dimension:
\begin{align}
 \Phi_4\longleftrightarrow{}&pJ[0,0;1,1; d-2],\notag\\
 \Phi_5\longleftrightarrow{}&\Bigl[
 J[1,0;-1,1; d-2]+2J[1,0;0,0; d-2]\notag\\
 &\hspace{15mm}+J[1,0;1,-1; d-2]
                  -p^2J[1,0;1,1; d-2]\Bigr],\notag\\
 \Phi_6\longleftrightarrow{}&\Bigl[
 J[0,1;-1,1; d-2]+2J[0,1;0,0; d-2]\notag\\
 &\hspace{15mm}+J[0,1;1,-1; d-2]
                  -p^2J[0,1;1,1; d-2]\Bigr],\notag\\
 \Phi_9\longleftrightarrow{}&\Bigl[
 J[0,0;0,1; d-2]+2J[0,0;1,0; d-2]\notag\\
 &\hspace{15mm}+J[0,0;2,-1; d-2]
                  -p^2J[0,0;2,1; d-2]\Bigr].
 \label{pp-dlog-J-combinations}
\end{align}
We use the relative normalization introduced in
Eq.~\eqref{eq:mixed-periods}; the measure and Gamma factors
are recorded in
Appendix~\ref{app:dimension-recurrence}.
Applying Eq.~\eqref{pp-dimension-recurrence} and the spatial IBPs to the
normalized combinations gives the physical-dimensional representatives
\begin{align}
 \Phi_4\longleftrightarrow{}&
 \delta pJ[1,0;1,2]-\delta pJ[0,1;1,2]
                         -\frac p\eps J[0,0;1,3],\notag\\
 \Phi_5\longleftrightarrow{}&\frac{2(2\eps-1)}\eps J[1,0;1,1],\notag\\
 \Phi_6\longleftrightarrow{}&\frac{2(2\eps-1)}\eps J[0,1;1,1],\notag\\
 \Phi_9\longleftrightarrow{}&-\frac1\eps J[0,0;0,3]
                    +\frac{4\eps-1}\eps J[0,0;1,2].
 \label{eq:top-reps}
\end{align}
These single-integral and radial representatives follow from the
dimension recurrence of Appendix~\ref{app:dimension-recurrence} combined
with the momentum IBP identities of Appendix~\ref{app:pp-spatial-ibps}.

The inverse map is
\begin{equation}
 J[n_1,n_2;b_1,b_2]\longmapsto
 \frac{J^{\mathrm{top}}_{n_1-1,n_2-1;b_1,b_2}}
 {(1+\delta\eps)_{n_1-1}(1+\delta\eps)_{n_2-1}}.
 \label{pp-inverse-time-map}
\end{equation}
For example, the inverse Pochhammer factors give
\begin{equation}
 pJ[0,0;2,2]\longmapsto
 \delta^2\eps^2pJ^{\mathrm{top}}_{-1,-1;2,2}.
\end{equation}
After the lift, we obtain the basis in the full $J^{\mathrm{top}}\oplus J^{\mathrm{sub}}$ family.
The unreduced combinations and
their exact simplifications are recorded in Appendix~\ref{app:pp-lifts}.

\subsection{Construction of the canonical differential equation}
\label{sec:pp-canonical-cde}

We order the constructed basis as the following block form, so that the canonical differential equation
becomes block triangular
\begin{equation}
 \bm f^{(++)}
 =(\bm f^{(++)}_C\mid\bm f^{(++)}_H\mid
   \bm f^{(++)}_V\mid f^{(++)}_{10})^T,
 \label{eq:block-order}
\end{equation}
The block vectors are
\begin{equation}
\begin{aligned}
 \bm f^{(++)}_C&=(f^{(++)}_1,f^{(++)}_2,f^{(++)}_3)^T,\\
 \bm f^{(++)}_H&=(f^{(++)}_4,f^{(++)}_5,f^{(++)}_6)^T,\\
 \bm f^{(++)}_V&=(f^{(++)}_7,f^{(++)}_8,f^{(++)}_9)^T.
\end{aligned}
\end{equation}
The contact block is defined in \eqref{eq:contact-basis}.  For the first block,
the basis are
\begin{subequations}
\label{eq:top-basis}
\begin{equation}
\begin{aligned}
 f^{(++)}_4&=\delta^2\eps p
 (J^{\mathrm{top}}_{0,-1;1,2}-J^{\mathrm{top}}_{-1,0;1,2}-J^{\mathrm{top}}_{-1,-1;1,3}),\\
 f^{(++)}_5&=2\delta(2\eps-1)J^{\mathrm{top}}_{0,-1;1,1},\\
 f^{(++)}_6&=2\delta(2\eps-1)J^{\mathrm{top}}_{-1,0;1,1}.
\end{aligned}
\end{equation}
The second block is
\begin{equation}
\begin{aligned}
 f^{(++)}_7&=-\delta\eps pJ^{\mathrm{top}}_{0,-1;1,2},\\
 f^{(++)}_8&=\delta\eps pJ^{\mathrm{top}}_{-1,0;1,2},\\
 f^{(++)}_9&=\delta^2\eps
 [-J^{\mathrm{top}}_{-1,-1;0,3}+(4\eps-1)J^{\mathrm{top}}_{-1,-1;1,2}].
\end{aligned}
\end{equation}
Finally, the last block is
\begin{equation}
 f^{(++)}_{10}=\delta^2\eps^2pJ^{\mathrm{top}}_{-1,-1;2,2}.
\end{equation}
\end{subequations}
Here we use the simplified expressions of $f^{(++)}_5$, $f^{(++)}_6$ and $f^{(++)}_9$ after IBP reduction.
The original complicated expressions from inverse map \eqref{pp-inverse-time-map} are recorded in Appendix~\ref{app:pp-lifts}.

After rescaling all external scales, the integral families obey the following scaling relations
\begin{equation}
\begin{aligned}
 J^{\mathrm{top}}_{a_1,a_2;b_1,b_2}(\lambda E_1,\lambda E_2,\lambda p)
 &=\lambda^{1-a_1-a_2-b_1-b_2-2(1+\delta)\eps}
 J^{\mathrm{top}}_{a_1,a_2;b_1,b_2}(E_1,E_2,p),\\[0.3em]
 J^{\mathrm{sub}}[n;b_1,b_2](\lambda k_t,\lambda p)
 &=\lambda^{2-n-b_1-b_2-2(1+\delta)\eps}
 J^{\mathrm{sub}}[n;b_1,b_2](k_t,p).
\end{aligned}
 \label{pp-integral-degrees}
\end{equation}
The integrals in the basis then have common scaling degree, $-2(1+\delta)\eps$.
The contact equation is already canonical:
\begin{equation}
 \dd\bm f^{(++)}_C
 =-2\eps[\dd\log p+\delta\dd\log k_t]\bm f^{(++)}_C.
 \label{pp-contact-dlog-cde}
\end{equation}
To obtain the differential equation, we differentiate the constructed
basis $\bm f^{(++)}$ with respect to each external scale $E_1,E_2,p$.
The derivative with respect to $p$ is evaluated as \eqref{pp-literal-K-derivative}.
Finally, we arrive at the canonical differential equation in the following form
\begin{equation}
 \dd\bm f^{(++)}
 =\eps\sum_{W\in\mathbb L_{++}}\widehat R_W(\delta)
 \dd\log W\,\bm f^{(++)}.
 \label{pp-three-scale-cde}
\end{equation}
where the CDE matrices are recorded in Appendix~\ref{app:ordered-residues}. And the letters are
\begin{equation}
 \mathbb L_{++}=\{p,E_1,E_2,p-E_1,p+E_1,p-E_2,p+E_2,k_t\}.
\end{equation}
It should be noticed that because we want to obtain the differential equation proportional to $\eps$,
so the matrices will contain $\delta$. Their sum satisfies
\begin{equation}
 \sum_W\widehat R_W=-2(1+\delta)\one_{10},
\end{equation}
in agreement with the mass dimension of basis.

\paragraph{Removing the overall scale}
After removing the overall scaling degree $-2(1+\delta)\eps$, we can introduce two dimensionless ratios
\begin{equation}
 x=\frac p{E_1},\qquad y=\frac p{E_2}.
\end{equation}
so the dimensionless basis vector can be written as 
\begin{equation}
 \bm f(x,y)=p^{2(1+\delta)\eps}\bm f^{(++)}(E_1,E_2,p).
 \label{eq:xy}
\end{equation}
Since $\sum_W\widehat R_W=-2(1+\delta)\one_{10}$, writing each letter as $p$
times a function of $x,y$ shows that this prefactor cancels the overall-scale
logarithm, which gives the two-variable CDE
\begin{equation}
 \dd\bm f(x,y)=\eps\sum_{W\in\mathbb L_{xy}}R_W(\delta)\dd\log W \bm f(x,y).
 \label{pp-xy-cde}
\end{equation}
where the alphabet is
\begin{equation}
 \mathbb L_{xy}=\{x,y,1-x,1+x,1-y,1+y,x+y\}.
 \label{pp-xy-alphabet}
\end{equation}
Also, the matrices $R_W$ can be obtained from $\widehat R_W$ by using \eqref{pp-xy-residue-map}.

\section{Boundary data and the MPL solution}
\label{sec:pp-solution}

A canonical equation organizes the solution as Chen iterated integrals
\cite{Chen:1977ena,Henn:2014qga}: each integration adds one logarithmic
kernel and one power of $\eps$. In order to obtain the full MPL solution,
we first need to determine the boundary vector for the differential equations. 

\subsection{The resolved large-energy boundary}

In the dimensionless variables of \eqref{eq:xy},
the divisors $x=0$, $y=0$, and $x+y=0$
meet at the large-energy corner, so this corner is a degenerate point.  An ordinary boundary value there
loses the ratio of the two external energies.  We therefore resolve
this corner by the blow-up
\begin{equation}
 x=uv,\qquad y=v,\qquad E_1=\frac p{uv},\qquad E_2=\frac p v.
 \label{pp-blowup-variables}
\end{equation}
Here $u=E_2/E_1$ refers to the ratio between two energies, while $v=p/E_2$ is the
second small parameter.  A blow-up of this kind was
used in \cite{Chen:2024glu} to determine the boundary conditions of massive
cosmological differential equations at tree level.  We approach the boundary point with $u,v\to0^+$,
corresponding to the hierarchy $E_1\gg E_2\gg p$.

From the change of variables $x=uv$, $y=v$, the CDE in \eqref{pp-xy-cde} becomes
\begin{equation}
 \dd\bm f=\eps\Omega_{uv}\bm f,\qquad
 \Omega_{uv}=\sum_{W\in\mathbb{L}_{uv}} \widetilde R_W(\delta)\,\dd\log W.
 \label{pp-uv-cde}
\end{equation}
and the corresponding alphabet is
\begin{equation}
 \mathbb{L}_{uv}=\{u,v,1-uv,1+uv,1-v,1+v,1+u\}.
 \label{pp-uv-letters}
\end{equation}
The matrices $\widetilde R_W$ are also given in
Appendix~\ref{app:corner-residues}.

Near the boundary point $(u,v)=(0,0)$, the CDE matrix can be written as
\begin{equation}
 \Omega_{uv}=\widetilde R_u\dd\log u+\widetilde R_v\dd\log v+\Omega^{\rm reg}_{uv},
 \qquad [\widetilde R_u,\widetilde R_v]=0.
 \label{eq:corner-residues}
\end{equation}
where the regular part means, for example $\dd\log(1+u)$, $\dd\log(1\pm v)$, and
$\dd\log(1\pm uv)$ are regular at the boundary point, which are collected in $\Omega^{\rm reg}_{uv}$.
They do not enter the leading indicial equations.
At leading power of u and v, the CDE reduces to the indicial system
\begin{equation}
 u\partial_u\bm f=\eps \widetilde R_u\bm f,\qquad
 v\partial_v\bm f=\eps \widetilde R_v\bm f.
 \label{eq:indicial}
\end{equation}
Since $\widetilde R_u$ and $\widetilde R_v$ commute, this system integrates
directly, and the leading behaviour near the boundary point is
\begin{equation}
 \bm f\sim u^{\eps \widetilde R_u}v^{\eps \widetilde R_v}\bm b(\eps,\delta).
 \label{eq:corner-ansatz}
\end{equation}
The constant vector $\bm b$ is the boundary vector we want to determine.
The indicial equations fix the eigenvalues of $\widetilde R_u$ and $\widetilde R_v$, which
behave as the leading powers of $u$ and $v$.  

Because $\widetilde R_u$ and $\widetilde R_v$ commute, they share a common eigenvector $\bm w$,
\begin{equation}
 \widetilde R_u\bm w=r\bm w,\qquad \widetilde R_v\bm w=s\bm w,
 \label{pp-common-eigenvectors}
\end{equation}
gives the solution $u^{r\eps}v^{s\eps}\bm w$ of
\eqref{eq:indicial}.  For the CDE matrices in
Appendix~\ref{app:corner-residues}, the eigenvalues are the roots of
\begin{equation}
 \begin{aligned}
 \det(r\one-\widetilde R_u)
 &=(r-\delta)^5(r-2\delta)^3(r-2-2\delta)(r-4-2\delta),\\
 \det(s\one-\widetilde R_v)
 &=(s-2\delta)^6(s-2-2\delta)^2(s-4-2\delta)^2.
 \end{aligned}
 \label{eq:charpoly}
\end{equation}
At generic $\delta$ there are ten common eigenvectors
$\bm w_1,\ldots,\bm w_{10}$, which form a basis, with six distinct pairs
$(r_A,s_A)$.  They are listed in Table~\ref{tab:corner-modes},
with unit vectors $\bm e_i$ in the order of \eqref{eq:block-order}.  Each $\bm w_A$ is labelled by its first
nonzero component, i.e. $\bm w_1=\bm e_1+\tfrac12\bm e_4$, $\bm w_2=\bm e_2-\tfrac12\bm e_9$  and etc.  
\begin{table}[H]
\centering
\small
\renewcommand{\arraystretch}{1.2}
\begin{tabular}{@{}lll@{}}
\toprule
$(r_A,s_A)$ & $A$ & Common eigenvector $\bm w_A$\\
\midrule
$(2\delta,2\delta)$ & $1$ & $\bm e_1+\tfrac12\bm e_4$\\
                    & $2$ & $\bm e_2-\tfrac12\bm e_9$\\
                    & $3$ & $\bm e_3-\tfrac12\bm e_{10}$\\
\addlinespace[2pt]
$(\delta,2\delta)$  & $4$ & $\bm e_4$\\
                    & $9$ & $\bm e_9$\\
                    & $10$ & $\bm e_{10}$\\
\addlinespace[2pt]
$(2+2\delta,2+2\delta)$ & $5$ & $\tfrac{\delta+2}{\delta}\bm e_5+\bm e_6+2\bm e_9$\\
$(\delta,2+2\delta)$    & $6$ & $\bm e_6-\delta\bm e_9$\\
$(4+2\delta,4+2\delta)$ & $7$ & $-\tfrac{\delta+4}{\delta}\bm e_7+\bm e_8+2\bm e_{10}$\\
$(\delta,4+2\delta)$    & $8$ & $\bm e_8-\tfrac{\delta}{2}\bm e_{10}$\\
\bottomrule
\end{tabular}
\caption{Common eigenvectors $\bm w_A$ of 
$\widetilde R_u,\widetilde R_v$ and their eigenvalue pairs $(r_A,s_A)$ at
generic $\delta$.  The coefficient of $\bm w_A$ in the boundary vector is
$\kappa_A$, Eq.~\eqref{eq:mode-expansion}.}
\label{tab:corner-modes}
\end{table}
\FloatBarrier

Since the $\bm w_A$ form a basis, we expand the boundary vector in it.
Inserting this expansion in \eqref{eq:corner-ansatz} gives
\begin{equation}
 \bm b(\eps,\delta)=\sum_{A=1}^{10}\kappa_A(\eps,\delta)\,\bm w_A,
 \qquad
 \bm f\sim\sum_{A=1}^{10}\kappa_A(\eps,\delta)\,
 u^{r_A\eps}v^{s_A\eps}\bm w_A.
 \label{eq:mode-expansion}
\end{equation}
Each term is one independent branch of the leading behaviour.  The
indicial equations fix $(r_A,s_A)$ and $\bm w_A$, but leave the ten
coefficients $\kappa_A$ undetermined.

In the next subsection, we determine the coefficients $\kappa_A$ by
matching the leading asymptotic terms of the basis integrals, obtained
by expansion by regions, to Eq.~\eqref{eq:mode-expansion}.
These coefficients fix the boundary vector $\bm b(\eps,\delta)$.

\subsection{Boundary vector from expansion by regions}
\label{sec:pp-regions}

We determine the coefficients $\kappa_A$ from the asymptotics of the
basis integrals in the limit $E_1\gg E_2\gg p$, i.e.\ $u,v\to0^+$, at
fixed regulators.  The asymptotic behaviors are computed by \emph{expansion by
regions}
\cite{Beneke:1997zp,Jantzen:2011regions,SemenovaSmirnovSmirnov:2019regions}.

\subsubsection{Expansion by regions}

For a general integral $J^{\mathrm{top}}_{a_1,a_2;b_1,b_2}$ define
\begin{equation}
 \alpha=a_1+\delta\eps,\qquad \beta=a_2+\delta\eps,
 \qquad \nu=\alpha+\beta+2,\qquad m=d-b_1-b_2.
 \label{pp-region-parameters}
\end{equation}
With $t_i=-\tau_i$, the Heaviside function restricts the times to
$t_2>t_1>0$.  Following the vacuum prescription, we rotate both time
contours, $t_i\to-\ii t_i$.  On the ordered domain
\begin{equation}
 z_3t_1+z_4t_2=k_t\,t_1+z_4(t_2-t_1)>0,
 \label{pp-ordered-positivity}
\end{equation}
so the rotated integral converges for every loop momentum, although
$z_3$ itself can be negative.  Using
\begin{equation}
 \ii^{a_1+a_2}(-\ii)^{\nu}=-\ee^{-\ii\pi\delta\eps},
\end{equation}
we obtain
\begin{equation}
 J^{\mathrm{top}}_{a_1,a_2;b_1,b_2}
 =-\ee^{-\ii\pi\delta\eps}\int[\dd^d\bm\ell]\,\frac{1}{z_1^{b_1}z_2^{b_2}}
 \int_0^\infty\dd t_1\int_0^\infty\dd t_2\;
 t_1^\alpha t_2^\beta\,\theta(t_2-t_1)\,\ee^{-z_3t_1-z_4t_2}.
 \label{eq:wick-rotation}
\end{equation}
The real exponentials select time scales inverse to the positive energy
combinations in the exponent, as specified in each region below.
In the case we're considering now, there exist three typical scales, so we expand the integrand of \eqref{eq:wick-rotation} in the three
loop-momentum regions
\begin{equation}
 z_1,z_2\sim p\quad\text{(soft)},\qquad
 z_1,z_2\sim E_2\quad\text{(semihard)},\qquad
 z_1,z_2\sim E_1\quad\text{(hard)}.
 \label{pp-three-regions}
\end{equation}
which provide different leading contributions.

\paragraph{Soft region}
For $z_1,z_2\sim p$, the loop energies are small compared with $E_1,E_2$,
so at leading power
\begin{equation}
 z_3\to E_1,\qquad z_4\to E_2 .
\end{equation}
The damping then selects $t_1\sim E_1^{-1}$ and $t_2\sim E_2^{-1}$.  On
this hierarchical time branch,
\begin{equation}
 \theta(t_2-t_1)\to1 .
\end{equation}
We retain $z_2=|\bm\ell+\bm p|$ exactly in the spatial bubble, and
\eqref{eq:wick-rotation} becomes
\begin{equation}
 J^{\mathrm{top},\mathrm{soft}}_{a_1,a_2;b_1,b_2}
 =-\ee^{-\ii\pi\delta\eps}\int[\dd^d\bm\ell]\,\frac{1}{z_1^{b_1}z_2^{b_2}}
 \int_0^\infty\dd t_1\,t_1^\alpha\ee^{-E_1t_1}
 \int_0^\infty\dd t_2\,t_2^\beta\ee^{-E_2t_2}.
\end{equation}
The two time integrals factorize, and the momentum integral is the
massless bubble \eqref{eq:contact-bubble}.  Thus
\begin{equation}
 J^{\mathrm{top},\mathrm{soft}}_{a_1,a_2;b_1,b_2}
 =-\ee^{-\ii\pi\delta\eps}\,\Gamma[\alpha+1]\Gamma[\beta+1]\,
 E_1^{-\alpha-1}E_2^{-\beta-1}p^m\mathcal B_{b_1b_2}(d).
 \label{eq:soft}
\end{equation}

\paragraph{Semihard region}
For $z_1\sim E_2$, we expand
\begin{equation}
 z_2\to z_1,\qquad z_3\to E_1,\qquad z_4\to E_2+2z_1 .
\end{equation}
The damping again selects $t_1\sim E_1^{-1}\ll t_2\sim E_2^{-1}$, so
$\theta(t_2-t_1)\to1$, and \eqref{eq:wick-rotation} becomes
\begin{equation}
\begin{aligned}
 J^{\mathrm{top},\mathrm{sh}}_{a_1,a_2;b_1,b_2}
 &=-\ee^{-\ii\pi\delta\eps}\int[\dd^d\bm\ell]\,\frac{1}{z_1^{b_1+b_2}}
 \int_0^\infty\dd t_1\,t_1^\alpha\ee^{-E_1t_1}
 \int_0^\infty\dd t_2\,t_2^\beta\ee^{-(E_2+2z_1)t_2}\\
 &=-\ee^{-\ii\pi\delta\eps}\,\Gamma[\alpha+1]\Gamma[\beta+1]\,
 E_1^{-\alpha-1}\int[\dd^d\bm\ell]\,\frac{(E_2+2z_1)^{-\beta-1}}{z_1^{b_1+b_2}}.
\end{aligned}
\end{equation}
The remaining integrand depends only on $z_1$, so the angular integration
gives
\begin{equation}
 \int[\dd^d\bm\ell]\,F(z_1)
 =c_d\int_0^\infty\dd z_1\,z_1^{d-1}F(z_1),\qquad
 c_d=\frac{\ee^{(1+2\delta)\gamma_E\eps}\,\Omega_{d-1}}{\pi^{(d-1)/2}}
 =\frac{2\sqrt\pi\,\ee^{(1+2\delta)\gamma_E\eps}}{\Gamma[d/2]}.
 \label{pp-radial-measure}
\end{equation}
With $z_1=E_2y/2$, the radial integral becomes a beta integral,
\begin{equation}
\begin{aligned}
 \int_0^\infty\dd z_1\,z_1^{m-1}(E_2+2z_1)^{-\beta-1}
 &=2^{-m}E_2^{m-\beta-1}\int_0^\infty\dd y\,\frac{y^{m-1}}{(1+y)^{\beta+1}}\\
 &=2^{-m}E_2^{m-\beta-1}
 \frac{\Gamma[m]\Gamma[\beta+1-m]}{\Gamma[\beta+1]}.
\end{aligned}
 \label{eq:semihard-beta}
\end{equation}
Therefore
\begin{equation}
 J^{\mathrm{top},\mathrm{sh}}_{a_1,a_2;b_1,b_2}
 =-\ee^{-\ii\pi\delta\eps}\,c_d\,2^{-m}\,
 \Gamma[\alpha+1]\Gamma[m]\Gamma[\beta+1-m]\,
 E_1^{-\alpha-1}E_2^{m-\beta-1}.
 \label{eq:semihard}
\end{equation}

\paragraph{Hard region}
For $z_1\sim E_1$, we expand
\begin{equation}
 z_2\to z_1,\qquad z_3\to E_1-2z_1,\qquad z_4\to2z_1 .
\end{equation}
Using \eqref{pp-radial-measure}, \eqref{eq:wick-rotation} becomes
\begin{equation}
 J^{\mathrm{top},\mathrm{hard}}_{a_1,a_2;b_1,b_2}
 =-\ee^{-\ii\pi\delta\eps}\,c_d\int_0^\infty\dd z_1\,z_1^{m-1}
 \int_0^\infty\dd t_1\int_0^\infty\dd t_2\;
 t_1^\alpha t_2^\beta\,\theta(t_2-t_1)\,\ee^{-(E_1-2z_1)t_1-2z_1t_2}.
 \label{pp-hard-start}
\end{equation}
Since $z_3$ changes sign at $z_1=E_1/2$, the two times must be treated
together.  We set
\begin{equation}
 t_2=\lambda,\qquad t_1=z\lambda,\qquad
 \dd t_1\dd t_2=\lambda\,\dd\lambda\,\dd z,\qquad 0<z<1,
 \label{pp-hard-time-variables}
\end{equation}
so that the ordering becomes $0<z<1$ and the exponent becomes
\begin{equation}
 (E_1-2z_1)t_1+2z_1t_2=\lambda\,[zE_1+2z_1(1-z)]>0 .
\end{equation}
Both times are therefore of order $E_1^{-1}$.  The $\lambda$ integral
gives
\begin{equation}
 \int_0^\infty\dd t_1\int_0^\infty\dd t_2\;
 t_1^\alpha t_2^\beta\,\theta(t_2-t_1)\,\ee^{-(E_1-2z_1)t_1-2z_1t_2}
 =\Gamma[\nu]\int_0^1
 \frac{z^\alpha\,\dd z}{[zE_1+2z_1(1-z)]^{\nu}}.
 \label{eq:hard-time}
\end{equation}
At fixed $z$, the substitution
\begin{equation}
 z_1=\frac{zE_1}{2(1-z)}\,y
\end{equation}
turns the radial integral into a beta integral,
\begin{equation}
\begin{aligned}
 \int_0^\infty\dd z_1\,\frac{z_1^{m-1}}{[zE_1+2z_1(1-z)]^{\nu}}
 &=\frac{(zE_1)^{m-\nu}}{[2(1-z)]^{m}}
 \int_0^\infty\dd y\,\frac{y^{m-1}}{(1+y)^{\nu}}\\
 &=2^{-m}(zE_1)^{m-\nu}(1-z)^{-m}
 \frac{\Gamma[m]\Gamma[\nu-m]}{\Gamma[\nu]}.
\end{aligned}
 \label{pp-hard-step-q}
\end{equation}
Using $\alpha+1-\nu=-\beta-1$, the remaining $z$ integral is
\begin{equation}
 \int_0^1\dd z\,z^{m-\beta-2}(1-z)^{-m}
 =\frac{\Gamma[m-\beta-1]\Gamma[1-m]}{\Gamma[-\beta]}.
 \label{pp-hard-step-z}
\end{equation}
Collecting the three factors,
\begin{equation}
 J^{\mathrm{top},\mathrm{hard}}_{a_1,a_2;b_1,b_2}
 =-\ee^{-\ii\pi\delta\eps}\,c_d\,2^{-m}
 \frac{\Gamma[m]\Gamma[\nu-m]\Gamma[m-\beta-1]\Gamma[1-m]}
 {\Gamma[-\beta]}\,E_1^{m-\nu}.
 \label{eq:hard}
\end{equation}

With
\begin{equation}
 E_1=\frac{p}{uv},\qquad E_2=\frac{p}{v},
\end{equation}
we can obtain the leading contributions corresponding to three regions
\begin{align}
 J^{\mathrm{top},\mathrm{soft}}&\propto
 u^{a_1+1+\delta\eps}v^{a_1+a_2+2+2\delta\eps},
 \label{eq:soft-scaling}\\
 J^{\mathrm{top},\mathrm{sh}}&\propto
 u^{a_1+1+\delta\eps}v^{a_1+a_2+b_1+b_2+2-d+2\delta\eps},
 \label{eq:semihard-scaling}\\
 J^{\mathrm{top},\mathrm{hard}}&\propto
 (uv)^{a_1+a_2+b_1+b_2+2-d+2\delta\eps}.
 \label{eq:hard-scaling}
\end{align}
With $d=3-2\eps$, the coefficients of $\eps$ in the powers of $(u,v)$
are $(\delta,2\delta)$, $(\delta,2+2\delta)$ and
$(2+2\delta,2+2\delta)$, respectively.

\subsubsection{Extraction of the coefficients}
\label{sec:mode-matching}

To isolate the coefficient $\kappa_A$, we multiply the leading behaviour
\eqref{eq:mode-expansion} by the dual rows of the ten eigenvectors,
\begin{equation}
 \begin{pmatrix}\bm w_1^\vee\\ \vdots\\ \bm w_{10}^\vee\end{pmatrix}
 =(\bm w_1\ \cdots\ \bm w_{10})^{-1},
 \qquad \bm w_A^\vee\bm w_B=\delta_{AB}.
 \label{pp-dual-rows}
\end{equation}
This gives
\begin{equation}
 \bm w_A^\vee\bm f
 \sim\sum_B \kappa_Bu^{r_B\eps}v^{s_B\eps}\bm w_A^\vee\bm w_B
 =\kappa_Au^{r_A\eps}v^{s_A\eps}.
 \label{eq:mode-projection}
\end{equation}
Thus $\kappa_A$ is the coefficient of $u^{r_A\eps}v^{s_A\eps}$ in the
combination $\bm w_A^\vee\bm f$, and this power selects the region through
\eqref{eq:soft-scaling}--\eqref{eq:hard-scaling}.
We now go through the eigenvalue pairs of
Table~\ref{tab:corner-modes}.

\paragraph{Contact pair $(2\delta,2\delta)$}
The dual rows are
\begin{equation}
 \bm w_1^\vee\bm f=f_1,\qquad \bm w_2^\vee\bm f=f_2,\qquad \bm w_3^\vee\bm f=f_3 .
\end{equation}
The contact integrals are known exactly from
\eqref{eq:contact-factor},
\begin{equation}
 \bm f_C=\left(\frac{uv}{1+u}\right)^{2\delta\eps}
 (\kappa_1,\kappa_2,\kappa_3)^T,
 \label{eq:contact-solution}
\end{equation}
with
\begin{equation}
\begin{aligned}
 \kappa_1&=-\ee^{-\ii\pi\delta\eps}\Gamma[1+2\delta\eps]
       \frac{(1-2\eps)^2}{\eps^2}\mathcal B_{11}(d),\\
 \kappa_2&=-\ee^{-\ii\pi\delta\eps}\Gamma[1+2\delta\eps]
       \frac{4\eps-1}\eps\mathcal B_{12}(d),\\
 \kappa_3&=-\ee^{-\ii\pi\delta\eps}\Gamma[1+2\delta\eps]\mathcal B_{22}(d).
\end{aligned}
\label{eq:contact-kappas}
\end{equation}

\paragraph{Soft pair $(\delta,2\delta)$}
The dual rows are
\begin{equation}
\begin{aligned}
 \bm w_4^\vee\bm f&=f_4-\tfrac12f_1,\\
 \bm w_9^\vee\bm f&=f_9+\tfrac12f_2+\delta(f_6-f_5),\\
 \bm w_{10}^\vee\bm f&=f_{10}+\tfrac12f_3+\tfrac\delta2(f_7+f_8).
\end{aligned}
\end{equation}
The power $u^{\delta\eps}v^{2\delta\eps}$ comes only from the soft
region.  In $\bm w_4^\vee\bm f$, the three integrals of $f_4$ have the soft
powers
\begin{equation}
 J^{\mathrm{top}}_{0,-1;1,2}\propto u^{1+\delta\eps}v^{1+2\delta\eps},\qquad
 J^{\mathrm{top}}_{-1,0;1,2}\propto u^{\delta\eps}v^{1+2\delta\eps},\qquad
 J^{\mathrm{top}}_{-1,-1;1,3}\propto u^{\delta\eps}v^{2\delta\eps},
\end{equation}
so only the last one contributes.  For $J^{\mathrm{top}}_{-1,-1;1,3}$,
\begin{equation}
 \alpha=\beta=-1+\delta\eps,\qquad m=-1-2\eps,
\end{equation}
and \eqref{eq:soft} gives
\begin{equation}
 f_4\supset-\delta^2\eps p^{1+2(1+\delta)\eps}
 J^{\mathrm{top},\mathrm{soft}}_{-1,-1;1,3}
 =\delta^2\eps\,\ee^{-\ii\pi\delta\eps}\Gamma[\delta\eps]^2
 \mathcal B_{13}(d)\,u^{\delta\eps}v^{2\delta\eps}.
\end{equation}
In $\bm w_9^\vee\bm f$, the first term of
\begin{equation}
 f_9=\delta^2\eps p^{2(1+\delta)\eps}\bigl[-J^{\mathrm{top}}_{-1,-1;0,3}
 +(4\eps-1)J^{\mathrm{top}}_{-1,-1;1,2}\bigr]
\end{equation}
has a scaleless soft bubble, $\mathcal B_{03}(d)=0$, so $\kappa_9$ comes
from the second term.  In $\bm w_{10}^\vee\bm f$, $\kappa_{10}$ comes from
\begin{equation}
 f_{10}=\delta^2\eps^2p^{1+2(1+\delta)\eps}J^{\mathrm{top}}_{-1,-1;2,2}.
\end{equation}
The remaining terms of
$\bm w_9^\vee\bm f$ and $\bm w_{10}^\vee\bm f$ do not contribute at this
power.  The results are
\begin{equation}
\begin{aligned}
 \kappa_4&=\delta^2\eps\,\ee^{-\ii\pi\delta\eps}
 \Gamma[\delta\eps]^2\mathcal B_{13}(d),\\
 \kappa_9&=\delta^2\eps(1-4\eps)\ee^{-\ii\pi\delta\eps}
 \Gamma[\delta\eps]^2\mathcal B_{12}(d),\\
 \kappa_{10}&=-\delta^2\eps^2\ee^{-\ii\pi\delta\eps}
 \Gamma[\delta\eps]^2\mathcal B_{22}(d).
\end{aligned}
\label{eq:soft-kappas}
\end{equation}

\paragraph{Semihard pair $(\delta,2+2\delta)$}
The dual row is
\begin{equation}
 \bm w_6^\vee\bm f=f_6-\frac{\delta}{\delta+2}f_5 .
\end{equation}
The power $u^{\delta\eps}v^{(2+2\delta)\eps}$ comes from the semihard
region.  It is supplied by
\begin{equation}
 f_6=2\delta(2\eps-1)p^{2(1+\delta)\eps}J^{\mathrm{top}}_{-1,0;1,1},\qquad
 \alpha=-1+\delta\eps,\quad \beta=\delta\eps,\quad m=1-2\eps,
\end{equation}
while the semihard term of $f_5$ starts one power of $u$ higher.
Inserting these indices in \eqref{eq:semihard} gives
\begin{equation}
 \kappa_6=\delta\,2^{2\eps}
 \frac{2\sqrt\pi\,\ee^{(1+2\delta)\gamma_E\eps}}{\Gamma[d/2]}
 \ee^{-\ii\pi\delta\eps}
 \Gamma[\delta\eps]\Gamma[(\delta+2)\eps]\Gamma[2-2\eps].
 \label{eq:semihard-kappa}
\end{equation}

\paragraph{Hard pair $(2+2\delta,2+2\delta)$}
The dual row is
\begin{equation}
 \bm w_5^\vee\bm f=\frac{\delta}{\delta+2}f_5 .
\end{equation}
The power $(uv)^{(2+2\delta)\eps}$ comes from the hard region of
\begin{equation}
 f_5=2\delta(2\eps-1)p^{2(1+\delta)\eps}J^{\mathrm{top}}_{0,-1;1,1},\qquad
 \alpha=\delta\eps,\quad \beta=-1+\delta\eps,\quad m=1-2\eps .
\end{equation}
Inserting these indices in \eqref{eq:hard} gives
\begin{equation}
 \kappa_5=\frac{\delta^2\,2^{2\eps}}{\delta+2}
 \frac{2\sqrt\pi\,\ee^{(1+2\delta)\gamma_E\eps}}{\Gamma[d/2]}
 \ee^{-\ii\pi\delta\eps}
 \frac{\Gamma[2-2\eps]\Gamma[2(1+\delta)\eps]\Gamma[2\eps]
 \Gamma[1-(\delta+2)\eps]}{\Gamma[1-\delta\eps]}.
 \label{eq:hard-kappas}
\end{equation}

\paragraph{The pairs $(\delta,4+2\delta)$ and $(4+2\delta,4+2\delta)$}
The dual rows are
\begin{equation}
 \bm w_7^\vee\bm f=-\frac{\delta}{\delta+4}f_7,\qquad
 \bm w_8^\vee\bm f=f_8+\frac{\delta}{\delta+4}f_7 .
\end{equation}
The three regions in Eqs.~\eqref{eq:soft-scaling}--\eqref{eq:hard-scaling}
give no contribution with exponent $(4+2\delta)\eps$ in $v$, so
\begin{equation}
 \kappa_7=\kappa_8=0 .
\end{equation}

\subsubsection{Exact boundary vector and its \texorpdfstring{$\eps$}{epsilon} expansion}

Summing the common eigenvectors with their coefficients as in
\eqref{eq:mode-expansion} gives
\begin{equation}
 \boxed{\bm b(\eps,\delta)=
 \begin{pmatrix}
 \kappa_1\\\kappa_2\\\kappa_3\\\kappa_4+\kappa_1/2\\(\delta+2)\kappa_5/\delta\\\kappa_6+\kappa_5\\
 0\\0\\\kappa_9-\kappa_2/2-\delta \kappa_6+2\kappa_5\\\kappa_{10}-\kappa_3/2
 \end{pmatrix}.}
 \label{eq:boundary-vector}
\end{equation}
Expanding the Gamma functions, the leading poles are
\begin{equation}
\begin{array}{c|rrrrrr}
 \kappa&\kappa_1&\kappa_2&\kappa_4&\kappa_6&\kappa_5\qquad&\kappa_9\\ \hline
 \lim_{\eps\to0}\eps^2\kappa
 &2&2&-2&\frac{4}{\delta+2}&\frac{\delta^2}{(\delta+1)(\delta+2)}&2
\end{array}
\end{equation}
while $\kappa_3,\kappa_{10}$ start at $\eps^0$.  At generic $\delta$,
\begin{align}
 \bm b^{[-2]}&=(2,2,0,-1,\delta/(1+\delta),
 (\delta+2)/(1+\delta),0,0,(1-\delta)/(1+\delta),0)^T,
 \label{pp-gpl-I0}\\
 \bm b^{[-1]}&=\bigl(-2\ii\pi\delta-4\log2,
 -2\ii\pi\delta-4\log2,0,\ii\pi\delta+2\log2,
 -\ii\pi\delta^2/(1+\delta),\notag\\
 &\quad-\ii\pi\delta(\delta+2)/(1+\delta),0,0,
 -\ii\pi\delta(1-\delta)/(1+\delta)-2\log2,0\bigr)^T.
 \label{pp-gpl-b-minus-one}
\end{align}
Here $\bm b=\sum_{j\geq-2}\eps^j\bm b^{[j]}$.
The higher coefficients needed through $\eps^1$ are specified compactly
in Appendix~\ref{app:pp-explicit-gpl}.

The normalization \eqref{eq:loop-measure} cancels the Euler constants which appear in
series expansions of $\eps$, this is the reason we use the normalization.  Using
\begin{equation}
 \log\Gamma[1+c\eps]=-\gamma_Ec\eps
 +\sum_{n\geq2}\frac{(-1)^n}{n}\zeta(n)c^n\eps^n,
 \label{eq:log-gamma}
\end{equation}
each boundary coefficient has weight $j+2$ at order $\eps^j$:
$\ii\pi$ and $\log2$ have weight one, and $\zeta(n)$ has weight $n$. Here the regulator $\delta$ can be
seen as a rational number, so it is assigned with weight zero.  
\subsection{Direct MPL transport}
\label{sec:pp-direct-gpl}

We use the boundary vector $\bm b$ from Section~\ref{sec:pp-regions}
as the initial data for the CDE. We first integrate in $v$ along $u=0$,
then in $u$ at fixed $v$, following
$\gamma:(0,0)\to(0,v)\to(u,v)$. To integrate along \(u=0\), we first factor
out \(u^{\epsilon\widetilde R_u}\) and then take the limit \(u\to0^+\).
The path avoids further singularities for $0<v<1$ and $0<uv<1$,
which is the chamber $E_1,E_2>p$.

We regularize the logarithms at the starting endpoints by subtracting
the divergent term at a lower cutoff $\eta$,
\begin{equation}
 \operatorname{Reg}\!\int_0^z\frac{\dd t}{t}
 =\lim_{\eta\to0^+}\left[\int_\eta^z\frac{\dd t}{t}+\log\eta\right]
 =\log z,\qquad z>0.
 \label{pp-gpl-endpoint-log}
\end{equation}

The MPL alphabets on the $v$ and $u$ segments are
$\{0,1,-1\}$ and $\{0,-1,1/v,-1/v\}$, respectively.
We use the convention
\begin{equation}
 G(w_1,\ldots,w_n;z)=\int_0^z\frac{\dd t}{t-w_1}
 G(w_2,\ldots,w_n;t),\qquad G(;z)=1,
 \label{pp-gpl-definition}
\end{equation}
with $G(0,\ldots,0;z)=\log^n z/n!$.
The matrices of the later $u$ segment stand to the left of those of the
earlier $v$ segment.

Write the Laurent solution as
\begin{equation}
 \bm f=\eps^{-2}\sum_{n\geq0}\eps^n\bm f^{(n)},\qquad
 \dd\bm f^{(0)}=0,\qquad
 \dd\bm f^{(n)}=\Omega_{uv}\bm f^{(n-1)}\quad(n\geq1).
 \label{eq:mpl-recursion}
\end{equation}
The index $n$ denotes the weight of $\bm f^{(n)}$, which multiplies
$\eps^{n-2}$. The leading coefficient is $\bm f^{(0)}=\bm b^{[-2]}$.
Using the endpoint prescription in Eq.~\eqref{pp-gpl-endpoint-log},
the next coefficient is
\begin{equation}
 \bm f^{(1)}=\bm b^{[-1]}+
 \int_\gamma\Omega_{uv}\bm b^{[-2]}
 =\bm b^{[-1]}+\sum_W \widetilde R_W\log W\,\bm b^{[-2]}.
 \label{eq:weight-one}
\end{equation}
Here $\bm b^{[-1]}$ is the constant term at this order, fixed by the
boundary vector, and the integral gives the dependence on $u$ and $v$.
For $i=3,7,8,10$, both the boundary term and the path integral
in \eqref{eq:weight-one} vanish.

\Needspace{12\baselineskip}
The components at this order are
\begin{equation}
\begin{gathered}
\begin{aligned}
 f^{(1)}_1=f^{(1)}_2
 &=-2\ii\pi\delta-4\log2+4\delta\log\frac{uv}{1+u},\\
 f^{(1)}_4&=\ii\pi\delta+2\log2-2\delta\log\frac{v}{1+v},\\
 f^{(1)}_5&=-\frac{\ii\pi\delta^2}{1+\delta}
 +2\delta\log\frac{uv}{1+u},\\
 f^{(1)}_6&=-\frac{\ii\pi\delta(\delta+2)}{1+\delta}
 +2\delta\log\frac{uv}{1+u}+4\log\frac{v}{1+v},\\
 f^{(1)}_9&=-\frac{\ii\pi\delta(1-\delta)}{1+\delta}
 -2\log2-2\delta\log\frac{v}{1+v}.
\end{aligned}\\[3pt]
 f^{(1)}_3=f^{(1)}_7=f^{(1)}_8=f^{(1)}_{10}=0.
\end{gathered}
\end{equation}
Appendix~\ref{app:pp-explicit-gpl} gives both the
finite MPL-word formula and the component expressions for
$\bm f^{(2)}$ and $\bm f^{(3)}$:
\begin{equation}
 \bm f=\eps^{-2}\bm f^{(0)}+\eps^{-1}\bm f^{(1)}
 +\bm f^{(2)}+\eps\bm f^{(3)}+O(\eps^2).
 \label{eq:laurent-series}
\end{equation}
Flatness and the regularized endpoint condition check the recursion
independently of a local power-series representation.

\subsection{Local series and numerical checks}

To check the analytic boundary vector, we use it to construct a local
Frobenius solution and compare the result with direct numerical
integration of the original integrals.  The series is expanded in
$u,v$ at fixed $\eps,\delta$, without expanding in $\eps$.
For each distinct common-residue eigenvalue pair $(r,s)$ at generic
$\delta$, write
\begin{equation}
 \bm\Phi_{r,s}(u,v)=u^{r\eps}v^{s\eps}
 \sum_{m,n\geq0}\bm h_{mn}^{(r,s)}u^mv^n.
 \label{pp-frobenius-ansatz}
\end{equation}
Writing the regular part in Eq.~\eqref{eq:corner-residues} as
$\Omega^{\rm reg}_{uv}=A_u^{\rm reg}\dd u+A_v^{\rm reg}\dd v$ and expanding
\(A_u^{\rm reg}=\sum A^{u}_{ab}u^av^b\) and
\(A_v^{\rm reg}=\sum A^{v}_{ab}u^av^b\) gives the compatible recursions
below, with the label $(r,s)$ suppressed on $\bm h_{mn}$:
\begin{align}
 [(r\eps+m)\one-\eps \widetilde R_u]\bm h_{mn}
 &={\eps}\sum_{a=0}^{m-1}\sum_{b=0}^{n}
 A^{u}_{ab}\bm h_{m-1-a,n-b},\notag\\
 [(s\eps+n)\one-\eps \widetilde R_v]\bm h_{mn}
 &={\eps}\sum_{a=0}^{m}\sum_{b=0}^{n-1}
 A^{v}_{ab}\bm h_{m-a,n-1-b}.
 \label{eq:frobenius-recursion}
\end{align}
The leading coefficients reproduce the boundary vector of
Eq.~\eqref{eq:boundary-vector}:
\begin{equation}
 \sum_{(r,s)}\bm h_{00}^{(r,s)}=\bm b(\eps,\delta).
 \label{eq:frobenius-seed}
\end{equation}
All higher coefficients follow from Eq.~\eqref{eq:frobenius-recursion}.
We compare the summed series $\bm f_N^{\rm Fr}$, truncated at
$m,n\leq N=150$, with direct evaluations of
Eqs.~\eqref{eq:contact-basis} and
\eqref{eq:top-basis} at $p=1$.
Each original integral is evaluated only in its convergence domain;
the three regulator windows in Table~\ref{tab:numerics}
cover all ten components.  At each regulator point, we report
\begin{equation}
 \begin{aligned}
 E(\eps,\delta)&=\max_{(u,v)}\max_i
 \frac{\bigl|f_i^{\rm dir}-f_{i,N}^{\rm Fr}\bigr|}
 {\bigl|f_{i,N}^{\rm Fr}\bigr|},\\
 (u,v)&\in\{(1/3,1/4),(1/5,1/6),(1/2,1/3),(1/4,1/2)\}.
 \end{aligned}
 \label{eq:rel-error}
\end{equation}
Over all 516 componentwise comparisons the largest relative error is
$1.51\times10^{-41}$.

\begin{table}[!htbp]
 \centering
 \footnotesize
 \renewcommand{\arraystretch}{0.85}
 \begin{tabular}{cccr}
 \toprule
 $\eps$&$\delta$&$\delta\eps$&$E(\eps,\delta)$\\
 \midrule
 \multicolumn{4}{l}{$0<\eps<\frac12,\ \delta>0$: \ $f_2,f_3,f_5,f_6,f_7,f_8,f_{10}$}\\
 $1/10$&$1/2$&$1/20$&$4.48\times10^{-48}$\\
 $1/10$&$3/2$&$3/20$&$8.22\times10^{-48}$\\
 $1/10$&$7/2$&$7/20$&$1.10\times10^{-46}$\\
 $3/10$&$1/2$&$3/20$&$1.03\times10^{-46}$\\
 $3/10$&$3/2$&$9/20$&$6.49\times10^{-45}$\\
 $3/10$&$7/2$&$21/20$&$5.85\times10^{-43}$\\
 $7/20$&$1/2$&$7/40$&$2.50\times10^{-46}$\\
 $7/20$&$3/2$&$21/40$&$4.88\times10^{-45}$\\
 $7/20$&$7/2$&$49/40$&$9.58\times10^{-42}$\\
 \midrule
 \multicolumn{4}{l}{$-\frac12<\eps<0,\ \delta<-1$: \ $f_3,f_4,f_7,f_8,f_9,f_{10}$}\\
 $-1/10$&$-3/2$&$3/20$&$4.37\times10^{-45}$\\
 $-1/10$&$-3$&$3/10$&$2.69\times10^{-47}$\\
 $-1/10$&$-9/2$&$9/20$&$2.48\times10^{-46}$\\
 $-1/5$&$-3/2$&$3/10$&$2.69\times10^{-47}$\\
 $-1/5$&$-3$&$3/5$&$2.01\times10^{-45}$\\
 $-1/5$&$-9/2$&$9/10$&$9.92\times10^{-44}$\\
 $-3/10$&$-3/2$&$9/20$&$2.30\times10^{-46}$\\
 $-3/10$&$-3$&$9/10$&$7.24\times10^{-44}$\\
 $-3/10$&$-9/2$&$27/20$&$1.51\times10^{-41}$\\
 \midrule
 \multicolumn{4}{l}{$\frac12<\eps<1,\ \delta>0$: \ $f_1,f_5,f_6$}\\
 $3/5$&$1/3$&$1/5$&$1.07\times10^{-45}$\\
 $3/5$&$7/6$&$7/10$&$1.56\times10^{-44}$\\
 $9/10$&$1/3$&$3/10$&$6.32\times10^{-45}$\\
 $9/10$&$7/6$&$21/20$&$4.17\times10^{-43}$\\
 \bottomrule
 \end{tabular}
 \caption{Observed relative differences \eqref{eq:rel-error}
 between the direct integrals and the Frobenius series ($N=150$),
 maximized over the four kinematic points and the components listed for
 each regulator window.  The direct integrals use 50 significant digits
 and the series 420-bit ball arithmetic; the series truncated at
 $N=125$ and $N=150$ differ by at most $4\times10^{-34}$, and recomputing
 107 direct integrals with 72 digits and a different subdivision changes
 them by at most $10^{-44}$.}
 \label{tab:numerics}
\end{table}
\FloatBarrier

\section{Dual-\texorpdfstring{$\theta$}{theta} representation of the \texorpdfstring{$(++)$}{(++)} family}
\label{sec:dual-theta}

We now return to the complete $(++)$ component in
Eq.~\eqref{bubble-ordered-component}, which contains both time orderings.
We retain the definitions of $z_1,z_2,p$, the regulated loop measure and
the phase convention of Eq.~\eqref{pp-ordered-integral}.  Including the
complementary ordering gives the dual-$\theta$ family.  Its canonical
basis and differential equation follow by combining the single-$\theta$
results for the two orderings.

\paragraph{Integral family and master integrals}
Set $\Delta=\tau_1-\tau_2$.  To describe both time orderings and their
relative sign, use the propagator of Eq.~\eqref{conf-sk} and its
sign-weighted counterpart,
\begin{equation}
\begin{aligned}
 G_{++,0}(z_1;\tau_1,\tau_2)&=G_{++}(z_1;\tau_1,\tau_2),\\
 G_{++,1}(z_1;\tau_1,\tau_2)
 &=\frac{-\theta(\Delta)\ee^{-\ii z_1\Delta}
          +\theta(-\Delta)\ee^{\ii z_1\Delta}}{2z_1}.
\end{aligned}
\label{dual-theta-modes}
\end{equation}
The additional subscript $0$ or $1$ labels a relative sign between the orderings;
it is not a Schwinger--Keldysh contour index.  Since the two internal
lines share the same time difference, at unequal times their products
obey the fold relations used in \texttt{dSIBP}~\cite{ChenTao:2026dsibp}:
\begin{equation}
\begin{aligned}
 G_{++,1}(z_1)G_{++,1}(z_2)
 &=G_{++,0}(z_1)G_{++,0}(z_2),\\
 G_{++,0}(z_1)G_{++,1}(z_2)
 &=G_{++,1}(z_1)G_{++,0}(z_2).
\end{aligned}
 \label{dual-theta-fold}
\end{equation}
The common arguments $(\tau_1,\tau_2)$ are suppressed in this equation.
Thus, if the two lines carry indices $n_1,n_2\in\{0,1\}$, only
their sum $n=(n_1+n_2)\bmod2$ is needed.  With the same
normalization and denominator powers as the single-$\theta$ family,
we define
\begin{equation}
\begin{aligned}
 \tilde{J}^{(n)}_{a_1,a_2;b_1,b_2}
 ={}&\ii^{a_1+a_2}\int_{-\infty}^0\dd\tau_1\dd\tau_2\,
 (-\tau_1)^{a_1+\delta\eps}(-\tau_2)^{a_2+\delta\eps}
 \ee^{\ii(E_1\tau_1+E_2\tau_2)}\\
 &\times\int[\dd^d\bm\ell]\,
 \frac{(-1)^n\theta(\Delta)\ee^{-\ii(z_1+z_2)\Delta}
       +\theta(-\Delta)\ee^{\ii(z_1+z_2)\Delta}}
      {z_1^{b_1}z_2^{b_2}},
 \qquad n\in\{0,1\}.
\end{aligned}
\label{dual-theta-family}
\end{equation}
In particular, $n=0$ includes both terms of the $(++)$ propagators,
while $n=1$ takes their difference with the sign convention above.
The equal-time terms generated by time IBPs belong to the same contact
family $J^{\mathrm{sub}}[n;b_1,b_2]$ as in Eq.~\eqref{pp-contact-family}.

To relate this family to Eq.~\eqref{pp-ordered-integral}, define the
complementary ordering by
\begin{equation}
 \overline{J}^{\mathrm{top}}_{a_1,a_2;b_1,b_2}(E_1,E_2,p)
 =J^{\mathrm{top}}_{a_2,a_1;b_1,b_2}(E_2,E_1,p).
 \label{eq:reverse-order}
\end{equation}
Exchanging the vertices therefore gives the second ordering without
introducing a new set of time or momentum indices.  The dual-$\theta$
integrals are the sum and difference
\begin{equation}
\begin{aligned}
 \tilde{J}^{(0)}_{a_1,a_2;b_1,b_2}
 &=\overline{J}^{\mathrm{top}}_{a_1,a_2;b_1,b_2}
   +J^{\mathrm{top}}_{a_1,a_2;b_1,b_2},\\
 \tilde{J}^{(1)}_{a_1,a_2;b_1,b_2}
 &=\overline{J}^{\mathrm{top}}_{a_1,a_2;b_1,b_2}
   -J^{\mathrm{top}}_{a_1,a_2;b_1,b_2}.
\end{aligned}
 \label{eq:ordering-pair}
\end{equation}
so that
$J^{\mathrm{top}}=\tfrac12(\tilde{J}^{(0)}-\tilde{J}^{(1)})$ and
$\overline{J}^{\mathrm{top}}=\tfrac12(\tilde{J}^{(0)}+\tilde{J}^{(1)})$.

The parity splitting of the loop IBP system follows
\cite{Chen:2026dqp}.  For massless banana subgraphs, the additional fold
relations in \texttt{dSIBP}~\cite{ChenTao:2026dsibp} merge the parity
classes with the same total parity
\begin{equation}
 p_{\rm top}=(b_1+b_2+n)\bmod2.
 \label{dual-theta-parity}
\end{equation}
\needspace{10\baselineskip}
Including the internal-line exchange symmetry, IBP reduction gives 17
masters.  The contact subsector is not split by parity, while the
topsector has seven masters in each parity block.  A master vector is
ordered as $\bm f^{(++)}_{\mathrm{raw},2\theta}=(\bm f^{(++)}_{\mathrm{raw},C}\mid\bm f^{(++)}_{\mathrm{raw},T,+}\mid\bm f^{(++)}_{\mathrm{raw},T,-})^T$,
where $+$ and $-$ denote $p_{\rm top}=0$ and $1$, respectively, and
\begin{equation}
\begin{aligned}
 \bm f^{(++)}_{\mathrm{raw},C}={}&\bigl(J^{\mathrm{sub}}[0;1,1],J^{\mathrm{sub}}[0;1,2],
 J^{\mathrm{sub}}[0;2,2]\bigr)^T,\\
 \bm f^{(++)}_{\mathrm{raw},T,+}={}&\bigl(
 \tilde{J}^{(0)}_{0,0;0,0},\tilde{J}^{(0)}_{0,1;0,0},
 \tilde{J}^{(0)}_{1,0;0,0},\tilde{J}^{(0)}_{0,0;0,2},
 \tilde{J}^{(0)}_{0,1;0,2},\tilde{J}^{(0)}_{0,0;0,4},
 \tilde{J}^{(0)}_{0,0;2,2}\bigr)^T,\\
 \bm f^{(++)}_{\mathrm{raw},T,-}={}&\bigl(
 \tilde{J}^{(1)}_{0,0;0,0},\tilde{J}^{(1)}_{0,0;0,2},
 \tilde{J}^{(0)}_{0,0;0,1},\tilde{J}^{(0)}_{0,1;0,1},
 \tilde{J}^{(0)}_{0,0;0,3},\tilde{J}^{(0)}_{0,0;0,5},
 \tilde{J}^{(0)}_{0,0;2,3}\bigr)^T .
\end{aligned}
\label{dual-theta-masters}
\end{equation}

\paragraph{Canonical basis in sector and parity blocks}
Collect the single-$\theta$ topsector basis of
Eq.~\eqref{eq:top-basis} as
\begin{equation}
 \bm f^{(++)}_T=(f^{(++)}_4,\ldots,f^{(++)}_{10})^T.
\end{equation}
The complementary ordering is obtained by evaluating the same vector
at $(E_2,E_1,p)$.  In this basis order, define the constant matrix
\begin{equation}
 S=\begin{pmatrix}
 1&0&0&0&0&0&0\\
 0&0&1&0&0&0&0\\
 0&1&0&0&0&0&0\\
 0&0&0&0&1&0&0\\
 0&0&0&1&0&0&0\\
 0&0&0&0&0&-1&0\\
 0&0&0&0&0&0&1
 \end{pmatrix},
 \qquad S^2=\one_7.
 \label{dual-theta-swap-matrix}
\end{equation}
The definite-parity canonical elements are then
\begin{equation}
 \bm f^{(++)}_{T,\pm}(E_1,E_2,p)
 =\frac12\Bigl[\bm f^{(++)}_T(E_1,E_2,p)
       \pm S\bm f^{(++)}_T(E_2,E_1,p)\Bigr].
 \label{dual-theta-parity-basis}
\end{equation}
The full canonical vector, in the same sector order as $\bm f^{(++)}_{\mathrm{raw},2\theta}$,
is
\begin{equation}
 \bm f^{(++)}_{2\theta}
 =(\bm f^{(++)}_C\mid\bm f^{(++)}_{T,+}\mid
   \bm f^{(++)}_{T,-})^T.
\end{equation}
The contact block is unchanged from
Eq.~\eqref{eq:contact-basis}.  On the 17-component vector
containing this contact block and the two seven-component ordering
vectors, Eq.~\eqref{dual-theta-parity-basis} is a constant invertible
change of basis.  It preserves the canonical form and gives the
dual-$\theta$ MPL solutions by the same combinations of the
single-$\theta$ solutions and their $E_1\leftrightarrow E_2$ images.

\paragraph{Block differential equation}
The alphabet is again
\begin{equation}
 \mathbb L_{++}=\{p,E_1,E_2,p-E_1,p+E_1,p-E_2,p+E_2,k_t\},
 \qquad k_t=E_1+E_2.
 \label{dual-theta-alphabet}
\end{equation}
Write the single-$\theta$ residues in their $3+7$ sector decomposition as
\begin{equation}
 \widehat R_W=\begin{pmatrix}
 \widehat R_W^{(CC)}&0\\[2pt]
 \widehat R_W^{(TC)}&\widehat R_W^{(TT)}
 \end{pmatrix},
 \qquad W\in\mathbb L_{++},
 \label{eq:residue-blocks}
\end{equation}
and let $\overline W$ denote the letter obtained by exchanging
$E_1\leftrightarrow E_2$.  The matrices in
Appendix~\ref{app:ordered-residues} obey
$S\widehat R_{\overline W}^{(TT)}S=\widehat R_W^{(TT)}$.
\needspace{14\baselineskip}
Consequently,
\begin{equation}
 \dd\bm f^{(++)}_{2\theta}
 =\eps\sum_{W\in\mathbb L_{++}}\widehat R_W^{(2\theta)}(\delta)\,
 \dd\log W\,\bm f^{(++)}_{2\theta},
 \label{dual-theta-cde}
\end{equation}
where every $17\times17$ residue matrix is explicitly
\begin{equation}
\widehat R_W^{(2\theta)}=\begin{pmatrix}
 \widehat R_W^{(CC)}&0&0\\[3pt]
 \dfrac{\widehat R_W^{(TC)}+S\widehat R_{\overline W}^{(TC)}}2
   &\widehat R_W^{(TT)}&0\\[6pt]
 \dfrac{\widehat R_W^{(TC)}-S\widehat R_{\overline W}^{(TC)}}2
   &0&\widehat R_W^{(TT)}
 \end{pmatrix}.
 \label{eq:dual-residues}
\end{equation}
Thus the two seven-dimensional topsector parity blocks decouple from one
another and receive separate sources only from the contact subsector.

\section{Full correlator results}
\label{sec:physical-integrals}
We sum the SK components to obtain the one-loop bubble contributions to
equal-time correlators on the future boundary of $\mathrm{dS}_4$.
We consider the conformally coupled $\phi^4$ and $\phi^3$ interactions
of Eq.~\eqref{eq:conformal-action} and the minimally coupled $\chi^3$ interaction.
For the conformally coupled scalar, the quartic and cubic vertex weights are
$(-\tau)^0$ and $(-\tau)^{-1}$, respectively, giving the ordered integrals
$J^{\mathrm{top}}_{0,0;1,1}$ and $J^{\mathrm{top}}_{-1,-1;1,1}$.

We consider the real coupling constant $\lambda$, so
$N_{\sigma_1\sigma_2}=-\sigma_1\sigma_2N$ with a real $N$.  Using the
families \eqref{pp-ordered-integral} and \eqref{pm-regulated-J-family}
and the time integrals \eqref{pm-time-integrals}, the four components
for the conformal interactions, with $a_1=a_2=a=n-4$ ($n=3,4$), combine into
\begin{equation}
 \mathcal A_{\rm bub}
 =\frac{N\pi^{(d-1)/2}}{4\,\ee^{(1+2\delta)\gamma_E\eps}}\,\mathcal S_{\phi^n}(\delta),
 \label{phys-A-bub}
\end{equation}
with
\begin{equation}
\begin{aligned}
 \mathcal S_{\phi^n}(\delta)={}&2\operatorname{Re}\Bigl\{(-1)^{a+1}
 \bigl[J^{\mathrm{top}}_{a,a;1,1}(E_1,E_2)+J^{\mathrm{top}}_{a,a;1,1}(E_2,E_1)\bigr]
 \Bigr\}\\
 &+2\,\Gamma[1+a+\delta\eps]^2J^{(+-)}[a+1,a+1;1,1].
\end{aligned}
 \label{phys-SK-sum}
\end{equation}
In the following we will concentrate on the $S$ part but not trace the normalization factor. The first term is $\mathcal A_{++}+\mathcal A_{--}$, which are complex conjugate with each other.
The second is $\mathcal A_{+-}+\mathcal A_{-+}$, whose phases
cancel.  We want to recognize the pole structure of $\delta$ and $\eps$ after the summation.

The family $J^{\mathrm{top}}$ in Eq.~\eqref{pp-ordered-integral} contains
only the $\theta(\tau_1-\tau_2)$ part of the $(++)$ component.  For the
equal vertex powers considered below, the complementary part with
$\theta(\tau_2-\tau_1)$ is obtained by exchanging $E_1$ and $E_2$.
In the variables of Eq.~\eqref{pp-blowup-variables}, this gives
\begin{equation}
 u'=\frac{E_1}{E_2}=\frac1u,\qquad v'=\frac{p}{E_1}=uv,
 \label{eq:crossing}
\end{equation}
so both time orderings follow from the same MPL solution by a variable
substitution.

We have also solved the mixed $(+-)$ system
\eqref{eq:mixed-cde} in MPLs.  Its explicit solution is provided
in the supplementary Mathematica files.  For the first few orders in
$\eps$ considered below, expanding the time-integrated integrands,
with the ultraviolet contributions treated analytically, gives the same
results as the MPL solutions.

\subsection{\texorpdfstring{$\phi^4$}{phi\textasciicircum 4} interaction}
\label{sec:physical-phi4}

Reducing $J^{\mathrm{top}}_{0,0;1,1}$ in the full $J^{\mathrm{top}}\oplus J^{\mathrm{sub}}$
family and expressing the masters through
\eqref{eq:xy} and \eqref{eq:top-basis}
gives
\begin{equation}
 J^{\mathrm{top}}_{0,0;1,1}
 =-\frac{\eps}{2(1-2\eps)k_t}\,p^{-2(1+\delta)\eps}
 \bigl(f_5+f_6\bigr).
 \label{phys-T0011}
\end{equation}
This is the sum of the two time IBPs,
$k_tJ^{\mathrm{top}}_{0,0;1,1}
=\delta\eps\bigl(J^{\mathrm{top}}_{-1,0;1,1}+J^{\mathrm{top}}_{0,-1;1,1}\bigr)$.
Neither the prefactor nor $f_5,f_6$ is singular at $\delta=0$, so the
limit can be taken order by order.

At $\delta=0$, the first two coefficients of
Eq.~\eqref{eq:laurent-series} give
\begin{equation}
 \bigl(f_5+f_6\bigr)\big|_{\delta=0}
 =\frac2{\eps^2}+\frac4\eps\bigl[G(0;v)-G(-1;v)\bigr]+O(\eps^0).
 \label{phys-f56-gpl}
\end{equation}
For $a=0$ no SK component is singular at $\delta=0$.  The mixed
component follows from
\begin{equation}
 J^{(+-)}[1,1;1,1]=-\frac{\eps}{2(1-2\eps)(E_1-E_2)}\,(f^{(+-)}_2+f^{(+-)}_3)
 \label{phys-Jm1111}
\end{equation}
with the MPL solution of the (+-) component; it is ultraviolet finite.
We set $p=1$; the general case follows from homogeneity,
$\mathcal S_{\phi^4}(E_1,E_2,p)=p^{-1-2\eps}\mathcal S_{\phi^4}(E_1/p,E_2/p,1)$.
Through $O(\eps^0)$, the three components are
\begin{equation}
\begin{aligned}
 k_tJ^{\mathrm{top}}_{0,0;1,1}(E_1,E_2)&=-\frac1\eps-2-2G(0;v)+2G(-1;v),\\
 k_tJ^{\mathrm{top}}_{0,0;1,1}(E_2,E_1)&=-\frac1\eps-2-2G(0;u)-2G(0;v)
 +2G(-\tfrac1v;u),\\
 2J^{(+-)}[1,1;1,1]&=\frac{4uv}{u-1}\Bigl[G(0;u)-G(-\tfrac1v;u)+G(-1;v)\Bigr].
\end{aligned}
 \label{eq:phi4-components}
\end{equation}
The second line is the first evaluated at $(u',v')$ of
Eq.~\eqref{eq:crossing}, rewritten with
$G(0;uv)=G(0;u)+G(0;v)$ and $G(-1;uv)=G(-\tfrac1v;u)$.
With $1/k_t=uv/(1+u)$, Eq.~\eqref{phys-SK-sum} gives
\begin{equation}
\begin{aligned}
 \mathcal S_{\phi^4}={}&\frac{4uv}{1+u}\Bigl[\frac1\eps+2+2G(0;v)\Bigr]\\
 &+\frac{8uv}{u^2-1}\Bigl[G(-1;v)+uG(0;u)-uG(-\tfrac1v;u)\Bigr]+O(\eps).
\end{aligned}
 \label{phys-phi4-gpl}
\end{equation}
In terms of the momenta, with $p$ restored,
\begin{equation}
 \boxed{\mathcal S_{\phi^4}
 =\frac{4}{k_t}\left[\frac1\eps+2
 -\frac{2\bigl(E_2\log(E_1+p)-E_1\log(E_2+p)\bigr)}{E_2-E_1}\right]
 +O(\eps).}
 \label{phys-phi4-result}
\end{equation}
Only a simple ultraviolet pole remains.  The result is symmetric in
$E_1\leftrightarrow E_2$ and regular at $E_1=E_2$: the second bracket in
Eq.~\eqref{phys-phi4-gpl} vanishes at $u=1$ and cancels the zero of
$u^2-1$, leaving a finite contribution.  The $O(\eps)$
coefficient follows in the same way from the weight-2 coefficients.

At $\delta=0$ the time integrals are also elementary.  Adding the four
components in \eqref{phys-SK-sum} at fixed loop momentum gives
\begin{equation}
 \mathcal S_{\phi^4}
 =\frac{4}{k_t}\,
 \frac{\ee^{\gamma_E\eps}2^{2\eps}\sqrt\pi}{\Gamma[\frac32-\eps]}
 p^{-2\eps}\int_1^\infty\dd t\,(t^2-1)^{-\eps}\,
 \frac{k_t/p+t}{(t+E_1/p)(t+E_2/p)}.
 \label{phys-phi4-integral}
\end{equation}
Its expansion reproduces \eqref{phys-phi4-result}, and each of the
three components agrees with its integral representation through
$O(\eps)$.

The total-energy point $k_t=0$ lies outside the physical region
$E_1,E_2\geq p$.  Analytically continuing the external energies to
$E_2=-E_1$, first with $E_1^2<p^2$, the residue follows from
\eqref{phys-phi4-integral} exactly in $\eps$:
\begin{equation}
 \lim_{k_t\to0}k_t\mathcal S_{\phi^4}
 =4\ee^{\gamma_E\eps}\,
 \frac{\Gamma[\eps]\Gamma[1-\eps]^2}{\Gamma[2-2\eps]}
 \,(p^2-E_1^2)^{-\eps}.
 \label{eq:phi4-residue}
\end{equation}
Except for the factor 4 from normalization, the remaining Gamma-function and
momentum dependence are those of the massless one-loop bubble in
$4-2\eps$ dimensions at $s=E_1^2-p^2$. The residue of the
total-energy pole reproduces the flat-space amplitude.  The cutoff result of
\cite{Chowdhury:2025simplicity} has the same residue, up to overall
normalization, with $1/\eps+2\leftrightarrow\ln4\Lambda^2$.

\subsection{\texorpdfstring{$\phi^3$}{phi\textasciicircum 3} interaction}
\label{sec:physical-phi3}
Before the explicit computation, it's necessary to mention that in our setup, we consider
the general kinematics configuration in which $E_1 \neq E_2 \neq p$, but for $\phi^3$ bubble,
we only have one external scale $E_1=E_2=p$. So we first compute
an off-shell extension of the $\phi^3$ bubble by treating the external
energies $E_1,E_2$ as independent of the spatial momentum magnitude
$p$.  The physical $\phi^3$  bubble is recovered by setting $E_1=E_2=p$ in the
general result below.

The same reduction gives
\begin{equation}
 J^{\mathrm{top}}_{-1,-1;1,1}
 =\frac{p^{-2(1+\delta)\eps}}{(1-2\eps)\bigl[1-2(1+\delta)\eps\bigr]}
 \left[-\frac{p}{\delta^2}f_4+\frac{E_1f_5+E_2f_6}{2\delta}\right].
 \label{phys-Tm1m1}
\end{equation}
Here the factors $1/\delta^2$ and $1/\delta$ survive.  At leading order,
\begin{equation}
 J^{\mathrm{top}}_{-1,-1;1,1}
 =\frac{1}{\eps^2}\left[\frac{p}{\delta^2}+\frac{E_2}{\delta}
 +\frac{E_1-E_2}{2(1+\delta)}\right]+O(\eps^{-1}).
 \label{phys-Tm1m1-leading}
\end{equation}
The poles at $\delta=0$ come from the late-time endpoint, where the
vertex weight $(-\tau)^{-1+\delta\eps}$ is not integrable at
$\delta=0$.  The double pole arises when both vertices approach
$\tau=0$, and the single pole when only the later one does.  The factor
$1/(1+\delta)$ reflects the overlap with the ultraviolet region, where
poles appear in $\eps+\rho=(1+\delta)\eps$.

The mixed component reduces to its canonical basis as
\begin{equation}
 J^{(+-)}[0,0;1,1]=\frac{\eps^2}{(1-2\eps)\bigl[1-2(1+\delta)\eps\bigr]}
 \Bigl[-p f^{(+-)}_1-\frac{\delta}{2}\bigl(E_1f^{(+-)}_2-E_2f^{(+-)}_3\bigr)\Bigr]
 \label{phys-Jm0011}
\end{equation}

\paragraph{Cancellation of the $\delta$ poles}
Expanding the ordered and mixed contributions in $\eps$ gives, at
leading order,
\begin{align}
 2\operatorname{Re}\bigl[J^{\mathrm{top}}_{-1,-1;1,1}(E_1,E_2)
 +J^{\mathrm{top}}_{-1,-1;1,1}(E_2,E_1)\bigr]
 &=\frac{1}{\eps^2}\left[\frac{4p}{\delta^2}+\frac{2k_t}{\delta}\right]
 +O(\eps^{-1}),\notag\\
 2\,\Gamma[\delta\eps]^2J^{(+-)}[0,0;1,1]
 &=\frac{1}{\eps^2}\left[-\frac{4p}{\delta^2}-\frac{2k_t}{\delta}
 +\frac{2k_t}{1+\delta}\right]+O(\eps^{-1}),
 \label{phys-leading-components}
\end{align}
so that
\begin{equation}
 \mathcal S_{\phi^3}(\delta)=\frac{2k_t}{\eps(\eps+\rho)}+O(\eps^{-1}).
 \label{phys-leading-sum}
\end{equation}
At this order, the poles at $\delta=0$ cancel between the ordered and mixed components.

At fixed loop momentum, the cancellation can be seen directly from the integrand.  Set
$t_i=-\tau_i>0$ and recall $\rho=\delta\eps$.  In an individual SK
component, a vertex approaching the future boundary gives
\begin{equation}
 \int_0^\eta\dd t\,t^{\rho-1}\ee^{\ii Et}
 =\frac{\eta^\rho}{\rho}+O(1),\qquad \rho\to0,
\end{equation}
where $\eta>0$ is a small fixed endpoint cutoff and $E$ is the
corresponding energy combination.  This produces a pole at $\delta=0$.
After summing the four SK components, the constant endpoint terms
cancel and the time-dependent factor becomes, on $0<t_1<t_2$,
\begin{equation}
 4\sin(E_1t_1)\sin\bigl[(E_2+s)t_2-st_1\bigr],\qquad s=z_1+z_2,
\end{equation}
up to common propagator denominators and the weights
$t_1^{\rho-1}t_2^{\rho-1}$.  Since
$\sin(E_1t_1)=E_1t_1+O(t_1^3)$, the extra factor of $t_1$ cancels
the dangerous $1/t_1$ at the latest vertex.  The endpoint integral is
therefore of the form
\begin{equation}
 \int_0^\eta\dd t\,t^{\rho-1}\sin(Et)
 =\frac{E\eta^{\rho+1}}{\rho+1}+O(\eta^{\rho+3}),
\end{equation}
which is regular at $\rho=0$.  When both times approach zero together,
the product of $\sin$ supplies two powers of time, removing the double
pole as well.  The other time ordering works in the same way.

The MPL solution through weight two verifies the cancellation after loop integration.
The coefficients of $\delta^{-2}$ and $\delta^{-1}$ in
$\mathcal S_{\phi^3}(\delta)$ vanish at orders $\eps^{-2}$,
$\eps^{-1}$ and $\eps^0$.

\paragraph{Result}
We set $\delta=0$ in the coefficients through $\eps^0$ and take $p=1$ for
simplicity. The general case follows from homogeneity,
$\mathcal S_{\phi^3}(E_1,E_2,p)=p^{1-2\eps}\mathcal S_{\phi^3}(E_1/p,E_2/p,1)$.
With $k_t=E_1+E_2$,
\begin{equation}
 \mathcal S_{\phi^3}
 =\frac{2k_t}{\eps^2}+\frac{4k_t(2-\log k_t)}{\eps}
 +\mathcal S^{(0)}_{\phi^3}+O(\eps),
 \label{phys-phi3-result}
\end{equation}
where
\begin{equation}
\boxed{\begin{aligned}
 \mathcal S^{(0)}_{\phi^3}={}&-\frac{4\pi^2}{3}
 +k_t\Bigl(24-16\log k_t+4\log^2k_t+\frac{\pi^2}{6}\Bigr)
\\
 &-2k_t\Bigl[\log^2\frac{k_t}{1+E_1}+\log^2\frac{k_t}{1+E_2}\Bigr]
 +(2+k_t)\log^2\frac{1+E_1}{1+E_2}\\
 &+4(1-E_1)\operatorname{Li}_2\frac{1-E_1}{1+E_2}
 +4(1-E_2)\operatorname{Li}_2\frac{1-E_2}{1+E_1}\\
 &+2(E_1-E_2)\Bigl[\operatorname{Li}_2\frac{E_2-E_1}{1+E_2}
 -\operatorname{Li}_2\frac{E_1-E_2}{1+E_1}\Bigr].
\end{aligned}}
 \label{phys-phi3-finite}
\end{equation}
The result is manifestly symmetric in
$E_1\leftrightarrow E_2$.

We now impose the on-shell condition to obtain the
$\phi^3$ bubble by setting $E_1=E_2=p$ in \eqref{phys-phi3-result}
\begin{equation}
\begin{aligned}
 \mathcal S_{\phi^3}\big|_{E_1=E_2=p}
 & =\frac{4p}{\eps^2}\,(2p)^{-2\eps}\ee^{4\eps}+(16-\pi^2)p+O(\eps)
 \\
 & =\frac{4p}{\eps^2}+\frac{8p}{\eps}\bigl[2-\log(2p)\bigr]
 \\
 &\quad+p\bigl[8\log^2(2p)-32\log(2p)+48-\pi^2\bigr]+O(\eps).
\end{aligned}
 \label{phys-phi3-two-point}
\end{equation}
The physical two point bubble result retains double and single poles in \(\epsilon\),
together with logarithmic dependence on the only independent external scale \(p\).

\subsection{Massless \texorpdfstring{$\chi^3$}{phi\textasciicircum 3} interaction}
\label{sec:physical-massless-phi3}

We now consider the cubic interaction of the massless minimally coupled
scalar $\chi$. In $\mathrm{dS}_4$ each vertex carries the weight
$\sqrt{-g}=(-H\tau)^{-4}$. As in Section~\ref{sec:physical-phi3},
we keep the external energies independent of $p$ until we impose
$E_1=E_2=p$. Using the propagators of Section~2 in
Eq.~\eqref{general-bubble} and factoring out the normalization in
Eq.~\eqref{phys-A-bub} gives
\begin{equation}
\begin{aligned}
 \mathcal S_{\chi^3}(\delta)={}&4\sum_{\sigma_1,\sigma_2=\pm}(-\sigma_1\sigma_2)
 \int_{-\infty}^0\dd\tau_1\dd\tau_2\,
 (-\tau_1)^{-4+\delta\eps}(-\tau_2)^{-4+\delta\eps}\\
 &\times(1-\ii\sigma_1E_1\tau_1)(1-\ii\sigma_2E_2\tau_2)\,
 \ee^{\ii\sigma_1E_1\tau_1+\ii\sigma_2E_2\tau_2}\\
 &\times\int[\dd^d\bm\ell]\,
 G^\chi_{\sigma_1\sigma_2}(z_1;\tau_1,\tau_2)\,
 G^\chi_{\sigma_1\sigma_2}(z_2;\tau_1,\tau_2),
\end{aligned}
 \label{phys-ml-integral}
\end{equation}
where we set $H=1$ and the normalizations $1/(2E_i^3)$ of the
bulk-to-boundary propagators are included in $N$.

Expanding the polynomial factors in the propagators expresses each time
ordering of the $(++)$ component as a sum of ordered integrals.
For the $(+-)$ component, the two time integrations factorize and give
a sum of mixed integrals. The index ranges in Eqs.~\eqref{pp-ordered-integral} and~\eqref{pm-regulated-J-family} are
\[
\begin{aligned}
 (++)\quad &J^{\mathrm{top}}_{a_1,a_2;b_1,b_2}
   && -4\le a_i\le-1, &\qquad 1\le b_i\le3,\\
 (+-)\quad &J^{(+-)}[n_1,n_2;b_1,b_2]
   && -3\le n_i\le0, &\qquad 1\le b_i\le3 .
\end{aligned}
\]
The other two SK components follow by complex conjugation.

At generic $\delta$, the Kira reduction gives $f_1,f_4,f_5,f_6$ for
each ordered contribution and $f^{(+-)}_1,f^{(+-)}_2,f^{(+-)}_3$ for
the mixed contribution. The coefficient of the contact integral $f_1$
is antisymmetric under $E_1\leftrightarrow E_2$, so its two contributions
cancel. The coefficients have at most a simple pole in $\eps$, and the
MPL solutions through weight three therefore determine the finite term.

At $p=1$, the leading term in the SK sum is
\begin{equation}
 \mathcal S_{\chi^3}
 =-\frac{1}{\eps^3}\left[\frac{16E_1^3E_2^3}{9\delta^2}
 +\frac{8(E_1^3+E_2^3)}{9\delta(1+\delta)}\right]+O(\eps^{-2}).
 \label{phys-ml-leading}
\end{equation}
The poles at $\delta=0$ survive the SK sum. At fixed loop momentum,
the sum at the latest vertex selects the imaginary part of the
bulk-to-boundary propagator. Its time dependence for $t=-\tau$ is
\begin{equation}
 \operatorname{Im}\bigl[(1+\ii Et)\ee^{-\ii Et}\bigr]=Et\cos(Et)-\sin(Et)
 =-\frac{(Et)^3}{3}+O(t^5).
\end{equation}
Including the vertex weight $t^{-4+\delta\eps}$ gives an integrand
proportional to $E^3 t^{-1+\delta\eps}$. The time integral thus has
a pole at $\delta=0$. When both vertices approach the boundary, the double
pole carries the factor $E_1^3E_2^3$. The soft loop integral supplies
the additional $1/\eps$ in Eq.~\eqref{phys-ml-leading}.
Through the finite term, the coefficients of $\delta^{-2}$ and
$(1+\delta)^{-1}$ remain proportional to $E_1^3E_2^3$ and
$E_1^3+E_2^3$, respectively.

The coefficients for general $E_1,E_2$, written in classical
polylogarithms up to $\operatorname{Li}_3$, are given in the
supplementary files.  At the physical point $E_1=E_2=p$ the logarithms
of $E_i-p$ cancel. Homogeneity fixes the overall power of $p$, leaving
coefficients that depend only on $\delta$,
\begin{equation}
 \mathcal S_{\chi^3}\big|_{E_1=E_2=p}
 =p^{3-2(1+\delta)\eps}\sum_{n=-3}^{0}\eps^n s_n(\delta)+O(\eps),
 \label{phys-ml-two-point}
\end{equation}
with $L=\log2$ and
\begin{equation}
\begin{aligned}
 s_{-3}={}&-\frac{16}{9\delta^2}-\frac{16}{9\delta(1+\delta)},\\
 s_{-2}={}&-\frac{32(1-L)}{9\delta^2}-\frac{320-192L}{27\delta}
 -\frac{76}{27(1+\delta)},\\
 s_{-1}={}&-\frac{4}{27\delta^2}\bigl(24L^2-48L+\pi^2\bigr)
 -\frac{4}{27\delta}\bigl(72L^2-208L+64+5\pi^2\bigr)
 +\frac{4(100-3\pi^2)}{81(1+\delta)}\\
 &-\frac{8}{81}\bigl(72L^2-297L+229+6\pi^2\bigr),\\
 s_{0}={}&\frac{8}{27\delta^2}\bigl(8L^3-24L^2+\pi^2L-\pi^2+26\zeta_3\bigr)\\
 &+\frac{16}{81\delta}\bigl(48L^3-192L^2+(96+9\pi^2)L-11\pi^2+93\zeta_3\bigr)
 +\frac{57\pi^2-616-720\zeta_3}{243(1+\delta)}\\
 &+\frac{2}{243}\bigl(1440L^3-7308L^2+(6600+360\pi^2)L-687\pi^2+1552+1584\zeta_3\bigr)\\
 &+\frac{2\delta}{81}\bigl(192L^3-1188L^2+(1832+48\pi^2)L-125\pi^2-416+96\zeta_3\bigr).
\end{aligned}
 \label{eq:chi3-coeffs}
\end{equation}

Equation~\eqref{phys-ml-two-point} gives the series expansion with pole structure, but not the finite 
physical result. The further renormalization is beyond the scope of this paper.  
For massless $\phi^4$ theory, related infrared and
late-time effects have been studied in the full theory
\cite{Beneke:2023wmt} and through matching to Soft de Sitter Effective
Theory~\cite{Beneke:2026sdset}, building on the effective description of Refs.~\cite{Cohen:2020php,Cohen:2021fzf}. 

\section{Conclusions and Outlook}
\label{sec:conclusions}

We have constructed canonical differential equations for the two-site
one-loop bubble integral family with monomial vertex time weights,
including the spatial loop-momentum integration.  The construction
covers the mixed component and the ordered component together with
the contact integrals generated by time IBPs.  For $d=3-2\eps$ and
$\rho_1=\rho_2=\delta\eps$, the Baikov dlog construction and IBP
reduction give an $\eps$-factorized system.  Matching the soft,
semihard and hard regions of the original integrals determines the
boundary vector analytically in terms of Gamma functions.

The canonical system and the exact boundary vector determine the MPL
expansion of every master integral in the regulated family to arbitrary
order in $\eps$.  The explicit expressions presented here extend through
weight three.
A constant recombination also gives the 17-element dual-$\theta$
canonical basis, whose topsector splits into two seven-dimensional
parity blocks.

The result concerns an integral family with variable time and momentum
powers and thus applies to both conformally coupled and massless
minimally coupled scalars.  In particular, expanding the finite
polynomial factors in the massless minimally coupled propagators
of Eq.~\eqref{massless-prop} expresses their bubble integrals as
combinations of family members with shifted indices.  Polynomial
numerators from local derivative interactions can be treated in the
same way.  These integrals can therefore be reduced to the canonical
masters and evaluated using their MPL solutions.

As explicit applications in $\mathrm{dS}_4$, we combine all
Schwinger--Keldysh components for the conformally coupled $\phi^3$
and $\phi^4$ bubbles and for the massless minimally coupled cubic
bubble. We give the results through the finite term in $\eps$.
For the conformally coupled quartic interaction, each SK component is regular at $\delta=0$.
For the conformally coupled cubic interaction, the $\delta$ poles cancel in the SK sum through $O(\eps^0)$.
Ultraviolet poles in $\eps$ remain in both cases. The bare minimally coupled cubic
bubble retains poles in both $\delta$ and $\eps$.
These examples show how the integral-family solution yields
the correlator contributions for specified interactions and kinematics.

\paragraph{Outlook}
\begin{itemize}
 \item We expect the same method to be useful for the one-loop triangle.
 The three vertices allow products of two Heaviside functions, and the
 additional external scales make the IBP reduction more involved, even
 for conformally coupled scalars. Our preliminary reductions suggest a
 master-integral count of order one hundred. The count remains to be
 checked for the different SK components, such as $(+++)$ and $(++-)$,
 which have different time orderings.

 \item Our construction of $\dd\log$ integrands uses the auxiliary
 momentum family obtained after the time integrations. For the triangle,
 it remains to be determined whether the corresponding family admits a
 complete basis of logarithmic integrands. The one-loop triangle studied in Ref.~\cite{McCune:2026elliptic}
 has elliptic master integrals, although the elliptic periods cancel in the
 flat-space wavefunction coefficient. It remains to determine
whether elliptic master integrals occur in the time-regulated triangle
family for an individual time ordering.
\end{itemize}

\section*{Acknowledgments}
Jiaqi Chen is supported by the National Natural Science Foundation of China (NSFC)
under Grant No. 12505094 and by the Science Foundation of China University of Petroleum,
Beijing (No. 2462025YJRC019).
Bo Feng is supported by the National Natural Science Foundation of China (NSFC) through Grants No. 12535003, No. 11935013, No. 11947301, and No. 12047502.

\medskip
\noindent\textbf{Statement on AI Use}\quad 
Several agent programs driven by different large language models were used in this work, mainly to assist with writing the computer programs. The authors have reviewed and verified all the results.

\bibliographystyle{JHEP}
\bibliography{references}

\appendix
\clearpage
\section{IBP identities and external-derivative rules}
\label{app:pp-ibp-identities}

For the phase-normalized families in
Eqs.~\eqref{pp-ordered-integral} and~\eqref{pp-contact-family}, we collect
the identities used in the reduction and in generating the differential
equations.  The energy denominators are those of
Eq.~\eqref{pp-energy-denominators}.

The analytic time regulators are introduced so that time IBPs can be
used; the $\ii0$ prescription controls the early-time endpoint.  The
time powers alone do not control the equal-time region
$\tau_1\to\tau_2$ of the ordered integrals, where the spatial integral
behaves as $|\tau_1-\tau_2|^{b_1+b_2-D}$.  We therefore also treat
$D$, $b_1$ and $b_2$ as independent analytic parameters: the equal-time
region requires $\Re D<\Re(b_1+b_2)+1$ and the soft loop regions
$\Re D>\Re b_i$, which hold together for large $\Re b_i$.  We derive the
identities where all boundary terms vanish and obtain them at the
required values by analytic continuation.

\subsection{Time IBP identities}
\label{app:pp-time-ibps}

The derivatives of $\theta(\tau_1-\tau_2)$ are opposite equal-time
delta functions.  They produce the contact terms in the two time IBPs:

\begin{align}
0&={}-(a_1+\delta\eps)J^{\mathrm{top}}_{a_1-1,a_2;b_1,b_2}
+E_1J^{\mathrm{top}}_{a_1,a_2;b_1,b_2}
\notag\\
&\quad-J^{\mathrm{top}}_{a_1,a_2;b_1-1,b_2}
-J^{\mathrm{top}}_{a_1,a_2;b_1,b_2-1}
+J^{\mathrm{sub}}[a_1+a_2;b_1,b_2],
\label{pp-time-ibp-1}\\
0&={}-(a_2+\delta\eps)J^{\mathrm{top}}_{a_1,a_2-1;b_1,b_2}
+E_2J^{\mathrm{top}}_{a_1,a_2;b_1,b_2}
\notag\\
&\quad+J^{\mathrm{top}}_{a_1,a_2;b_1-1,b_2}
+J^{\mathrm{top}}_{a_1,a_2;b_1,b_2-1}
-J^{\mathrm{sub}}[a_1+a_2;b_1,b_2].
\label{pp-time-ibp-2}
\end{align}

The contact family \eqref{pp-contact-family} closes under its own IBPs.
Its time identity is
\begin{equation}
0=-(n+2\delta\eps)J^{\mathrm{sub}}[n-1;b_1,b_2]
 +k_tJ^{\mathrm{sub}}[n;b_1,b_2].
 \label{pp-contact-time-ibp}
\end{equation}

\subsection{Spatial IBP identities and exchange symmetry}
\label{app:pp-spatial-ibps}

Spatial IBPs are generated from
\begin{equation}
 0=\int\dd^d\bm\ell\,
 \partial_{\ell^\mu}\!\left[v^\mu\,
 \frac{\ee^{\ii z_3\tau_1+\ii z_4\tau_2}}
 {z_1^{b_1}z_2^{b_2}}\right],
 \label{pp-momentum-ibp}
\end{equation}
with boundary terms understood by dimensional and analytic continuation \cite{Chetyrkin:1981qh}.
For later convenience, we define the time-index difference operator
\begin{equation}
 \Delta_aJ^{\mathrm{top}}_{b_1,b_2}
 \equiv
 J^{\mathrm{top}}_{a_1+1,a_2;b_1,b_2}
 -J^{\mathrm{top}}_{a_1,a_2+1;b_1,b_2}.
 \label{eq:time-shift}
\end{equation}
Using
\begin{equation}
 \bm\ell\cdot(\bm\ell+\bm p)
 =\frac{z_1^2+z_2^2-p^2}{2},
 \qquad
 \partial_{\ell^\mu}(z_1+z_2)
 =\frac{\ell_\mu}{z_1}+\frac{\ell_\mu+p_\mu}{z_2},
\end{equation}
the choices of IBP vector $v^\mu=\ell^\mu$ and $v^\mu=\ell^\mu+p^\mu$ give
\begin{align}
0&={}\left(d-b_1-\frac{b_2}{2}\right)
 J^{\mathrm{top}}_{a_1,a_2;b_1,b_2}
 -\frac{b_2}{2}J^{\mathrm{top}}_{a_1,a_2;b_1-2,b_2+2}
 +\frac{b_2p^2}{2}J^{\mathrm{top}}_{a_1,a_2;b_1,b_2+2}
\nonumber\\
&+\Delta_a\bigg(
 J^{\mathrm{top}}_{b_1-1,b_2}
 +\frac12J^{\mathrm{top}}_{b_1-2,b_2+1}
 +\frac12J^{\mathrm{top}}_{b_1,b_2-1}
 -\frac{p^2}{2}J^{\mathrm{top}}_{b_1,b_2+1}
 \bigg),
 \label{pp-momentum-ibp-p}\\[1mm]
0&={}\left(d-b_2-\frac{b_1}{2}\right)
 J^{\mathrm{top}}_{a_1,a_2;b_1,b_2}
 -\frac{b_1}{2}J^{\mathrm{top}}_{a_1,a_2;b_1+2,b_2-2}
 +\frac{b_1p^2}{2}J^{\mathrm{top}}_{a_1,a_2;b_1+2,b_2}
\nonumber\\
&+\Delta_a\bigg(
 \frac12J^{\mathrm{top}}_{b_1-1,b_2}
 +\frac12J^{\mathrm{top}}_{b_1+1,b_2-2}
 -\frac{p^2}{2}J^{\mathrm{top}}_{b_1+1,b_2}
 +J^{\mathrm{top}}_{b_1,b_2-1}
 \bigg).
 \label{pp-momentum-ibp-pk}
\end{align}
The momentum IBPs shift the time and momentum indices but do not generate contact integrals.

The contact family obeys the corresponding spatial identities:
\begin{align}
0&={}\left(d-b_1-\frac{b_2}{2}\right)
 J^{\mathrm{sub}}[n;b_1,b_2]
 -\frac{b_2}{2}J^{\mathrm{sub}}[n;b_1-2,b_2+2]
 +\frac{b_2p^2}{2}J^{\mathrm{sub}}[n;b_1,b_2+2],
\label{pp-contact-momentum-ibp-p}\\
0&={}\left(d-b_2-\frac{b_1}{2}\right)
 J^{\mathrm{sub}}[n;b_1,b_2]
 -\frac{b_1}{2}J^{\mathrm{sub}}[n;b_1+2,b_2-2]
 +\frac{b_1p^2}{2}J^{\mathrm{sub}}[n;b_1+2,b_2].
\label{pp-contact-momentum-ibp-pk}
\end{align}

\begingroup
\setlength{\abovedisplayskip}{5pt plus 1pt minus 1pt}
\setlength{\belowdisplayskip}{5pt plus 1pt minus 1pt}
\setlength{\abovedisplayshortskip}{3pt plus 1pt}
\setlength{\belowdisplayshortskip}{4pt plus 1pt minus 1pt}
\setlength{\jot}{3pt}
For comparison, let $F$ be the integrand of the auxiliary family in
Eq.~\eqref{pp-J-family}.  Its momentum IBPs, generated by the same vectors
$\bm\ell$ and $\bm\ell+\bm p$, are
\begin{align}
0&={}\int[\dd^d\bm\ell]F\Biggl[d-b_1
 -\frac{b_2(z_1^2+z_2^2-p^2)}{2z_2^2}\notag\\
&\qquad
 +\left(\frac{n_1+\delta\eps}{z_3}
       -\frac{n_2+\delta\eps}{z_4}\right)
 \left(z_1+\frac{z_1^2+z_2^2-p^2}{2z_2}\right)\Biggr],\notag\\[1mm]
0&={}\int[\dd^d\bm\ell]F\Biggl[d-b_2
 -\frac{b_1(z_1^2+z_2^2-p^2)}{2z_1^2}\notag\\
&\qquad
 +\left(\frac{n_1+\delta\eps}{z_3}
       -\frac{n_2+\delta\eps}{z_4}\right)
 \left(z_2+\frac{z_1^2+z_2^2-p^2}{2z_1}\right)\Biggr].
 \label{pp-J-momentum-ibps}
\end{align}
In Eq.~\eqref{pp-J-momentum-ibps}, set $a_i=n_i-1$ and multiply by
$(1+\delta\eps)_{a_1}(1+\delta\eps)_{a_2}$. With
$(n_i+\delta\eps)(1+\delta\eps)_{a_i}=(1+\delta\eps)_{a_i+1}$,
the energy-denominator term becomes
\begin{equation}
\begin{aligned}
 & (1+\delta\eps)_{a_1}(1+\delta\eps)_{a_2}
 \int[\dd^d\bm\ell]F
 \left(\frac{n_1+\delta\eps}{z_3}-\frac{n_2+\delta\eps}{z_4}\right)\\
 &\quad= (1+\delta\eps)_{a_1+1}(1+\delta\eps)_{a_2}
 J[n_1+1,n_2;b_1,b_2]\\
 &\qquad -(1+\delta\eps)_{a_1}(1+\delta\eps)_{a_2+1}
 J[n_1,n_2+1;b_1,b_2]\\
 &\quad\longleftrightarrow
 J^{\mathrm{top}}_{a_1+1,a_2;b_1,b_2}
 -J^{\mathrm{top}}_{a_1,a_2+1;b_1,b_2}
 =\Delta_a J^{\mathrm{top}}_{b_1,b_2}.
\end{aligned}
\label{eq:ibp-map}
\postdisplaypenalty=10000
\end{equation}
The last line follows from Eq.~\eqref{pp-top-pairing}. Applying the same map
to the spatial shifts recovers Eqs.~\eqref{pp-momentum-ibp-p}
and~\eqref{pp-momentum-ibp-pk} term by term. We can therefore derive the momentum 
IBP relations in the auxiliary family and use Eq. (5.16) to carry them back to the ordered top sector.

There is also an exchange symmetry. Under the loop-momentum redefinition
$\boldsymbol{\ell}\rightarrow-\boldsymbol{\ell}-\boldsymbol{p}$, one has
$z_1\leftrightarrow z_2$, and hence
\begin{equation}
    J^{\mathrm{top}}_{a_1,a_2;b_1,b_2}
    =
    J^{\mathrm{top}}_{a_1,a_2;b_2,b_1},
    \qquad
    J^{\mathrm{sub}}[n;b_1,b_2]
    =
    J^{\mathrm{sub}}[n;b_2,b_1].
\end{equation}

\subsection{External-derivative rules}
\label{app:derivatives}

At fixed loop momentum and fixed external-momentum direction, the
shifted momentum magnitude obeys
\begin{equation}
 \partial_p z_2=\frac{p^2+z_2^2-z_1^2}{2p z_2}.
 \label{pp-literal-K-derivative}
\end{equation}

Using the energy derivatives in Eq.~\eqref{pp-energy-derivatives}
and the time IBPs \eqref{pp-time-ibp-1}--\eqref{pp-time-ibp-2}
makes the contact source explicit before any momentum reduction:
\begin{align}
 \partial_{E_1}J^{\mathrm{top}}_{a_1,a_2;b_1,b_2}
&={}-\frac{a_1+1+\delta\eps}{E_1}
 J^{\mathrm{top}}_{a_1,a_2;b_1,b_2}
 -\frac{1}{E_1}J^{\mathrm{top}}_{a_1+1,a_2;b_1-1,b_2}
 -\frac{1}{E_1}J^{\mathrm{top}}_{a_1+1,a_2;b_1,b_2-1}
\nonumber\\
&+\frac{1}{E_1}J^{\mathrm{sub}}[a_1+a_2+1;b_1,b_2],
\label{eq:E1-derivative}\\
 \partial_{E_2}J^{\mathrm{top}}_{a_1,a_2;b_1,b_2}
&={}-\frac{a_2+1+\delta\eps}{E_2}
 J^{\mathrm{top}}_{a_1,a_2;b_1,b_2}
 +\frac{1}{E_2}J^{\mathrm{top}}_{a_1,a_2+1;b_1-1,b_2}
 +\frac{1}{E_2}J^{\mathrm{top}}_{a_1,a_2+1;b_1,b_2-1}
\nonumber\\
&-\frac{1}{E_2}J^{\mathrm{sub}}[a_1+a_2+1;b_1,b_2].
\label{eq:E2-derivative}
\end{align}

The $p$ derivative is also taken directly, at fixed direction of
$\bm p$.  With the time-index difference already defined above,
\begin{align}
 \partial_pJ^{\mathrm{top}}_{a_1,a_2;b_1,b_2}
 &={}-\frac{b_2}{2p}\left(
 J^{\mathrm{top}}_{a_1,a_2;b_1,b_2}
 +p^2J^{\mathrm{top}}_{a_1,a_2;b_1,b_2+2}
 -J^{\mathrm{top}}_{a_1,a_2;b_1-2,b_2+2}\right)\notag\\
 &+\frac1{2p}\Delta_a\left(
 J^{\mathrm{top}}_{b_1,b_2-1}+p^2J^{\mathrm{top}}_{b_1,b_2+1}
 -J^{\mathrm{top}}_{b_1-2,b_2+1}\right),\notag\\
 \partial_pJ^{\mathrm{sub}}[n;b_1,b_2]
 &={}-\frac{b_2}{2p}\left(
 J^{\mathrm{sub}}[n;b_1,b_2]+p^2J^{\mathrm{sub}}[n;b_1,b_2+2]
 -J^{\mathrm{sub}}[n;b_1-2,b_2+2]\right).
 \label{pp-K-direct-derivatives}
\end{align}
\endgroup

\Needspace{24\baselineskip}
\section{Baikov representation and dimension recurrence}
\label{app:dimension-recurrence}

This appendix derives the Baikov measure and the dimension-shift identity
used in the main text, and records the normalization of the dlog integrals.

The spatial integrand depends on the loop momentum only through
$z_1=|\bm\ell|$ and $z_2=|\bm\ell+\bm p|$, with one independent external
spatial vector $\bm p$, $p=|\bm p|>0$.  Decomposing $\bm\ell$ along $\bm p$
and integrating the transverse angles, the measure becomes
\begin{equation}
\begin{aligned}
 \dd^D\bm\ell
 &\longrightarrow
 \frac{\Omega_{D-2}}{p}\,\dd z_1\,\dd z_2\,z_1 z_2\,
 \left[\frac{\mathcal G(\bm\ell,\bm p)}{p^2}\right]^{(D-3)/2},\\[3pt]
 \Omega_{D-2}&=\frac{2\pi^{(D-1)/2}}{\Gamma[(D-1)/2]},
\end{aligned}
 \label{baikov-gram-measure}
\end{equation}
where the Gram determinant of the loop and external momenta is
\begin{equation}
 \mathcal G(\bm\ell,\bm p)
 =\det\begin{pmatrix}\bm\ell^2&\bm\ell\cdot\bm p\\
 \bm\ell\cdot\bm p&p^2\end{pmatrix}
 =p^2\bm\ell^2-(\bm\ell\cdot\bm p)^2 .
 \label{baikov-gram-determinant}
\end{equation}
With one loop momentum and one external vector the Baikov power is
$(D-3)/2$; the energy $E_1,E_2$ are parameters of the integrand, not
additional vectors in the spatial Gram matrix.  Using the Baikov polynomial
$\mathcal P=4\mathcal G$ and the domain $\mathcal D_p$ of
Eq.~\eqref{pm-baikov-family}, the complete measure in the normalization of
Eq.~\eqref{eq:loop-measure} is
\begin{equation}
 \int[\dd^D\bm\ell]F
 =\frac{2^{4-D}\ee^{(1+2\delta)\gamma_E\eps}p^{2-D}}
 {\Gamma[(D-1)/2]}
 \int_{\mathcal D_p}\dd z_1\,\dd z_2\,
 z_1 z_2\,\mathcal P^{(D-3)/2}F .
 \label{baikov-complete-measure}
\end{equation}
Substituting the integrand of Eq.~\eqref{pm-regulated-J-family} gives
Eq.~\eqref{pm-baikov-family}, including its prefactor $\mathcal N_D p^{2-D}$.
The same spatial measure applies to the ordered energy denominators and to
the contact bubble. Here the power of Baikov polynomial $P$ is 
$\mathcal P^{-\eps}$ at $D=d$ and $\mathcal P^{-1-\eps}$ at $D=d-2$.

For a systematic treatment of dimensional recurrence relations, see Ref.~\cite{Tarasov:1996br}.  In the
Baikov representation \cite{Baikov:1996rk}, see also
Ref.~\cite{Chen:2026dqp} for de Sitter loop integrals, the shift follows from
a single total derivative with respect to $\bm\ell^2$ at fixed
$\bm\ell\cdot\bm p$.  For $\nu=(D-3)/2$ and the factor
\begin{equation}
 F_{n_1n_2b_1b_2}
 =\frac{1}{z_3^{n_1+\delta\eps}z_4^{n_2+\delta\eps}z_1^{b_1}z_2^{b_2}},
 \label{pp-indexed-factor}
\end{equation}
one has $\partial_{\bm\ell^2}\mathcal P=4p^2$, so in a domain where the
boundary term at $\mathcal P=0$ vanishes ($\Re\nu>0$),
\begin{align}
 0&={}\int\dd(\bm\ell^2)\,\dd(\bm\ell\cdot\bm p)\,
 \partial_{\bm\ell^2}\!\left(\mathcal P^\nu F\right)\nonumber\\
 &={}4\nu p^2\int\mathcal P^{\nu-1}F
 +\int\mathcal P^\nu\,\partial_{\bm\ell^2}F .
 \label{eq:dimension-ibp}
\end{align}
The identity is continued meromorphically in the dimension and indices
before setting $D=3-2\eps$ and $\rho_1=\rho_2=\delta\eps$, where
$\mathcal P^{-\eps}$ does not, in general, vanish at $\mathcal P=0$.

With the ordered denominators $z_3=E_1-z_1-z_2$, $z_4=E_2+z_1+z_2$ of
Eq.~\eqref{pp-energy-denominators},
$\partial_{\bm\ell^2}(z_1+z_2)=\tfrac12(1/z_1+1/z_2)$, so the derivative acts
on the indexed factor as
\begin{align}
 \partial_{\bm\ell^2}F_{n_1n_2b_1b_2}
 &={}\frac{n_1+\delta\eps}{2}\bigl(
 F_{n_1+1,n_2,b_1+1,b_2}
 +F_{n_1+1,n_2,b_1,b_2+1}\bigr)
 \nonumber\\
 &-\frac{n_2+\delta\eps}{2}\bigl(
 F_{n_1,n_2+1,b_1+1,b_2}
 +F_{n_1,n_2+1,b_1,b_2+1}\bigr)
 \nonumber\\
 &-\frac{b_1}{2}F_{n_1,n_2,b_1+2,b_2}
 -\frac{b_2}{2}F_{n_1,n_2,b_1,b_2+2}.
 \label{pp-F-derivative}
\end{align}
With the measure in Eq.~\eqref{eq:loop-measure}, the Baikov
normalizations obey $\mathcal N_{D-2}/\mathcal N_D=4\nu$, and the common
$\ee^{(1+2\delta)\gamma_E\eps}$ cancels between adjacent dimensions.  The
explicit recurrence is
\begingroup
\setlength{\fboxsep}{6pt}
\begin{equation}
\boxed{\begin{aligned}
 -J[n_1,n_2;b_1,b_2; D-2]
 &={}\frac{n_1+\delta\eps}{2}
 J[n_1+1,n_2;b_1+1,b_2; D]\\
 &+\frac{n_1+\delta\eps}{2}
 J[n_1+1,n_2;b_1,b_2+1; D]\\
 &-\frac{n_2+\delta\eps}{2}
 J[n_1,n_2+1;b_1+1,b_2; D]\\
 &-\frac{n_2+\delta\eps}{2}
 J[n_1,n_2+1;b_1,b_2+1; D]\\
 &-\frac{b_1}{2}J[n_1,n_2;b_1+2,b_2; D]\\
 &-\frac{b_2}{2}J[n_1,n_2;b_1,b_2+2; D].
\end{aligned}}
 \label{pp-dimension-recurrence}
\end{equation}
\endgroup
The coefficients $p$ and $1$ in Eq.~\eqref{pp-dlog-J-combinations}
are fixed by matching the rational Baikov kernels to the chosen
logarithmic forms. For the $(+-)$ component, $z_3=E_1+z_1+z_2$, so the two terms
proportional to $n_1+\delta\eps$ on the right-hand side should change the sign.

\paragraph{Normalization of the dlog integrals}
The auxiliary family has the Baikov representation
\eqref{pm-baikov-family} with the ordered energy denominators.  In the
two dimensions used for the construction, its scalar prefactors are
\begin{align}
 \mathcal N_d p^{2-d}
 &=\frac2p\,
 \frac{\ee^{(1+2\delta)\gamma_E\eps}}{\Gamma[1-\eps]}
 (4p^2)^\eps,\notag\\
 \mathcal N_{d-2}p^{4-d}
 &=8p\,
 \frac{\ee^{(1+2\delta)\gamma_E\eps}}{\Gamma[-\eps]}
 (4p^2)^\eps.
 \label{eq:baikov-norm}
\end{align}
The Baikov powers are $\mathcal P^{-\eps}$ and
$\mathcal P^{-1-\eps}$, respectively.  The lower-dimensional prefactor
vanishes at $\eps=0$, and the chosen relative normalization uses
\begin{equation}
\begin{gathered}
 \Gamma[1-\eps]=-\eps\Gamma[-\eps],\\[4pt]
 -\frac1\eps\mathcal N_{d-2}p^{4-d}
 =8p\,\frac{\ee^{(1+2\delta)\gamma_E\eps}}{\Gamma[1-\eps]}
 (4p^2)^\eps.
\end{gathered}
 \label{eq:gamma-shift}
\end{equation}
Thus the chosen factor $-1/\eps$ converts the inverse Gamma factor to
one whose value at $\eps=0$ is unity, compensating the extra power of
$\eps$ in the lower-dimensional measure independently of the energy
regulator.

We normalize the integral associated with each $\Phi_i$ to have the
common Baikov representation
\begin{equation}
 \frac{\ee^{(1+2\delta)\gamma_E\eps}}{\Gamma[1-\eps]}
 \int_{\mathcal D_p}
 \left(\frac{4p^2}{\mathcal P}\right)^\eps
 (z_3)^{-\delta\eps}(z_4)^{-\delta\eps}\Phi_i.
 \label{eq:period-norm}
\end{equation}
Its rational kernel at $\eps=0$ is $\Phi_i$.  The Euler-normalized
Gamma factor has the uniformly weighted expansion
\begin{equation}
 \frac{\ee^{\gamma_E\eps}}{\Gamma[1-\eps]}
 =\exp\left[-\sum_{n=2}^\infty\frac{\zeta(n)}n\eps^n\right].
\end{equation}
The additional Euler factor in our loop convention cancels the common
Euler terms of the Gamma functions from two time integrations.  More explicitly,
\begin{equation}
 \frac{\ee^{(1+2\delta)\gamma_E\eps}\Gamma[1+\delta\eps]^2}
 {\Gamma[1-\eps]}
 =\exp\left[\sum_{n=2}^\infty
 \frac{2(-\delta)^n-1}{n}\zeta(n)\eps^n\right].
 \label{eq:gamma-factor}
\end{equation}
We omit this common normalization when matching the indices of the two families.
\section{Simplifications of three ordered basis integrals}
\label{app:pp-lifts}

Three ordered basis integrals admit the following shorter representations in the
\(J^{\mathrm{top}}\oplus J^{\mathrm{sub}}\) family.
\begin{align}
 f^{(++)}_5&={}\frac\delta2\bigl(
 -2J^{\mathrm{top}}_{0,-1;-1,3}
 +2J^{\mathrm{top}}_{0,-1;1,1}
 +2p^2J^{\mathrm{top}}_{0,-1;1,3}
 -2J^{\mathrm{top}}_{0,0;-1,2}
 \notag\\[-1mm]
 &\quad
 -6J^{\mathrm{top}}_{0,0;0,1}
 +2p^2J^{\mathrm{top}}_{0,0;1,2}
 +2J^{\mathrm{top}}_{1,-1;-1,2}
 +6J^{\mathrm{top}}_{1,-1;0,1}
 -2p^2J^{\mathrm{top}}_{1,-1;1,2}
 \bigr)
 \notag\\
 &={}2\delta(2\eps-1)J^{\mathrm{top}}_{0,-1;1,1},
 \label{eq:lift-f5}\\
 f^{(++)}_6&={}\frac\delta2\bigl(
 -2J^{\mathrm{top}}_{-1,0;-1,3}
 +2J^{\mathrm{top}}_{-1,0;1,1}
 +2p^2J^{\mathrm{top}}_{-1,0;1,3}
 -2J^{\mathrm{top}}_{-1,1;-1,2}
 \notag\\[-1mm]
 &\quad
 -6J^{\mathrm{top}}_{-1,1;0,1}
 +2p^2J^{\mathrm{top}}_{-1,1;1,2}
 +2J^{\mathrm{top}}_{0,0;-1,2}
 +6J^{\mathrm{top}}_{0,0;0,1}
 -2p^2J^{\mathrm{top}}_{0,0;1,2}
 \bigr)
 \notag\\
 &={}2\delta(2\eps-1)J^{\mathrm{top}}_{-1,0;1,1}.
 \label{eq:lift-f6}
\end{align}

\begin{align}
f^{(++)}_9&=-\frac{\delta^2\eps}{2}\Bigl(
2J^{\mathrm{top}}_{-1,-1;-1,4}+3J^{\mathrm{top}}_{-1,-1;0,3}
-J^{\mathrm{top}}_{-1,-1;1,2}\notag\\
&\quad-2p^2J^{\mathrm{top}}_{-1,-1;1,4}-p^2J^{\mathrm{top}}_{-1,-1;2,3}
+J^{\mathrm{top}}_{-1,0;-1,3}+4J^{\mathrm{top}}_{-1,0;0,2}\notag\\
&\quad+3J^{\mathrm{top}}_{-1,0;1,1}-p^2J^{\mathrm{top}}_{-1,0;1,3}
-p^2J^{\mathrm{top}}_{-1,0;2,2}\notag\\
&\quad-J^{\mathrm{top}}_{0,-1;-1,3}-4J^{\mathrm{top}}_{0,-1;0,2}
-3J^{\mathrm{top}}_{0,-1;1,1}+p^2J^{\mathrm{top}}_{0,-1;1,3}
+p^2J^{\mathrm{top}}_{0,-1;2,2}\Bigr)\notag\\
&={}\delta^2\eps[-J^{\mathrm{top}}_{-1,-1;0,3}
+(4\eps-1)J^{\mathrm{top}}_{-1,-1;1,2}].
 \label{eq:radial-lift}
\end{align}
These simplifications follow from IBP reduction in $d-2$ dimensions and involve no contact integrals.

\Needspace{20\baselineskip}
\section{CDE matrices}
\label{app:residue-matrices}
This appendix collects the residue matrices of the three connections used
in the main text: the three-scale connection of the mixed sector, the
three-scale connection of the ordered sector, and the ordered connection
in the blow-up variables.

\subsection{CDE matrices for the $(+-)$ component}
\label{app:mixed-residues}
These residues act on the unscaled integral vector $(f^{(+-)}_1,\ldots,f^{(+-)}_7)^T$ in
\eqref{eq:mixed-cde}.  Their entries are functions of $\delta$ only.
\begingroup\setlength{\arraycolsep}{4pt}\renewcommand{\arraystretch}{1.15}
\begin{equation}
R^{(+-)}_{p}=\begin{pmatrix}
-2 & 0 & 0 & 0 & 0 & 0 & 0\\
0 & 0 & 0 & 0 & 0 & 0 & 0\\
0 & 0 & 0 & 0 & 0 & 0 & 0\\
0 & 0 & 0 & 2 & 0 & 0 & 0\\
0 & 0 & 0 & 0 & 2 & 0 & 0\\
0 & 2\delta & -2\delta & 0 & 0 & -2 & 0\\
0 & 0 & 0 & -2\delta & -2\delta & 0 & -2
\end{pmatrix}
\end{equation}

\begin{equation}
R^{(+-)}_{E_1}=\begin{pmatrix}
0 & 0 & 0 & 0 & 0 & 0 & 0\\
0 & 0 & 0 & 0 & 0 & 0 & 0\\
0 & 0 & 0 & 0 & 0 & 0 & 0\\
0 & 0 & 0 & 0 & 0 & 0 & 0\\
0 & 0 & 0 & 0 & 0 & 0 & 0\\
0 & 0 & 0 & 0 & 0 & 0 & 0\\
0 & 0 & 0 & 2\delta & 0 & 0 & -\delta
\end{pmatrix}
\end{equation}

\begin{equation}
R^{(+-)}_{E_2}=\begin{pmatrix}
0 & 0 & 0 & 0 & 0 & 0 & 0\\
0 & 0 & 0 & 0 & 0 & 0 & 0\\
0 & 0 & 0 & 0 & 0 & 0 & 0\\
0 & 0 & 0 & 0 & 0 & 0 & 0\\
0 & 0 & 0 & 0 & 0 & 0 & 0\\
0 & 0 & 0 & 0 & 0 & 0 & 0\\
0 & 0 & 0 & 0 & 2\delta & 0 & -\delta
\end{pmatrix}
\end{equation}

\begin{equation}
R^{(+-)}_{p-E_1}=\frac{1}{4}\begin{pmatrix}
-2\delta & \delta(\delta+2) & -\delta^{2} & 0 & 0 & 0 & 0\\
4 & -2(\delta+2) & 2\delta & 0 & 0 & 0 & 0\\
0 & 0 & 0 & 0 & 0 & 0 & 0\\
0 & 2-\delta & \delta & -2(\delta+4) & -2\delta & 2 & 0\\
0 & 0 & 0 & 0 & 0 & 0 & 0\\
2\delta & -4\delta & 0 & 2\delta(\delta+4) & 2\delta^{2} & -2\delta & 0\\
0 & 0 & 0 & 0 & 0 & 0 & 0
\end{pmatrix}
\end{equation}

\begin{equation}
R^{(+-)}_{p+E_1}=\frac{1}{4}\begin{pmatrix}
-2\delta & -\delta(\delta+2) & \delta^{2} & 0 & 0 & 0 & 0\\
-4 & -2(\delta+2) & 2\delta & 0 & 0 & 0 & 0\\
0 & 0 & 0 & 0 & 0 & 0 & 0\\
0 & \delta-2 & -\delta & -2(\delta+4) & -2\delta & -2 & 0\\
0 & 0 & 0 & 0 & 0 & 0 & 0\\
-2\delta & -4\delta & 0 & -2\delta(\delta+4) & -2\delta^{2} & -2\delta & 0\\
0 & 0 & 0 & 0 & 0 & 0 & 0
\end{pmatrix}
\end{equation}

\begin{equation}
R^{(+-)}_{p-E_2}=\frac{1}{4}\begin{pmatrix}
-2\delta & \delta^{2} & -\delta(\delta+2) & 0 & 0 & 0 & 0\\
0 & 0 & 0 & 0 & 0 & 0 & 0\\
-4 & 2\delta & -2(\delta+2) & 0 & 0 & 0 & 0\\
0 & 0 & 0 & 0 & 0 & 0 & 0\\
0 & -\delta & \delta-2 & -2\delta & -2(\delta+4) & 2 & 0\\
2\delta & 0 & 4\delta & 2\delta^{2} & 2\delta(\delta+4) & -2\delta & 0\\
0 & 0 & 0 & 0 & 0 & 0 & 0
\end{pmatrix}
\end{equation}

\begin{equation}
R^{(+-)}_{p+E_2}=\frac{1}{4}\begin{pmatrix}
-2\delta & -\delta^{2} & \delta(\delta+2) & 0 & 0 & 0 & 0\\
0 & 0 & 0 & 0 & 0 & 0 & 0\\
4 & 2\delta & -2(\delta+2) & 0 & 0 & 0 & 0\\
0 & 0 & 0 & 0 & 0 & 0 & 0\\
0 & \delta & 2-\delta & -2\delta & -2(\delta+4) & -2 & 0\\
-2\delta & 0 & 4\delta & -2\delta^{2} & -2\delta(\delta+4) & -2\delta & 0\\
0 & 0 & 0 & 0 & 0 & 0 & 0
\end{pmatrix}
\end{equation}

\begin{equation}
R^{(+-)}_{E_1-E_2}=\begin{pmatrix}
0 & 0 & 0 & 0 & 0 & 0 & 0\\
0 & -\delta & -\delta & 0 & 0 & 0 & 0\\
0 & -\delta & -\delta & 0 & 0 & 0 & 0\\
0 & 0 & 0 & -\delta & \delta & 0 & 0\\
0 & 0 & 0 & \delta & -\delta & 0 & 0\\
0 & 0 & 0 & 0 & 0 & 0 & 0\\
0 & 0 & 0 & 0 & 0 & 0 & 0
\end{pmatrix}
\end{equation}
\endgroup

\Needspace{20\baselineskip}
\subsection{CDE matrices for the $(++)$ component}
\label{app:ordered-residues}
The matrices $\widehat R_W$ below are the residues of the dimensionful
equation \eqref{pp-three-scale-cde}, in the order
$(f^{(++)}_1,f^{(++)}_2,f^{(++)}_3\mid f^{(++)}_4,f^{(++)}_5,f^{(++)}_6\mid
f^{(++)}_7,f^{(++)}_8,f^{(++)}_9\mid f^{(++)}_{10})$.
They satisfy $\sum_W\widehat R_W=-2(1+\delta)\one_{10}$.
\begingroup\setlength{\arraycolsep}{4pt}\renewcommand{\arraystretch}{1.15}
\begin{equation}
\widehat R_{p}=\begin{pmatrix}
-2 & 0 & 0 & 0 & 0 & 0 & 0 & 0 & 0 & 0\\
0 & -2 & 0 & 0 & 0 & 0 & 0 & 0 & 0 & 0\\
0 & 0 & -2 & 0 & 0 & 0 & 0 & 0 & 0 & 0\\
0 & 0 & 0 & -2 & 0 & 0 & 0 & 0 & 0 & 0\\
0 & 0 & 0 & 0 & 0 & 0 & 0 & 0 & 0 & 0\\
0 & 0 & 0 & 0 & 0 & 0 & 0 & 0 & 0 & 0\\
0 & 0 & 0 & 0 & 0 & 0 & 2 & 0 & 0 & 0\\
0 & 0 & 0 & 0 & 0 & 0 & 0 & 2 & 0 & 0\\
0 & 0 & 0 & 0 & 2\delta & -2\delta & 0 & 0 & -2 & 0\\
0 & 0 & 0 & 0 & 0 & 0 & -2\delta & -2\delta & 0 & -2
\end{pmatrix}
\end{equation}

\begin{equation}
\widehat R_{E_1}=\frac{1}{2}\begin{pmatrix}
0 & 0 & 0 & 0 & 0 & 0 & 0 & 0 & 0 & 0\\
0 & 0 & 0 & 0 & 0 & 0 & 0 & 0 & 0 & 0\\
0 & 0 & 0 & 0 & 0 & 0 & 0 & 0 & 0 & 0\\
0 & 0 & 0 & 0 & 0 & 0 & 0 & 0 & 0 & 0\\
0 & 0 & 0 & 0 & 0 & 0 & 0 & 0 & 0 & 0\\
0 & 0 & 0 & 0 & 0 & 0 & 0 & 0 & 0 & 0\\
0 & 0 & 0 & 0 & 0 & 0 & 0 & 0 & 0 & 0\\
0 & 0 & 0 & 0 & 0 & 0 & 0 & 0 & 0 & 0\\
0 & 0 & 0 & 0 & 0 & 0 & 0 & 0 & 0 & 0\\
0 & 0 & \delta & 0 & 0 & 0 & 4\delta & 0 & 0 & -2\delta
\end{pmatrix}
\end{equation}

\begin{equation}
\widehat R_{E_2}=\frac{1}{2}\begin{pmatrix}
0 & 0 & 0 & 0 & 0 & 0 & 0 & 0 & 0 & 0\\
0 & 0 & 0 & 0 & 0 & 0 & 0 & 0 & 0 & 0\\
0 & 0 & 0 & 0 & 0 & 0 & 0 & 0 & 0 & 0\\
0 & 0 & 0 & 0 & 0 & 0 & 0 & 0 & 0 & 0\\
0 & 0 & 0 & 0 & 0 & 0 & 0 & 0 & 0 & 0\\
0 & 0 & 0 & 0 & 0 & 0 & 0 & 0 & 0 & 0\\
0 & 0 & 0 & 0 & 0 & 0 & 0 & 0 & 0 & 0\\
0 & 0 & 0 & 0 & 0 & 0 & 0 & 0 & 0 & 0\\
0 & 0 & 0 & 0 & 0 & 0 & 0 & 0 & 0 & 0\\
0 & 0 & -\delta & 0 & 0 & 0 & 0 & 4\delta & 0 & -2\delta
\end{pmatrix}
\end{equation}

\begin{equation}
\widehat R_{p-E_1}=\frac{1}{4}\begin{pmatrix}
0 & 0 & 0 & 0 & 0 & 0 & 0 & 0 & 0 & 0\\
0 & 0 & 0 & 0 & 0 & 0 & 0 & 0 & 0 & 0\\
0 & 0 & 0 & 0 & 0 & 0 & 0 & 0 & 0 & 0\\
-\delta & 0 & 0 & -2\delta & -\delta(\delta+2) & \delta^{2} & 0 & 0 & 0 & 0\\
-2 & 0 & 0 & -4 & -2(\delta+2) & 2\delta & 0 & 0 & 0 & 0\\
0 & 0 & 0 & 0 & 0 & 0 & 0 & 0 & 0 & 0\\
0 & 1 & 0 & 0 & \delta-2 & -\delta & -2(\delta+4) & -2\delta & -2 & 0\\
0 & 0 & 0 & 0 & 0 & 0 & 0 & 0 & 0 & 0\\
-\delta & \delta & 0 & -2\delta & -4\delta & 0 & -2\delta(\delta+4) & -2\delta^{2} & -2\delta & 0\\
0 & 0 & 0 & 0 & 0 & 0 & 0 & 0 & 0 & 0
\end{pmatrix}
\end{equation}

\begin{equation}
\widehat R_{p+E_1}=\frac{1}{4}\begin{pmatrix}
0 & 0 & 0 & 0 & 0 & 0 & 0 & 0 & 0 & 0\\
0 & 0 & 0 & 0 & 0 & 0 & 0 & 0 & 0 & 0\\
0 & 0 & 0 & 0 & 0 & 0 & 0 & 0 & 0 & 0\\
-\delta & 0 & 0 & -2\delta & \delta(\delta+2) & -\delta^{2} & 0 & 0 & 0 & 0\\
2 & 0 & 0 & 4 & -2(\delta+2) & 2\delta & 0 & 0 & 0 & 0\\
0 & 0 & 0 & 0 & 0 & 0 & 0 & 0 & 0 & 0\\
0 & -1 & 0 & 0 & 2-\delta & \delta & -2(\delta+4) & -2\delta & 2 & 0\\
0 & 0 & 0 & 0 & 0 & 0 & 0 & 0 & 0 & 0\\
\delta & \delta & 0 & 2\delta & -4\delta & 0 & 2\delta(\delta+4) & 2\delta^{2} & -2\delta & 0\\
0 & 0 & 0 & 0 & 0 & 0 & 0 & 0 & 0 & 0
\end{pmatrix}
\end{equation}

\begin{equation}
\widehat R_{p-E_2}=\frac{1}{4}\begin{pmatrix}
0 & 0 & 0 & 0 & 0 & 0 & 0 & 0 & 0 & 0\\
0 & 0 & 0 & 0 & 0 & 0 & 0 & 0 & 0 & 0\\
0 & 0 & 0 & 0 & 0 & 0 & 0 & 0 & 0 & 0\\
\delta & 0 & 0 & -2\delta & \delta^{2} & -\delta(\delta+2) & 0 & 0 & 0 & 0\\
0 & 0 & 0 & 0 & 0 & 0 & 0 & 0 & 0 & 0\\
2 & 0 & 0 & -4 & 2\delta & -2(\delta+2) & 0 & 0 & 0 & 0\\
0 & 0 & 0 & 0 & 0 & 0 & 0 & 0 & 0 & 0\\
0 & 1 & 0 & 0 & -\delta & \delta-2 & -2\delta & -2(\delta+4) & 2 & 0\\
-\delta & -\delta & 0 & 2\delta & 0 & 4\delta & 2\delta^{2} & 2\delta(\delta+4) & -2\delta & 0\\
0 & 0 & 0 & 0 & 0 & 0 & 0 & 0 & 0 & 0
\end{pmatrix}
\end{equation}

\begin{equation}
\widehat R_{p+E_2}=\frac{1}{4}\begin{pmatrix}
0 & 0 & 0 & 0 & 0 & 0 & 0 & 0 & 0 & 0\\
0 & 0 & 0 & 0 & 0 & 0 & 0 & 0 & 0 & 0\\
0 & 0 & 0 & 0 & 0 & 0 & 0 & 0 & 0 & 0\\
\delta & 0 & 0 & -2\delta & -\delta^{2} & \delta(\delta+2) & 0 & 0 & 0 & 0\\
0 & 0 & 0 & 0 & 0 & 0 & 0 & 0 & 0 & 0\\
-2 & 0 & 0 & 4 & 2\delta & -2(\delta+2) & 0 & 0 & 0 & 0\\
0 & 0 & 0 & 0 & 0 & 0 & 0 & 0 & 0 & 0\\
0 & -1 & 0 & 0 & \delta & 2-\delta & -2\delta & -2(\delta+4) & -2 & 0\\
\delta & -\delta & 0 & -2\delta & 0 & 4\delta & -2\delta^{2} & -2\delta(\delta+4) & -2\delta & 0\\
0 & 0 & 0 & 0 & 0 & 0 & 0 & 0 & 0 & 0
\end{pmatrix}
\end{equation}

\begin{equation}
\widehat R_{E_1+E_2}=\begin{pmatrix}
-2\delta & 0 & 0 & 0 & 0 & 0 & 0 & 0 & 0 & 0\\
0 & -2\delta & 0 & 0 & 0 & 0 & 0 & 0 & 0 & 0\\
0 & 0 & -2\delta & 0 & 0 & 0 & 0 & 0 & 0 & 0\\
0 & 0 & 0 & 0 & 0 & 0 & 0 & 0 & 0 & 0\\
0 & 0 & 0 & 0 & -\delta & -\delta & 0 & 0 & 0 & 0\\
0 & 0 & 0 & 0 & -\delta & -\delta & 0 & 0 & 0 & 0\\
0 & 0 & 0 & 0 & 0 & 0 & -\delta & \delta & 0 & 0\\
0 & 0 & 0 & 0 & 0 & 0 & \delta & -\delta & 0 & 0\\
0 & 0 & 0 & 0 & 0 & 0 & 0 & 0 & 0 & 0\\
0 & 0 & 0 & 0 & 0 & 0 & 0 & 0 & 0 & 0
\end{pmatrix}
\end{equation}
\endgroup

For the dimensionless variables $x=p/E_1$, $y=p/E_2$, the matrices $R_W$,
$W\in\mathbb L_{xy}$, in Eq.~\eqref{pp-xy-cde} are obtained from the change of variables.
Writing each letter of $\mathbb L_{++}$ as $p$ times a
function of $x,y$, the prefactor $p^{2(1+\delta)\eps}$ in
Eq.~\eqref{eq:xy} cancels the total $\dd\log p$ term, so
$\widehat R_p$ does not enter and
\begin{equation}
\begin{aligned}
 R_x&=-\widehat R_{E_1}-\widehat R_{p-E_1}-\widehat R_{p+E_1}-\widehat R_{E_1+E_2},\\
 R_y&=-\widehat R_{E_2}-\widehat R_{p-E_2}-\widehat R_{p+E_2}-\widehat R_{E_1+E_2},\\
 R_{1\pm x}&=\widehat R_{p\pm E_1},\qquad
 R_{1\pm y}=\widehat R_{p\pm E_2},\\
 R_{x+y}&=\widehat R_{E_1+E_2}.
\end{aligned}
 \label{pp-xy-residue-map}
\end{equation}
The constant sign between $p-E_1$ and $1-x$, and likewise for $y$,
does not affect the logarithmic differential.

\Needspace{20\baselineskip}
\subsection{Blow-up matrices}
\label{app:corner-residues}
These are the full-connection matrices in \eqref{pp-uv-cde}.
Only $\widetilde R_u,\widetilde R_v$ are singular at the resolved corner;
the remaining five matrices are retained for MPL transport.
Since $x=uv$, $y=v$ and $x+y=v(1+u)$, they follow directly from the
matrices $R_W$ of \eqref{pp-xy-cde}:
\begin{equation}
\begin{aligned}
 \widetilde R_u&=R_x,\qquad
 \widetilde R_v=R_x+R_y+R_{x+y},\qquad
 \widetilde R_{1+u}=R_{x+y},\\
 \widetilde R_{1\pm uv}&=R_{1\pm x},\qquad
 \widetilde R_{1\pm v}=R_{1\pm y}.
\end{aligned}
 \label{pp-uv-residue-map}
\end{equation}
For convenience we list them explicitly.
\begingroup\setlength{\arraycolsep}{4pt}\renewcommand{\arraystretch}{1.15}
\begin{equation}
\widetilde R_{u}=\frac{1}{2}\begin{pmatrix}
4\delta & 0 & 0 & 0 & 0 & 0 & 0 & 0 & 0 & 0\\
0 & 4\delta & 0 & 0 & 0 & 0 & 0 & 0 & 0 & 0\\
0 & 0 & 4\delta & 0 & 0 & 0 & 0 & 0 & 0 & 0\\
\delta & 0 & 0 & 2\delta & 0 & 0 & 0 & 0 & 0 & 0\\
0 & 0 & 0 & 0 & 4(\delta+1) & 0 & 0 & 0 & 0 & 0\\
0 & 0 & 0 & 0 & 2\delta & 2\delta & 0 & 0 & 0 & 0\\
0 & 0 & 0 & 0 & 0 & 0 & 4(\delta+2) & 0 & 0 & 0\\
0 & 0 & 0 & 0 & 0 & 0 & -2\delta & 2\delta & 0 & 0\\
0 & -\delta & 0 & 0 & 4\delta & 0 & 0 & 0 & 2\delta & 0\\
0 & 0 & -\delta & 0 & 0 & 0 & -4\delta & 0 & 0 & 2\delta
\end{pmatrix}
\end{equation}

\begin{equation}
\widetilde R_{v}=\begin{pmatrix}
2\delta & 0 & 0 & 0 & 0 & 0 & 0 & 0 & 0 & 0\\
0 & 2\delta & 0 & 0 & 0 & 0 & 0 & 0 & 0 & 0\\
0 & 0 & 2\delta & 0 & 0 & 0 & 0 & 0 & 0 & 0\\
0 & 0 & 0 & 2\delta & 0 & 0 & 0 & 0 & 0 & 0\\
0 & 0 & 0 & 0 & 2(\delta+1) & 0 & 0 & 0 & 0 & 0\\
0 & 0 & 0 & 0 & 0 & 2(\delta+1) & 0 & 0 & 0 & 0\\
0 & 0 & 0 & 0 & 0 & 0 & 2(\delta+2) & 0 & 0 & 0\\
0 & 0 & 0 & 0 & 0 & 0 & 0 & 2(\delta+2) & 0 & 0\\
0 & 0 & 0 & 0 & 2\delta & -2\delta & 0 & 0 & 2\delta & 0\\
0 & 0 & 0 & 0 & 0 & 0 & -2\delta & -2\delta & 0 & 2\delta
\end{pmatrix}
\end{equation}

\begin{equation}
\widetilde R_{1-uv}=\frac{1}{4}\begin{pmatrix}
0 & 0 & 0 & 0 & 0 & 0 & 0 & 0 & 0 & 0\\
0 & 0 & 0 & 0 & 0 & 0 & 0 & 0 & 0 & 0\\
0 & 0 & 0 & 0 & 0 & 0 & 0 & 0 & 0 & 0\\
-\delta & 0 & 0 & -2\delta & -\delta(\delta+2) & \delta^{2} & 0 & 0 & 0 & 0\\
-2 & 0 & 0 & -4 & -2(\delta+2) & 2\delta & 0 & 0 & 0 & 0\\
0 & 0 & 0 & 0 & 0 & 0 & 0 & 0 & 0 & 0\\
0 & 1 & 0 & 0 & \delta-2 & -\delta & -2(\delta+4) & -2\delta & -2 & 0\\
0 & 0 & 0 & 0 & 0 & 0 & 0 & 0 & 0 & 0\\
-\delta & \delta & 0 & -2\delta & -4\delta & 0 & -2\delta(\delta+4) & -2\delta^{2} & -2\delta & 0\\
0 & 0 & 0 & 0 & 0 & 0 & 0 & 0 & 0 & 0
\end{pmatrix}
\end{equation}

\begin{equation}
\widetilde R_{1+uv}=\frac{1}{4}\begin{pmatrix}
0 & 0 & 0 & 0 & 0 & 0 & 0 & 0 & 0 & 0\\
0 & 0 & 0 & 0 & 0 & 0 & 0 & 0 & 0 & 0\\
0 & 0 & 0 & 0 & 0 & 0 & 0 & 0 & 0 & 0\\
-\delta & 0 & 0 & -2\delta & \delta(\delta+2) & -\delta^{2} & 0 & 0 & 0 & 0\\
2 & 0 & 0 & 4 & -2(\delta+2) & 2\delta & 0 & 0 & 0 & 0\\
0 & 0 & 0 & 0 & 0 & 0 & 0 & 0 & 0 & 0\\
0 & -1 & 0 & 0 & 2-\delta & \delta & -2(\delta+4) & -2\delta & 2 & 0\\
0 & 0 & 0 & 0 & 0 & 0 & 0 & 0 & 0 & 0\\
\delta & \delta & 0 & 2\delta & -4\delta & 0 & 2\delta(\delta+4) & 2\delta^{2} & -2\delta & 0\\
0 & 0 & 0 & 0 & 0 & 0 & 0 & 0 & 0 & 0
\end{pmatrix}
\end{equation}

\begin{equation}
\widetilde R_{1-v}=\frac{1}{4}\begin{pmatrix}
0 & 0 & 0 & 0 & 0 & 0 & 0 & 0 & 0 & 0\\
0 & 0 & 0 & 0 & 0 & 0 & 0 & 0 & 0 & 0\\
0 & 0 & 0 & 0 & 0 & 0 & 0 & 0 & 0 & 0\\
\delta & 0 & 0 & -2\delta & \delta^{2} & -\delta(\delta+2) & 0 & 0 & 0 & 0\\
0 & 0 & 0 & 0 & 0 & 0 & 0 & 0 & 0 & 0\\
2 & 0 & 0 & -4 & 2\delta & -2(\delta+2) & 0 & 0 & 0 & 0\\
0 & 0 & 0 & 0 & 0 & 0 & 0 & 0 & 0 & 0\\
0 & 1 & 0 & 0 & -\delta & \delta-2 & -2\delta & -2(\delta+4) & 2 & 0\\
-\delta & -\delta & 0 & 2\delta & 0 & 4\delta & 2\delta^{2} & 2\delta(\delta+4) & -2\delta & 0\\
0 & 0 & 0 & 0 & 0 & 0 & 0 & 0 & 0 & 0
\end{pmatrix}
\end{equation}

\begin{equation}
\widetilde R_{1+v}=\frac{1}{4}\begin{pmatrix}
0 & 0 & 0 & 0 & 0 & 0 & 0 & 0 & 0 & 0\\
0 & 0 & 0 & 0 & 0 & 0 & 0 & 0 & 0 & 0\\
0 & 0 & 0 & 0 & 0 & 0 & 0 & 0 & 0 & 0\\
\delta & 0 & 0 & -2\delta & -\delta^{2} & \delta(\delta+2) & 0 & 0 & 0 & 0\\
0 & 0 & 0 & 0 & 0 & 0 & 0 & 0 & 0 & 0\\
-2 & 0 & 0 & 4 & 2\delta & -2(\delta+2) & 0 & 0 & 0 & 0\\
0 & 0 & 0 & 0 & 0 & 0 & 0 & 0 & 0 & 0\\
0 & -1 & 0 & 0 & \delta & 2-\delta & -2\delta & -2(\delta+4) & -2 & 0\\
\delta & -\delta & 0 & -2\delta & 0 & 4\delta & -2\delta^{2} & -2\delta(\delta+4) & -2\delta & 0\\
0 & 0 & 0 & 0 & 0 & 0 & 0 & 0 & 0 & 0
\end{pmatrix}
\end{equation}

\begin{equation}
\widetilde R_{1+u}=\begin{pmatrix}
-2\delta & 0 & 0 & 0 & 0 & 0 & 0 & 0 & 0 & 0\\
0 & -2\delta & 0 & 0 & 0 & 0 & 0 & 0 & 0 & 0\\
0 & 0 & -2\delta & 0 & 0 & 0 & 0 & 0 & 0 & 0\\
0 & 0 & 0 & 0 & 0 & 0 & 0 & 0 & 0 & 0\\
0 & 0 & 0 & 0 & -\delta & -\delta & 0 & 0 & 0 & 0\\
0 & 0 & 0 & 0 & -\delta & -\delta & 0 & 0 & 0 & 0\\
0 & 0 & 0 & 0 & 0 & 0 & -\delta & \delta & 0 & 0\\
0 & 0 & 0 & 0 & 0 & 0 & \delta & -\delta & 0 & 0\\
0 & 0 & 0 & 0 & 0 & 0 & 0 & 0 & 0 & 0\\
0 & 0 & 0 & 0 & 0 & 0 & 0 & 0 & 0 & 0
\end{pmatrix}
\end{equation}
\endgroup

\Needspace{20\baselineskip}
\section{Compact MPL coefficients through weight three}
\label{app:pp-explicit-gpl}
\raggedbottom

We first state the finite word sum and its boundary coefficients, then
list every component at orders two and three.  Only residue
matrices already specified in Appendix~\ref{app:corner-residues} and
the exact boundary vector are needed.  For the two path segments use
the dictionaries
\begingroup
\renewcommand{\arraystretch}{1.6}
\setlength{\arraycolsep}{6pt}
\begin{equation}
\begin{array}{c|cccc}
 w&0&-1&1/v&-1/v\\ \hline
 \widetilde R^u_w&\widetilde R_u&\widetilde R_{1+u}&\widetilde R_{1-uv}&\widetilde R_{1+uv}
\end{array}
\qquad
\begin{array}{c|ccc}
 \bar w&0&1&-1\\ \hline
 \widetilde R^v_{\bar w}&\widetilde R_v&\widetilde R_{1-v}&\widetilde R_{1+v}\rlap{.}
\end{array}
\end{equation}
\endgroup
These superscripts label segment kernels, not new connection matrices.
For $n=0,1,2,3$ the analytic answer is
\begin{equation}
\begin{aligned}
 \bm f^{(n)}&={}
 \sum_{j+j'\leq n}\;
 \sum_{w_1,\ldots,w_j\in\{0,-1,1/v,-1/v\}}\;
 \sum_{\bar w_1,\ldots,\bar w_{j'}\in\{0,1,-1\}}\\
 &\quad G(w_1,\ldots,w_j;u)G(\bar w_1,\ldots,\bar w_{j'};v)
 \widetilde R^u_{w_1}\cdots \widetilde R^u_{w_j}
 \widetilde R^v_{\bar w_1}\cdots \widetilde R^v_{\bar w_{j'}}\bm b^{[n-2-j-j']}.
\end{aligned}
\label{pp-finite-gpl-word-sum}
\end{equation}
Empty words and matrix products are one and the identity, respectively.
The word lengths $j,j'$ are non-negative.  All sums are finite; setting $n=2$
or $n=3$ gives the two higher
orders directly, including their boundary constants.  The order of the
matrices is essential.  For instance, the mixed length-two term is
$G(w;u)G(\bar w;v)\widetilde R^u_w\widetilde R^v_{\bar w}\bm b^{[-2]}$, not the reversed product.

The boundary coefficients follow from the expansions summarized in
Table~\ref{tab:kappas}:
\begin{equation}
 \kappa(\eps)=A\eps^p\exp\left[\eps L
 +\frac{\zeta(2)}2C_2\eps^2-\frac{\zeta(3)}3C_3\eps^3
 +O(\eps^4)\right].
 \label{eq:kappa-series}
\end{equation}
\begin{table}[b]
\centering
\small
\begin{tabular}{c c c l l l}
\toprule
$\kappa$&$p$&$A$&$L$&$C_2$&$C_3$\\
\midrule
$\kappa_1$&$-2$&$2$&$-\ii\pi\delta-2\log2$&$4\delta^2+1$&$8\delta^3+13$\\
$\kappa_2$&$-2$&$2$&$-\ii\pi\delta-2\log2$&$4\delta^2-7$&$8\delta^3+49$\\
$\kappa_3$&$0$&$-\pi^2$&$-\ii\pi\delta+2\log2$&$4\delta^2+5$&$8\delta^3+1$\\
$\kappa_4$&$-2$&$-2$&$-\ii\pi\delta-2\log2$&$2\delta^2+1$&$2\delta^3+13$\\
$\kappa_6$&$-2$&$\frac4{\delta+2}$&$-\ii\pi\delta$&$2\delta^2+4\delta+5$&$2\delta^3+6\delta^2+12\delta+7$\\
$\kappa_5$&$-2$&$\frac{\delta^2}{(\delta+1)(\delta+2)}$&$-\ii\pi\delta$&$4\delta^2+12\delta+13$&$8\delta^3+18\delta^2+12\delta+7$\\
$\kappa_9$&$-2$&$2$&$-\ii\pi\delta-2\log2$&$2\delta^2-7$&$2\delta^3+49$\\
$\kappa_{10}$&$0$&$-\pi^2$&$-\ii\pi\delta+2\log2$&$2\delta^2+5$&$2\delta^3+1$\\
\bottomrule
\end{tabular}
\caption{Parameters of the expansion
\eqref{eq:kappa-series} for the eight nonzero mode
coefficients $\kappa_A$ entering the boundary vector
\eqref{eq:boundary-vector}.}
\label{tab:kappas}
\end{table}
Explicitly, the four multipliers of $A\eps^p$ are
\begin{equation}
 1,\qquad L,\qquad
 \frac{L^2+\zeta(2)C_2}{2},\qquad
 \frac{L^3+3L\zeta(2)C_2-2\zeta(3)C_3}{6}.
\end{equation}
Insert these scalar coefficients into
\eqref{eq:boundary-vector} to obtain
$\bm b^{[-2]},\bm b^{[-1]},\bm b^{[0]},\bm b^{[1]}$.
For the two rows with $p=0$, only the first two multipliers are needed
through $\eps^1$.  The table follows by Gamma shifts and duplication from
\eqref{eq:contact-kappas},
\eqref{eq:soft-kappas},
\eqref{eq:semihard-kappa}, and
\eqref{eq:hard-kappas}.

\subsection{Contact components}

The first two contact components are elementary at these orders:
\begin{align}
 f^{(2)}_1&={}\left[-\ii\pi\delta-2\log2
 +2\delta\log\frac{uv}{1+u}\right]^2
 +(1+4\delta^2)\zeta(2),\notag\\
 f^{(3)}_1&={}\frac13\left[-\ii\pi\delta-2\log2
 +2\delta\log\frac{uv}{1+u}\right]^3\notag\\
 &+(1+4\delta^2)\zeta(2)
 \left[-\ii\pi\delta-2\log2+2\delta\log\frac{uv}{1+u}\right]
 -\frac23(13+8\delta^3)\zeta(3).
\end{align}
\begin{align}
 f^{(2)}_2&={}\left[-\ii\pi\delta-2\log2
 +2\delta\log\frac{uv}{1+u}\right]^2
 +(-7+4\delta^2)\zeta(2),\notag\\
 f^{(3)}_2&={}\frac13\left[-\ii\pi\delta-2\log2
 +2\delta\log\frac{uv}{1+u}\right]^3\notag\\
 &+(-7+4\delta^2)\zeta(2)
 \left[-\ii\pi\delta-2\log2+2\delta\log\frac{uv}{1+u}\right]
 -\frac23(49+8\delta^3)\zeta(3).
\end{align}
The third starts at weight two:
\begin{equation}
 f^{(2)}_3=-\pi^2,\qquad
 f^{(3)}_3=-\pi^2\left[-\ii\pi\delta+2\log2
 +2\delta\log\frac{uv}{1+u}\right].
\end{equation}

\subsection{The second-order components}

The contact components are given above.  The remaining seven components
are listed explicitly below.  Repeated-letter MPLs may equivalently be
replaced by
$G(w,\ldots,w;z)=G(w;z)^n/n!$ for a word of length $n$; retaining a few such words shortens the
formulas.  Products at different endpoints are kept as products.
All logarithms are on the real branch for $0<v<1$, $0<uv<1$.
\begingroup\small\allowdisplaybreaks[0]

\begin{align}
 f^{(2)}_{4}&={}-2 \delta  (\delta +1) G\bigl(-1,-1;v\bigr)+2 \delta  (2 \delta +1) G\bigl(-1,0;v\bigr)\notag\\
 &\quad{}+2 \delta ^2 (-G\bigl(0,-1;u\bigr)+G\bigl(0,0;u\bigr)+G\bigl(\frac{1}{v},-1;u\bigr)-G\bigl(\frac{1}{v},0;u\bigr))\notag\\
 &\quad{}+4 \delta ^2 (G\bigl(0,-1;v\bigr)-G\bigl(0,0;v\bigr))+2 \delta  (G\bigl(1,-1;v\bigr)-G\bigl(1,0;v\bigr))\notag\\
 &\quad{}+\frac{1}{12} \bigl(\delta ^2 \bigl(24 \log u \log(1+v)-24 \log(1+v) \log(1-uv)+24 \ii \pi  \log v\notag\\
 &\quad{}-24 \ii \pi  \log(1+v)+6 \pi ^2\bigr)+\delta  (-24 \log 2 \log (1-v)+48 \log 2 \log v\notag\\
 &\quad{}-24 \log 2 \log(1+v)-24 \ii \pi  \log 2)-\pi ^2-24 \log^2 2\bigr)\label{pp-explicit-I2-4}
\end{align}

\begin{align}
 f^{(2)}_{5}&={}4 \delta ^2 (G\bigl(-1,-1;u\bigr)-G\bigl(-1,0;u\bigr))+4 \delta  (\delta +1) (-G\bigl(0,-1;u\bigr)+G\bigl(0,0;u\bigr)\notag\\
 &\quad{}+G\bigl(0,0;v\bigr))+4 \delta  (G\bigl(\frac{1}{v},-1;u\bigr)-G\bigl(\frac{1}{v},0;u\bigr))\notag\\
 &\quad{}+\frac{1}{12 (\delta +1)}\bigl[\bigl((\delta  (48 \delta +96)+48) \delta  \log v+(-24 \ii \pi  \delta -24 \ii \pi ) \delta ^2\bigr) \log \bigl(\frac{u v}{u+1}\bigr)\notag\\
 &\quad{}+\delta  (48 \delta +48) \log(1+u) \log(1+v)+(-48 \delta -48) \delta  \log(1+v) \log(1-uv)\notag\\
 &\quad{}+\delta  ((-48 \delta -96) \delta -48) \log ^2(v)+\delta  \bigl(\delta  \bigl(12 \pi ^2-2 \pi ^2 \delta \bigr)+13 \pi ^2\bigr)\bigr]\label{pp-explicit-I2-5}
\end{align}

\begin{align}
 f^{(2)}_{6}&={}4 \delta ^2 (G\bigl(-1,-1;u\bigr)-G\bigl(-1,0;u\bigr)-G\bigl(0,-1;u\bigr)+G\bigl(0,0;u\bigr))+4 (\delta +1) G\bigl(-1,-1;v\bigr)\notag\\
 &\quad{}-4 (2 \delta +1) G\bigl(-1,0;v\bigr)-8 (\delta +1) G\bigl(0,-1;v\bigr)+4 (\delta +1) (\delta +2) G\bigl(0,0;v\bigr)\notag\\
 &\quad{}+4 (G\bigl(1,-1;v\bigr)-G\bigl(1,0;v\bigr))\notag\\
 &\quad{}+\frac{1}{12 (\delta +1)}\bigl[\bigl((-24 \ii \pi  \delta -24 \ii \pi ) \delta ^2+(\delta  (48 \delta +96)+48) \delta  \log v\notag\\
 &\quad{}+(-48 \delta -48) \delta  \log(1+v)\bigr) \log \bigl(\frac{u v}{u+1}\bigr)+\delta  \bigl(13 \pi ^2-2 \pi ^2 \delta ^2\bigr)\notag\\
 &\quad{}+(-48 \delta -48) \log 2 \log (1-v)\notag\\
 &\quad{}+((48 \delta +48) \log 2+\delta  (48 \ii \pi  \delta +48 \ii \pi )) \log(1+v)\notag\\
 &\quad{}+\log v (\delta  ((-48 \delta -96) \delta -48) \log v+\delta  (48 \delta +48) \log(1+v)\notag\\
 &\quad{}+\delta  (-48 \ii \pi  \delta -48 \ii \pi ))+10 \pi ^2\bigr]\label{pp-explicit-I2-6}
\end{align}

\begin{align}
 f^{(2)}_{7}&={}0\label{pp-explicit-I2-7}
\end{align}

\begin{align}
 f^{(2)}_{8}&={}2 (-G\bigl(-1,-1;v\bigr)+G\bigl(-1,0;v\bigr)+G\bigl(1,-1;v\bigr)-G\bigl(1,0;v\bigr))\notag\\
 &\quad{}-2 \log 2 (\log (1-v)-\log(1+v))\label{pp-explicit-I2-8}
\end{align}

\begin{align}
 f^{(2)}_{9}&={}-2 \delta  (\delta +2) G\bigl(-1,-1;v\bigr)+4 \delta  (\delta +1) G\bigl(-1,0;v\bigr)\notag\\
 &\quad{}+2 \delta ^2 (-G\bigl(0,-1;u\bigr)+G\bigl(0,0;u\bigr)+G\bigl(\frac{1}{v},-1;u\bigr)-G\bigl(\frac{1}{v},0;u\bigr))\notag\\
 &\quad{}+4 \delta  (\delta +2) (G\bigl(0,-1;v\bigr)-G\bigl(0,0;v\bigr))+4 \delta  (G\bigl(1,0;v\bigr)-G\bigl(1,-1;v\bigr))\notag\\
 &\quad{}+\frac{1}{12 (\delta +1)}\bigl[\delta  \bigl(\delta  \bigl(\delta  \bigl(24 \log u \log(1+v)-24 \log(1+v) \log(1-uv)+24 \ii \pi  \log v\notag\\
 &\quad{}-24 \ii \pi  \log(1+v)+6 \pi ^2\bigr)+24 \log u \log(1+v)-24 \log(1+v) \log(1-uv)\notag\\
 &\quad{}+48 \log 2 \log (1-v)+(-48 \log 2+24 \ii \pi ) \log v-24 \ii \pi  \log(1+v)-6 \pi ^2\notag\\
 &\quad{}+24 \ii \pi  \log 2\bigr)+48 \log 2 \log (1-v)-48 \log 2 \log v-17 \pi ^2+\log 2 (24 \log 2+24 \ii \pi )\bigr)\notag\\
 &\quad{}-7 \pi ^2+24 \log^2 2\bigr]\label{pp-explicit-I2-9}
\end{align}

\begin{align}
 f^{(2)}_{10}&={}-\frac{\pi ^2}{2}\label{pp-explicit-I2-10}
\end{align}

\endgroup

\subsection{The third-order components}

The nonzero coefficients $\kappa_A$ share the phase
$\exp(-\ii\pi\delta\eps)$.  We extract its contributions below only
to avoid printing the same complex terms repeatedly.  Multiplication by
$\exp(\ii\pi\delta\eps)$ gives at third order
\begin{equation}
 f_i^{(3)}+\ii\pi\delta f_i^{(2)}
 -\frac{\pi^2\delta^2}{2}f_i^{(1)}
 -\frac{\ii\pi^3\delta^3}{6}f_i^{(0)}.
\end{equation}
For real $\delta$, this combination is real in the stated chamber.  The expressions below
have been solved for $f_i^{(3)}$; every lower-order component on their
right-hand sides is already explicit.
\begingroup\small

\begin{align}
 f^{(3)}_{4}&={}-\ii\pi\delta f^{(2)}_{4}+\frac{\pi^2\delta^2}{2}f^{(1)}_{4}+\frac{\ii\pi^3\delta^3}{6}f^{(0)}_{4}+2 \delta  (\delta +1)^2 G\bigl(-1,-1,-1;v\bigr)\notag\\
 &\quad{}-2 \delta  (\delta +1) (2 \delta +1) G\bigl(-1,-1,0;v\bigr)-2 \delta  \bigl(2 \delta ^2+3 \delta +2\bigr) G\bigl(-1,0,-1;v\bigr)\notag\\
 &\quad{}+4 \delta  \bigl(2 \delta ^2+2 \delta +1\bigr) G\bigl(-1,0,0;v\bigr)+2 \delta  (G\bigl(-1,1,-1;v\bigr)-G\bigl(-1,1,0;v\bigr))\notag\\
 &\quad{}+4 \delta ^3 (G\bigl(0,-1,-1;u\bigr)-G\bigl(0,-1,0;u\bigr)-G\bigl(\frac{1}{v},-1,-1;u\bigr)+G\bigl(\frac{1}{v},-1,0;u\bigr))\notag\\
 &\quad{}-4 \delta ^2 (\delta +1) G\bigl(0,-1,-1;v\bigr)+4 \delta ^2 (2 \delta +1) G\bigl(0,-1,0;v\bigr)\notag\\
 &\quad{}+6 \delta ^3 (G\bigl(0,0,0;u\bigr)-G\bigl(0,0,-1;u\bigr))+8 \delta ^3 (G\bigl(0,0,-1;v\bigr)-G\bigl(0,0,0;v\bigr))\notag\\
 &\quad{}+4 \delta ^2 (G\bigl(0,1,-1;v\bigr)-G\bigl(0,1,0;v\bigr))+2 \delta ^3 (G\bigl(0,\frac{1}{v},-1;u\bigr)-G\bigl(0,\frac{1}{v},0;u\bigr))\notag\\
 &\quad{}+2 \delta  (\delta +1) (-G\bigl(1,-1,-1;v\bigr)-G\bigl(1,1,-1;v\bigr)+G\bigl(1,1,0;v\bigr))\notag\\
 &\quad{}+2 \delta  (2 \delta +1) G\bigl(1,-1,0;v\bigr)+2 \delta  (3 \delta +2) G\bigl(1,0,-1;v\bigr)\notag\\
 &\quad{}-4 \delta  (2 \delta +1) G\bigl(1,0,0;v\bigr)+2 \delta ^2 (-G\bigl(-\frac{1}{v},0,-1;u\bigr)+G\bigl(-\frac{1}{v},0,0;u\bigr)\notag\\
 &\quad{}+G\bigl(-\frac{1}{v},\frac{1}{v},-1;u\bigr)-G\bigl(-\frac{1}{v},\frac{1}{v},0;u\bigr))+2 \delta ^2 (3 \delta +1) (G\bigl(\frac{1}{v},0,-1;u\bigr)-G\bigl(\frac{1}{v},0,0;u\bigr))\notag\\
 &\quad{}+2 \delta ^2 (\delta +1) (G\bigl(\frac{1}{v},\frac{1}{v},0;u\bigr)-G\bigl(\frac{1}{v},\frac{1}{v},-1;u\bigr))-2 \delta  G\bigl(-1,1;v\bigr) \log 2\notag\\
 &\quad{}+8 \delta ^2 G\bigl(0,0;v\bigr) \log 2-4 \delta ^2 G\bigl(0,1;v\bigr) \log 2+2 \delta  (\delta +1) G\bigl(1,1;v\bigr) \log 2\notag\\
 &\quad{}+2 \delta ^2 G\bigl(-\frac{1}{v},0;u\bigr) (\log v+\log 2)-4 \delta ^2 G\bigl(0,-1;u\bigr) (\delta  \log v-\log 2)\notag\\
 &\quad{}+\delta ^2 G\bigl(\frac{1}{v},0;u\bigr) (\delta  (-4 \log v-2 \log(1+v))-2 \log v+2 \log 2)\notag\\
 &\quad{}-2 \delta ^2 G\bigl(-\frac{1}{v},-1;u\bigr) (\log v-\log(1+v)+\log 2)+2 \delta ^2 G\bigl(\frac{1}{v},-1;u\bigr) (2 \delta  \log v+\log v\notag\\
 &\quad{}-\log(1+v)-\log 2)-2 \delta ^3 G\bigl(0,\frac{1}{v};u\bigr) \log(1+v)-2 \delta ^2 G\bigl(-\frac{1}{v},\frac{1}{v};u\bigr) \log(1+v)\notag\\
 &\quad{}+2 \delta ^2 (\delta +1) G\bigl(\frac{1}{v},\frac{1}{v};u\bigr) \log(1+v)+\delta ^2 G\bigl(0,0;u\bigr) (\delta  (4 \log v+2 \log(1+v))\notag\\
 &\quad{}-4 \log 2)+\delta ^2 G\bigl(-1,0;v\bigr) (\delta  (4 \log u-4 \log(1-uv))-2 \log(1-uv)+2 \log u\notag\\
 &\quad{}-4 \log 2)+2 \delta  (\delta +1) G\bigl(-1,-1;v\bigr) (\delta  \log(1-uv)-\delta  \log u+\log 2)\notag\\
 &\quad{}+\delta  G\bigl(1,-1;v\bigr) (\delta  (2 \log u-2 \log(1+uv))-2 \log 2)\notag\\
 &\quad{}-2 \delta ^2 G\bigl(1,0;v\bigr) (-\log(1+uv)+\log u+2 \log 2)\notag\\
 &\quad{}+\delta ^2 G\bigl(0,-1;v\bigr) (\delta  (4 \log u-4 \log(1-uv))-2 \log(1-uv)+2 \log(1+uv)-4 \log 2)\notag\\
 &\quad{}+\frac{1}{6} \bigl(\delta ^3 \bigl(-2 \pi ^2 \log(1-uv)+2 \pi ^2 \log u+2 \pi ^2 \log(1+v)-8 \zeta(3)\bigr)\notag\\
 &\quad{}+\delta ^2 \bigl(\bigl(-12 \log 2 \log u-2 \pi ^2\bigr) \log (1-v)-12 \log 2 \log u \log(1+v)\notag\\
 &\quad{}+\bigl(12 \log 2 \log(1+v)-2 \pi ^2\bigr) \log(1-uv)+\bigl(12 \log 2 \log (1-v)+2 \pi ^2\bigr) \log(1+uv)\notag\\
 &\quad{}+2 \pi ^2 \log(1+v)\bigr)+\delta  \bigl(\bigl(12 \log^2 2-2 \pi ^2\bigr) \log (1-v)+\bigl(-\pi ^2-24 \log^2 2\bigr) \log v\notag\\
 &\quad{}+\bigl(3 \pi ^2+12 \log^2 2\bigr) \log(1+v)\bigr)+26 \zeta(3)+8 \log^3 2+\pi ^2 \log 2\bigr)\label{pp-explicit-I3-4}
\end{align}

\begin{align}
 f^{(3)}_{5}&={}-\ii\pi\delta f^{(2)}_{5}+\frac{\pi^2\delta^2}{2}f^{(1)}_{5}+\frac{\ii\pi^3\delta^3}{6}f^{(0)}_{5}\notag\\
 &\quad{}+8 \delta ^3 (G\bigl(-1,-1,0;u\bigr)-G\bigl(-1,-1,-1;u\bigr))+4 \delta ^2 (2 \delta +1) (G\bigl(-1,0,-1;u\bigr)\notag\\
 &\quad{}-G\bigl(-1,0,0;u\bigr))+4 \delta ^2 (G\bigl(-1,\frac{1}{v},0;u\bigr)-G\bigl(-1,\frac{1}{v},-1;u\bigr))\notag\\
 &\quad{}+8 \delta ^2 (\delta +1) (G\bigl(0,-1,-1;u\bigr)-G\bigl(0,-1,0;u\bigr))+8 \delta  (\delta +1)^2 (-G\bigl(0,0,-1;u\bigr)\notag\\
 &\quad{}+G\bigl(0,0,0;u\bigr)+G\bigl(0,0,0;v\bigr))+8 \delta  (\delta +1) (G\bigl(0,\frac{1}{v},-1;u\bigr)-G\bigl(0,\frac{1}{v},0;u\bigr))\notag\\
 &\quad{}+4 \delta  (G\bigl(-\frac{1}{v},0,-1;u\bigr)-G\bigl(-\frac{1}{v},0,0;u\bigr)-G\bigl(-\frac{1}{v},\frac{1}{v},-1;u\bigr)+G\bigl(-\frac{1}{v},\frac{1}{v},0;u\bigr))\notag\\
 &\quad{}+8 \delta ^2 (G\bigl(\frac{1}{v},-1,0;u\bigr)-G\bigl(\frac{1}{v},-1,-1;u\bigr))+4 \delta  (3 \delta +1) (G\bigl(\frac{1}{v},0,-1;u\bigr)\notag\\
 &\quad{}-G\bigl(\frac{1}{v},0,0;u\bigr))+4 \delta  (\delta +1) (G\bigl(\frac{1}{v},\frac{1}{v},0;u\bigr)-G\bigl(\frac{1}{v},\frac{1}{v},-1;u\bigr))\notag\\
 &\quad{}+8 \delta  (\delta +1)^2 G\bigl(0,0;u\bigr) \log v-4 \delta  G\bigl(-\frac{1}{v},0;u\bigr) (\log v+\log 2)\notag\\
 &\quad{}+8 \delta  (\delta +1)^2 G\bigl(0,0;v\bigr) \bigl(\log \bigl(\frac{u v}{u+1}\bigr)-\log v\bigr)\notag\\
 &\quad{}+\delta  G\bigl(\frac{1}{v},0;u\bigr) (\delta  (-8 \log v-4 \log(1+v))-4 \log v+4 \log 2)\notag\\
 &\quad{}+4 \delta  G\bigl(-\frac{1}{v},-1;u\bigr) (\log v-\log(1+v)+\log 2)+8 \delta ^2 G\bigl(-1,-1;u\bigr) (\delta  \log v+\log v\notag\\
 &\quad{}-\log(1+v))-8 \delta  (\delta +1) G\bigl(0,-1;u\bigr) (\delta  \log v+\log v-\log(1+v))\notag\\
 &\quad{}+4 \delta  G\bigl(\frac{1}{v},-1;u\bigr) (2 \delta  \log v+\log v-\log(1+v)-\log 2)-4 \delta ^2 G\bigl(-1,0;u\bigr) (2 \delta  \log v\notag\\
 &\quad{}+2 \log v-\log(1+v))+4 \delta ^2 G\bigl(-1,\frac{1}{v};u\bigr) \log(1+v)\notag\\
 &\quad{}-8 \delta  (\delta +1) G\bigl(0,\frac{1}{v};u\bigr) \log(1+v)+4 \delta  G\bigl(-\frac{1}{v},\frac{1}{v};u\bigr) \log(1+v)\notag\\
 &\quad{}+4 \delta  (\delta +1) G\bigl(\frac{1}{v},\frac{1}{v};u\bigr) \log(1+v)+4 \delta  (\delta +1) G\bigl(-1,-1;v\bigr) (\log(1-uv)\notag\\
 &\quad{}-\log(1+u))-4 \delta  (2 \delta +1) G\bigl(-1,0;v\bigr) (\log(1-uv)-\log(1+u))\notag\\
 &\quad{}+\delta  G\bigl(0,-1;v\bigr) (\delta  (8 \log(1+u)-8 \log(1-uv))-4 \log(1-uv)-4 \log(1+uv)\notag\\
 &\quad{}+8 \log(1+u))-4 \delta  G\bigl(1,-1;v\bigr) (\log(1+u)-\log(1+uv))\notag\\
 &\quad{}+4 \delta  G\bigl(1,0;v\bigr) (\log(1+u)-\log(1+uv))+\frac{1}{6 (\delta +1)}\bigl[-16 \delta ^4 \zeta(3)-36 \delta ^3 \zeta(3)\notag\\
 &\quad{}+\bigl(24 \delta ^2+24 \delta \bigr) \log 2 \log(1+v) \log(1-uv)+\bigl(\bigl(-24 \delta ^2-24 \delta \bigr) \log 2 \log (1-v)\notag\\
 &\quad{}-4 \pi ^2 \delta ^2-4 \pi ^2 \delta \bigr) \log(1+uv)-24 \delta ^2 \zeta(3)\notag\\
 &\quad{}+\log(1+u) \bigl(4 \pi ^2 \delta ^3+\bigl(24 \delta ^2+24 \delta \bigr) \log 2 \log (1-v)\notag\\
 &\quad{}+\bigl(-24 \delta ^2-24 \delta \bigr) \log 2 \log(1+v)+12 \pi ^2 \delta ^2+8 \pi ^2 \delta \bigr)\notag\\
 &\quad{}+\bigl(-4 \pi ^2 \delta ^3-8 \pi ^2 \delta ^2-4 \pi ^2 \delta \bigr) \log(1-uv)+\bigl(4 \pi ^2 \delta ^4+16 \pi ^2 \delta ^3+25 \pi ^2 \delta ^2\notag\\
 &\quad{}+13 \pi ^2 \delta \bigr) \log \bigl(\frac{u v}{u+1}\bigr)-14 \delta  \zeta(3)\bigr]\label{pp-explicit-I3-5}
\end{align}

\begin{align}
 f^{(3)}_{6}&={}-\ii\pi\delta f^{(2)}_{6}+\frac{\pi^2\delta^2}{2}f^{(1)}_{6}+\frac{\ii\pi^3\delta^3}{6}f^{(0)}_{6}+8 \delta ^3 (-G\bigl(-1,-1,-1;u\bigr)\notag\\
 &\quad{}+G\bigl(-1,-1,0;u\bigr)+G\bigl(0,-1,-1;u\bigr)-G\bigl(0,-1,0;u\bigr))-4 (\delta +1)^2 G\bigl(-1,-1,-1;v\bigr)\notag\\
 &\quad{}+4 (\delta +1) (2 \delta +1) G\bigl(-1,-1,0;v\bigr)+4 \delta ^2 (2 \delta +1) (G\bigl(-1,0,-1;u\bigr)\notag\\
 &\quad{}-G\bigl(-1,0,0;u\bigr)-G\bigl(0,0,-1;u\bigr)+G\bigl(0,0,0;u\bigr))+4 \bigl(2 \delta ^2+3 \delta +2\bigr) G\bigl(-1,0,-1;v\bigr)\notag\\
 &\quad{}-8 \bigl(2 \delta ^2+2 \delta +1\bigr) G\bigl(-1,0,0;v\bigr)+4 (G\bigl(-1,1,0;v\bigr)-G\bigl(-1,1,-1;v\bigr))\notag\\
 &\quad{}+4 \delta ^2 (-G\bigl(-1,\frac{1}{v},-1;u\bigr)+G\bigl(-1,\frac{1}{v},0;u\bigr)+G\bigl(0,\frac{1}{v},-1;u\bigr)-G\bigl(0,\frac{1}{v},0;u\bigr))\notag\\
 &\quad{}+8 (\delta +1)^2 G\bigl(0,-1,-1;v\bigr)-8 (\delta +1) (2 \delta +1) G\bigl(0,-1,0;v\bigr)\notag\\
 &\quad{}-16 (\delta +1)^2 G\bigl(0,0,-1;v\bigr)+8 (\delta +1)^2 (\delta +2) G\bigl(0,0,0;v\bigr)\notag\\
 &\quad{}+8 (\delta +1) (G\bigl(0,1,-1;v\bigr)-G\bigl(0,1,0;v\bigr))+4 (\delta +1) (-G\bigl(1,-1,-1;v\bigr)\notag\\
 &\quad{}-G\bigl(1,1,-1;v\bigr)+G\bigl(1,1,0;v\bigr))+4 (2 \delta +1) G\bigl(1,-1,0;v\bigr)\notag\\
 &\quad{}+4 (3 \delta +2) G\bigl(1,0,-1;v\bigr)-8 (2 \delta +1) G\bigl(1,0,0;v\bigr)+4 G\bigl(-1,1;v\bigr) \log 2\notag\\
 &\quad{}-8 (\delta +1) G\bigl(0,1;v\bigr) \log 2+4 (\delta +1) G\bigl(1,1;v\bigr) \log 2\notag\\
 &\quad{}+G\bigl(1,-1;v\bigr) (\delta  (4 \log u-4 \log(1+u))-4 \log 2)-4 \delta  G\bigl(1,0;v\bigr) (\log u-\log(1+u)\notag\\
 &\quad{}+2 \log 2)-8 \delta  (\delta +1)^2 G\bigl(0,0;v\bigr) (\log(1+u)-\log u)-4 (\delta +1) G\bigl(-1,-1;v\bigr) (\notag\\
 &\quad{}-\delta  \log u+\delta  \log(1+u)+\log 2)+8 (\delta +1) G\bigl(0,-1;v\bigr) (-\delta  \log u+\delta  \log(1+u)\notag\\
 &\quad{}+\log 2)+\delta  G\bigl(-1,0;v\bigr) (\delta  (8 \log(1+u)-8 \log u)-4 \log u+4 \log(1+u)+8 \log 2)\notag\\
 &\quad{}+8 \delta ^2 G\bigl(-1,-1;u\bigr) (\delta  \log v+\log v-\log(1+v))-8 \delta ^2 G\bigl(0,-1;u\bigr) (\delta  \log v+\log v\notag\\
 &\quad{}-\log(1+v))-4 \delta ^2 G\bigl(-1,0;u\bigr) (2 \delta  \log v+2 \log v-\log(1+v))\notag\\
 &\quad{}+4 \delta ^2 G\bigl(0,0;u\bigr) (2 \delta  \log v+2 \log v-\log(1+v))+4 \delta ^2 G\bigl(-1,\frac{1}{v};u\bigr) \log(1+v)\notag\\
 &\quad{}-4 \delta ^2 G\bigl(0,\frac{1}{v};u\bigr) \log(1+v)+\frac{1}{6 (\delta +1)}\bigl[-16 \delta ^4 \zeta(3)-20 \delta ^3 \zeta(3)\notag\\
 &\quad{}+\bigl(-4 \pi ^2 \delta ^2-8 \pi ^2 \delta +(24 \delta +24) \log^2 2-4 \pi ^2\bigr) \log (1-v)-48 \delta ^2 \zeta(3)\notag\\
 &\quad{}+\bigl(-4 \pi ^2 \delta ^3-8 \pi ^2 \delta ^2-10 \pi ^2 \delta +(-24 \delta -24) \log^2 2-6 \pi ^2\bigr) \log(1+v)\notag\\
 &\quad{}+\log(1+u) \bigl(-4 \pi ^2 \delta ^4-12 \pi ^2 \delta ^3+\bigl(24 \delta ^2+24 \delta \bigr) \log 2 \log (1-v)\notag\\
 &\quad{}+\bigl(-24 \delta ^2-24 \delta \bigr) \log 2 \log(1+v)-13 \pi ^2 \delta ^2-5 \pi ^2 \delta \bigr)\notag\\
 &\quad{}+\log u \bigl(4 \pi ^2 \delta ^4+12 \pi ^2 \delta ^3+\bigl(-24 \delta ^2-24 \delta \bigr) \log 2 \log (1-v)\notag\\
 &\quad{}+\bigl(24 \delta ^2+24 \delta \bigr) \log 2 \log(1+v)+13 \pi ^2 \delta ^2+5 \pi ^2 \delta \bigr)\notag\\
 &\quad{}+\bigl(4 \pi ^2 \delta ^4+16 \pi ^2 \delta ^3+25 \pi ^2 \delta ^2+23 \pi ^2 \delta +10 \pi ^2\bigr) \log v-62 \delta  \zeta(3)\notag\\
 &\quad{}-28 \zeta(3)\bigr]\label{pp-explicit-I3-6}
\end{align}

\begin{align}
 f^{(3)}_{7}&={}2 \delta  (-G\bigl(-\frac{1}{v},0,-1;u\bigr)+G\bigl(-\frac{1}{v},0,0;u\bigr)+G\bigl(-\frac{1}{v},\frac{1}{v},-1;u\bigr)-G\bigl(-\frac{1}{v},\frac{1}{v},0;u\bigr)\notag\\
 &\quad{}+G\bigl(\frac{1}{v},0,-1;u\bigr)-G\bigl(\frac{1}{v},0,0;u\bigr)-G\bigl(\frac{1}{v},\frac{1}{v},-1;u\bigr)+G\bigl(\frac{1}{v},\frac{1}{v},0;u\bigr))\notag\\
 &\quad{}+2 \delta  G\bigl(-\frac{1}{v},0;u\bigr) (\log v+\log 2)-2 \delta  G\bigl(\frac{1}{v},0;u\bigr) (\log v+\log 2)\notag\\
 &\quad{}-2 \delta  G\bigl(-\frac{1}{v},-1;u\bigr) (\log v-\log(1+v)+\log 2)+2 \delta  G\bigl(\frac{1}{v},-1;u\bigr) (\log v-\log(1+v)\notag\\
 &\quad{}+\log 2)-2 \delta  G\bigl(-\frac{1}{v},\frac{1}{v};u\bigr) \log(1+v)+2 \delta  G\bigl(\frac{1}{v},\frac{1}{v};u\bigr) \log(1+v)\notag\\
 &\quad{}+2 \delta  G\bigl(-1,-1;v\bigr) (\log(1-uv)-\log(1+u))-2 \delta  G\bigl(-1,0;v\bigr) (\log(1-uv)\notag\\
 &\quad{}-\log(1+u))+2 \delta  G\bigl(1,-1;v\bigr) (\log(1+u)-\log(1+uv))\notag\\
 &\quad{}-2 \delta  G\bigl(1,0;v\bigr) (\log(1+u)-\log(1+uv))-2 \delta  G\bigl(0,-1;v\bigr) (\log(1-uv)\notag\\
 &\quad{}-\log(1+uv))-\frac{1}{3} \delta  \bigl(6 \log 2 \log(1+u) \log (1-v)-6 \log 2 \log (1-v) \log(1+uv)\notag\\
 &\quad{}-6 \log 2 \log(1+u) \log(1+v)+6 \log 2 \log(1+v) \log(1-uv)+\pi ^2 \log(1-uv)\notag\\
 &\quad{}-\pi ^2 \log(1+uv)\bigr)\label{pp-explicit-I3-7}
\end{align}

\begin{align}
 f^{(3)}_{8}&={}-\ii\pi\delta f^{(2)}_{8}+2 (2 \delta +3) G\bigl(-1,-1,-1;v\bigr)-6 (\delta +1) G\bigl(-1,-1,0;v\bigr)\notag\\
 &\quad{}+2 (3 \delta +2) (G\bigl(1,0,-1;v\bigr)-G\bigl(-1,0,-1;v\bigr))+4 (2 \delta +1) (G\bigl(-1,0,0;v\bigr)\notag\\
 &\quad{}-G\bigl(1,0,0;v\bigr))+2 (G\bigl(-1,1,0;v\bigr)-G\bigl(-1,1,-1;v\bigr))+4 (\delta +2) (-G\bigl(0,-1,-1;v\bigr)\notag\\
 &\quad{}+G\bigl(0,-1,0;v\bigr)+G\bigl(0,1,-1;v\bigr)-G\bigl(0,1,0;v\bigr))-2 (\delta -1) G\bigl(1,-1,-1;v\bigr)\notag\\
 &\quad{}+2 (2 \delta -1) G\bigl(1,-1,0;v\bigr)+2 (\delta +3) (G\bigl(1,1,0;v\bigr)-G\bigl(1,1,-1;v\bigr))\notag\\
 &\quad{}+2 G\bigl(-1,1;v\bigr) \log 2+4 (\delta +2) G\bigl(0,-1;v\bigr) \log 2-4 (\delta +2) G\bigl(0,1;v\bigr) \log 2\notag\\
 &\quad{}+2 (\delta +3) G\bigl(1,1;v\bigr) \log 2+G\bigl(1,-1;v\bigr) (\delta  (2 \log u-2 \log(1+u))-6 \log 2)\notag\\
 &\quad{}+2 \delta  G\bigl(-1,0;v\bigr) (\log u-\log(1+u)+2 \log 2)-2 \delta  G\bigl(1,0;v\bigr) (\log u-\log(1+u)\notag\\
 &\quad{}+2 \log 2)+G\bigl(-1,-1;v\bigr) (\delta  (-2 \log u+2 \log(1+u)-2 \log 2)-2 \log 2)\notag\\
 &\quad{}+\frac{1}{3} (\log (1-v)-\log(1+v)) \bigl(-6 \delta  \log 2 \log u+6 \delta  \log 2 \log(1+u)-\pi ^2 \delta -3 \pi ^2\notag\\
 &\quad{}+6 \log^2 2\bigr)\label{pp-explicit-I3-8}
\end{align}

\begin{align}
 f^{(3)}_{9}&={}-\ii\pi\delta f^{(2)}_{9}+\frac{\pi^2\delta^2}{2}f^{(1)}_{9}+\frac{\ii\pi^3\delta^3}{6}f^{(0)}_{9}+2 \delta  (\delta +2)^2 G\bigl(-1,-1,-1;v\bigr)\notag\\
 &\quad{}+4 \delta  (\delta +1) (\delta +2) (-G\bigl(-1,-1,0;v\bigr)-G\bigl(-1,0,-1;v\bigr))\notag\\
 &\quad{}+8 \delta  (\delta +1)^2 G\bigl(-1,0,0;v\bigr)+4 \delta ^3 (G\bigl(0,-1,-1;u\bigr)-G\bigl(0,-1,0;u\bigr)\notag\\
 &\quad{}-G\bigl(\frac{1}{v},-1,-1;u\bigr)+G\bigl(\frac{1}{v},-1,0;u\bigr))-4 \delta  \bigl(\delta ^2+4 \delta +2\bigr) G\bigl(0,-1,-1;v\bigr)\notag\\
 &\quad{}+8 \delta  \bigl(\delta ^2+3 \delta +1\bigr) G\bigl(0,-1,0;v\bigr)+2 \delta ^2 (3 \delta +4) (G\bigl(0,0,0;u\bigr)-G\bigl(0,0,-1;u\bigr))\notag\\
 &\quad{}+8 \delta  \bigl(\delta ^2+4 \delta +2\bigr) (G\bigl(0,0,-1;v\bigr)-G\bigl(0,0,0;v\bigr))+8 \delta  (\delta +1) (G\bigl(0,1,0;v\bigr)\notag\\
 &\quad{}-G\bigl(0,1,-1;v\bigr))+2 \delta ^2 (\delta +4) (G\bigl(0,\frac{1}{v},-1;u\bigr)-G\bigl(0,\frac{1}{v},0;u\bigr))\notag\\
 &\quad{}+4 \delta ^2 (G\bigl(1,-1,-1;v\bigr)+G\bigl(-\frac{1}{v},0,-1;u\bigr)-G\bigl(-\frac{1}{v},0,0;u\bigr)-G\bigl(-\frac{1}{v},\frac{1}{v},-1;u\bigr)\notag\\
 &\quad{}+G\bigl(-\frac{1}{v},\frac{1}{v},0;u\bigr))-8 \delta ^2 G\bigl(1,-1,0;v\bigr)-4 \delta  (3 \delta +2) G\bigl(1,0,-1;v\bigr)\notag\\
 &\quad{}+8 \delta  (2 \delta +1) G\bigl(1,0,0;v\bigr)+4 \delta  (\delta +2) (G\bigl(1,1,-1;v\bigr)-G\bigl(1,1,0;v\bigr))\notag\\
 &\quad{}+2 \delta ^2 (3 \delta +2) (G\bigl(\frac{1}{v},0,-1;u\bigr)-G\bigl(\frac{1}{v},0,0;u\bigr))+2 \delta ^2 (\delta +2) (G\bigl(\frac{1}{v},\frac{1}{v},0;u\bigr)\notag\\
 &\quad{}-G\bigl(\frac{1}{v},\frac{1}{v},-1;u\bigr))-8 \delta ^2 G\bigl(0,0;v\bigr) \log 2+8 \delta  (\delta +1) G\bigl(0,1;v\bigr) \log 2\notag\\
 &\quad{}-4 \delta  (\delta +2) G\bigl(1,1;v\bigr) \log 2-4 \delta ^2 G\bigl(-\frac{1}{v},0;u\bigr) (\log v+\log 2)\notag\\
 &\quad{}+\delta ^2 G\bigl(\frac{1}{v},0;u\bigr) (\delta  (-4 \log v-2 \log(1+v))-4 \log v)-4 \delta ^2 G\bigl(0,-1;u\bigr) (\delta  \log v\notag\\
 &\quad{}+2 \log v-2 \log(1+v)+\log 2)+4 \delta ^2 G\bigl(-\frac{1}{v},-1;u\bigr) (\log v-\log(1+v)+\log 2)\notag\\
 &\quad{}+4 \delta ^2 G\bigl(\frac{1}{v},-1;u\bigr) (\delta  \log v+\log v-\log(1+v))-2 \delta ^2 (\delta +4) G\bigl(0,\frac{1}{v};u\bigr) \log(1+v)\notag\\
 &\quad{}+4 \delta ^2 G\bigl(-\frac{1}{v},\frac{1}{v};u\bigr) \log(1+v)+2 \delta ^2 (\delta +2) G\bigl(\frac{1}{v},\frac{1}{v};u\bigr) \log(1+v)\notag\\
 &\quad{}+\delta ^2 G\bigl(0,0;u\bigr) (\delta  (4 \log v+2 \log(1+v))+8 \log v+4 \log 2)\notag\\
 &\quad{}+2 \delta ^2 (\delta +2) G\bigl(-1,-1;v\bigr) (\log(1-uv)-\log u)\notag\\
 &\quad{}-4 \delta ^2 (\delta +1) G\bigl(-1,0;v\bigr) (\log(1-uv)-\log u)\notag\\
 &\quad{}+\delta  G\bigl(0,-1;v\bigr) (\delta  (\delta  (4 \log u-4 \log(1-uv))-4 \log(1-uv)-4 \log(1+uv)\notag\\
 &\quad{}+8 \log u)-8 \log 2)+4 \delta ^2 G\bigl(1,0;v\bigr) (-\log(1+uv)+\log u+2 \log 2)\notag\\
 &\quad{}+\delta  G\bigl(1,-1;v\bigr) (\delta  (4 \log(1+uv)-4 \log u)+8 \log 2)\notag\\
 &\quad{}+\frac{1}{6 (\delta +1)}\bigl[\delta ^4 \bigl(-2 \pi ^2 \log(1-uv)+2 \pi ^2 \log u+2 \pi ^2 \log(1+v)-8 \zeta(3)\bigr)\notag\\
 &\quad{}+\delta ^3 \bigl(\bigl(24 \log 2 \log u+4 \pi ^2\bigr) \log (1-v)-6 \pi ^2 \log(1-uv)\notag\\
 &\quad{}+\bigl(-24 \log 2 \log (1-v)-4 \pi ^2\bigr) \log(1+uv)+10 \pi ^2 \log u+6 \pi ^2 \log(1+v)\notag\\
 &\quad{}+32 \zeta(3)\bigr)+\delta ^2 \bigl(\bigl(24 \log 2 \log u+12 \pi ^2-24 \log^2 2\bigr) \log (1-v)-4 \pi ^2 \log(1-uv)\notag\\
 &\quad{}+\bigl(-24 \log 2 \log (1-v)-4 \pi ^2\bigr) \log(1+uv)+8 \pi ^2 \log u+\bigl(24 \log^2 2-17 \pi ^2\bigr) \log v\notag\\
 &\quad{}+13 \pi ^2 \log(1+v)+48 \zeta(3)\bigr)+\delta  \bigl(\bigl(8 \pi ^2-24 \log^2 2\bigr) \log (1-v)\notag\\
 &\quad{}+\bigl(24 \log^2 2-17 \pi ^2\bigr) \log v+9 \pi ^2 \log(1+v)-70 \zeta(3)-8 \log^3 2+7 \pi ^2 \log 2\bigr)\notag\\
 &\quad{}-98 \zeta(3)-8 \log^3 2+7 \pi ^2 \log 2\bigr]\label{pp-explicit-I3-9}
\end{align}

\begin{align}
 f^{(3)}_{10}&={}-\ii\pi\delta f^{(2)}_{10}+4 \delta  (G\bigl(0,-1,-1;v\bigr)-G\bigl(0,-1,0;v\bigr)-G\bigl(0,1,-1;v\bigr)\notag\\
 &\quad{}+G\bigl(0,1,0;v\bigr))-4 \delta  G\bigl(0,-1;v\bigr) \log 2+4 \delta  G\bigl(0,1;v\bigr) \log 2-\pi ^2 (\delta  \log v+\log 2)\label{pp-explicit-I3-10}
\end{align}

\endgroup

The expressions follow from the finite word sum
\eqref{pp-finite-gpl-word-sum} by collecting identical words and equal
coefficients.

\end{document}